\documentclass[11pt]{article}
\pdfoutput=1

\usepackage{jheppub}
\usepackage[T1]{fontenc}
\usepackage[normalem]{ulem}
\usepackage{jheppub}
\usepackage{url,comment}
\usepackage{times}
\usepackage{latexsym}
\usepackage{graphicx, graphics, hyperref, amsmath, amssymb, slashed, color, bbm,amsthm, array}
\usepackage[usenames,dvipsnames]{xcolor}
 \usepackage{subfigure}
\usepackage{mdwlist}
\usepackage{multirow}
\usepackage[e]{esvect}
\usepackage{slashed}

\usepackage{cancel}
\usepackage{pstricks}
\usepackage[toc,page]{appendix}
\usepackage{enumerate}
\usepackage{ascmac}

\usepackage{comment}

\allowdisplaybreaks

\title{
Consistency of EFT illuminated via relative entropy: 
\\
A case study in scalar field theory
}
\author{
  Daiki Ueda$^{1,2}$, and Kazuhiro Tatsumi$^{3}$
}
\affiliation{\vspace{2mm} 
$^1$Physics Department, Technion \text{--} Israel Institute of Technology,
Technion city, Haifa 3200003, Israel
\\
$^2$Theoretical Physics Department, CERN, 1211 Geneva 23, Switzerland
\\
$^3$
Mizuho Research \& Technologies, Ltd., 2-3 Kanda-Nishikicho, Chiyoda-ku, Tokyo, 101-8443, Japan 
}
\abstract{
Relative entropy is a non-negative quantity and offers a powerful means of achieving a unified understanding of fundamental properties in physics, including the second law of thermodynamics and positivity bounds on effective field theories (EFTs).
We analyze the relative entropy in scalar field theories and show that the non-negativity of relative entropy is potentially violated in perturbative calculations based on operator and loop expansions.
Conversely, this suggests that the consistency of the EFT description in the scalar field theory can be identified by the sign of the relative entropy.
In fact, we revisit an EFT of single-field inflation and present a relation between its non-linear parameter $f_{\rm NL}$ and the consistency condition of the EFT description derived from the relative entropy method. 
We find that interesting regions of $f_{\rm NL}$ that are observationally allowed can be constrained from the relative entropy by imposing the consistency of the EFT description when the EFT is generated via the interaction with heavy fields in UV theories.
}
\emailAdd{daiki.ueda@cern.ch}
\emailAdd{kazuhiro.tatsumi@mizuho-rt.co.jp}

\begin{document}
\maketitle

\section{Introduction}
\label{sec:introduction}
One objective of scientific inquiry is to provide accurate predictions of physical phenomena. To this end, essential properties of physical theories, such as unitarity, causality, locality, and symmetry, are often employed to confirm the consistency of these predictions~\cite{Lee:1977eg,PhysRevLett.38.883,PhysRevD.7.3111,Chanowitz:1985hj,Adams:2006sv}.
Among them, the unitarity frequently appears in particle physics because the $S$-matrix of the particle scattering process satisfies it.
Therefore, unitarity is commonly utilized to make significant implications for various phenomenologies. This includes predicting new physics scales via perturbative unitarity~\cite{Lee:1977eg,PhysRevLett.38.883,Corbett:2014ora,Chang:2019vez} and constraining the sign of Wilson coefficients in various effective field theories (EFTs)~\cite{Adams:2006sv,Remmen:2019cyz,Remmen:2020vts,Remmen:2020uze,Pham:1985cr,Ananthanarayan:1994hf,Pennington:1994kc}.
Some examples of highly successful applications include the prediction of the Higgs boson mass as a scale of unitarity violation~\cite{Lee:1977eg,PhysRevLett.38.883,PhysRevD.7.3111,Chanowitz:1985hj}, estimations of the scale of the new physics realizing non-Gaussianity~\cite{Maldacena:2002vr} with $|f_{\rm NL}|\gtrsim 1$ in inflation models~\cite{Grall:2020tqc,Kim:2021pbr}, and positivity bounds on the Wilson coefficients in the chiral perturbation theory~\cite{Pham:1985cr,Ananthanarayan:1994hf,Pennington:1994kc}, the Standard Model EFT~\cite{Remmen:2019cyz,Remmen:2020vts,Remmen:2020uze}, and the EFT of inflation~\cite{Cheung:2007st,Baumann:2015nta,Grall:2021xxm,Freytsis:2022aho}.

The instances of success mentioned above are grounded in the unitarity of the $S$-matrix.
However, the unitarity of physical theories could manifest in quantities other than the $S$-matrix. 
A new method \cite{Cao:2022iqh,Cao:2022ajt} has been proposed to assess the consistency of physical quantities, such as Wilson coefficients of EFTs, with underlying UV theories.
This method focuses on the non-negative relative entropy~\cite{10.1214/aoms/1177729694,10.2996/kmj/1138844604,RevModPhys.50.221} that arises from the unitarity of the theory. 
The relative entropy, also known as Kullback-Leibler divergence, is a non-negative quantity defined by two probability distribution functions. It is zero only when the two probability distribution functions are equal.
The relative entropy, in brief, measures the disparity between two probability distributions. 
In Refs.~\cite{Cao:2022ajt,Cao:2022iqh}, it was suggested that the relative entropy can be utilized to quantify the disparity between theories with and without interactions between heavy and light degrees of freedom, {\it i.e.,} a transfer of UV information to EFTs consisting of light degrees of freedom through the interactions. 
There, a canonical distribution function is considered for the probability distribution function of the theory. In this case, the non-negativity of relative entropy holds when the Hamiltonian of systems respects Hermiticity, {\it i.e.}, unitarity for the time evolution of systems. 
Refs.~\cite{Cao:2022ajt,Cao:2022iqh} indicate that the non-negativity of relative entropy and the consideration of the $S$-matrix lead to several common outcomes, such as positivity bounds on the SMEFT gauge bosonic operator~\cite{Remmen:2019cyz} and adherence to the mild weak gravity conjecture~\cite{Vafa:2005ui,Arkani-Hamed:2006emk,Kats:2006xp,Cheung:2018cwt,Cheung:2019cwi,Loges:2019jzs,Reall:2019sah,Goon:2019faz,Bellazzini:2019xts,Hamada:2018dde,Arkani-Hamed:2021ajd,Bittar:2024xuc}. 
One notable aspect of this relative entropy consideration is that it presents constraints on the EFT in a different approach than the $S$ matrix consideration. 
This motivates applying entropy considerations to various phenomena to reveal possible connections and differences between relative entropy considerations and $S$-matrix considerations and to understand when relative entropy methods can have more vital implications.

For several reasons, EFTs consisting of scalar fields could be candidates for adapting the relative entropy method. 
The first reason is that while scalar field theories are relevant to exciting phenomena such as inflation and phase transitions, considerations of relative entropy have not yet been comprehensively adapted. 
As will be explained later, the EFT consisting of scalar fields provides an excellent example of considering a relation between the effectiveness of perturbative calculations and the unitarity appearing as non-negativity of relative entropy, because the tree-level effects may be more dominant in the corrections from UV theory to the EFT\footnote{For instance, loop-level effects may dominate in the EFT derived from a UV theory, where fermion fields are the only heavy degrees of freedom.
However, if heavy scalar fields are present in the UV theory, tree-level effects from heavy scalar fields dominate under naive loop expansions.
See Sec.~\ref{sec:weak} for details.
}.
In the scattering amplitude considerations, the breakdown of perturbative calculations (where the validity of such calculations is often judged by the negligible contribution of loop-level terms relative to tree-level terms) can lead to a violation of inequalities stemming from the unitarity. However, the relation between the breakdown of the non-negativity of relative entropy and the breakdown of perturbative calculations is still unclear. It will be essential to clarify the relation and its phenomenological consequences in the theory of scalar fields.  
Second, inflation models described by EFTs involving scalar fields would provide a valuable example to validate the applicability of relative entropy calculated for the background field, especially since the inflationary background field violates Lorentz invariance.
For instance, in conventional positivity bounds, Lorentz symmetry is an essential factor in determining the behavior of the scattering amplitude, and several techniques have been made in previous studies to derive bounds on the inflationary EFT~\cite{Baumann:2015nta,Grall:2021xxm,Freytsis:2022aho}. 
On the other hand, the Lorentz invariance is not an essential factor in the relative entropy method, as it has been adapted to non-relativistic spin systems, for instance. 
Hence, the relative entropy method might be more potent in inflation models\footnote{The purity for a cosmological model has also been studied in past works, {\it e.g.}, Ref.~\cite{Colas:2024ysu}.}. 
Third, it has been pointed out that in a single-field inflationary EFT, assuming a particular class of UV theories~\cite{Kim:2021pbr} (two scalar field theories), the scattering amplitudes at high energy scales violate the unitarity. 
The single-field inflationary EFT model~\cite{Garriga:1999vw,Chen:2006nt} is one of the attractive models because of its potential to generate observable non-Gaussianity. 
On the other hand, it is pointed out that if one considers the two scalar field theories as UV theories of their EFTs, the non-Gaussianity of $|f_{\rm NL}|\gtrsim 1$ can only be realized by turning on interactions that violate the unitarity at high energy scales.
Revisiting these point with considerations of relative entropy would lead to understanding the relation and differences between considerations of relative entropy and perturbative unitarity based on scattering amplitudes. 
For the above reasons, state-of-the-art research would encourage considering entropy in scalar field EFTs, including inflation models.

This paper examines the role of relative entropy in scalar field theories with and without interactions between heavy and light fields. We will demonstrate that the non-negativity of relative entropy can be violated in perturbative calculations based on operator and loop expansions.
As a byproduct, we highlight that the relative entropy can be utilized to identify theories for which the perturbative calculations are invalid. 
As will be explained, the EFT expansions are valid for EFTs derived from weakly coupled theories, and the sign of the relative entropy can help determine the theories for which the EFT description is valid. 
When applying this concept to two-scalar inflation, we consider two typical cases: one with a renormalizable potential respecting U(1) symmetry and another with a modified potential.
We observe that the former U(1)-symmetric case corresponds to a weakly coupled theory, whereas the modified case can be interpreted as a strongly coupled one.
Furthermore, we provide constraints on the Wilson coefficients, focusing on higher-dimensional operators in a specific EFT of a single scalar field derived from general weakly coupled theories. Comparing these constraints with the parameter domain suggested by the non-linear parameter $f_{\rm NL}$ allowed by observations~\cite{Planck:2019kim}, we argue that the sign of the non-linear parameter $f_{\rm NL}$ can be severely constrained under the valid EFT description.

The rest of the paper is organized as follows.
In Sec.~\ref{sec:entropy}, we briefly review Refs.~\cite{Cao:2022iqh,Cao:2022ajt} and then illustrate how to calculate the relative entropy in three typical cases of scalar field theories.
In particular, the definition of the weakly coupled theory related to the valid EFT description is presented in Sec.~\ref{sec:weak}.
In Sec.~\ref{sec:quasi}, we apply the results of Sec.~\ref{sec:entropy} to an EFT of single-field inflation, and consequences of the non-negativity of relative entropy are examined numerically and analytically in three UV theories, including a general theory, with an emphasis on the validity of the EFT description.
In Sec.~\ref{sec:non-G}, as a phenomenological application example of results of Sec.~\ref{sec:entropy}, we present a relation between the non-linear parameter in the EFT of single-field inflation and the parameter regions where the relative entropy indicates the validity of the EFT description.
%
%
The main findings and outlook of our analysis are summarized in Sec.~\ref{sec:summary}.
The Appendix contains procedures of field redefinitions to make calculations easier (Appendix~\ref{app:red}), perturbative calculations of $\widetilde{\phi}_{1,\lambda}$ (Appendix~\ref{sec:phi1}), the calculations of the effective actions up to the sixth order of the perturbative expansion (Appendix~\ref{app:B}), and a derivation of eq.~\eqref{eq:delWgN} (Appendix~\ref{sec:delv}), as well as a relation between different bases of field (Appendix~\ref{app:bases}).

\section{Preliminaries}
\label{sec:entropy}
We will commence with a brief review of Refs.~\cite{Cao:2022iqh,Cao:2022ajt} to enhance our understanding of the forthcoming sections. 
Then, we will illustrate how to calculate the relative entropy in theories where perturbative calculations are valid (detailed conditions of the valid perturbation theory are summarized in Sec.~\ref{sec:weak}) with a specific focus on scalar field theories.
Drawing upon the results and employing a methodology based on relative entropy~\cite{Cao:2022iqh,Cao:2022ajt}, we aim to assess whether the EFT originates from a UV theory and to determine the parameter regions within which the EFT description remains valid.

\subsection{Non-negativity of relative entropy arising in unitary theories}
The relative entropy, defined by two probability distribution functions $\rho_{\rm R}$ and $\rho_{\rm T}$, is given as:
\begin{align}
    S(\rho_{\rm R}||\rho_{\rm T}):={\rm Tr}\left[
    \rho_{\rm R}\ln \rho_{\rm R}-\rho_{\rm R} \ln \rho_{\rm T}
    \right],\label{eq:rel0}
\end{align}
where $\rho_{\rm R}$ and $\rho_{\rm T}$ denote the probability distribution functions and $\rho_{\rm R}=\rho_{\rm R}^{\dagger}$, $\rho_{\rm T}=\rho_{\rm T}^{\dagger}$, and ${\rm Tr}\left[\rho_{\rm R}\right]={\rm Tr}\left[\rho_{\rm T}\right]=1$ hold. 
One of the essential properties of relative entropy is that it is non-negative: 
\begin{align}
    S(\rho_{\rm R}||\rho_{\rm T})\geq 0,\label{eq:non}
\end{align}
where the equality holds if and only if $\rho_{\rm R}=\rho_{\rm T}$.
In reference to eq.~\eqref{eq:non}, the relative entropy serves as a measure of information regarding the distinction between two entities as defined by $\rho_{\rm R}$ and $\rho_{\rm T}$.

In Refs.~\cite{Cao:2022ajt, Cao:2022iqh}, a proposal was introduced to quantify the transfer of information from a UV theory to an EFT using the relative entropy. 
We adhere to Refs.~\cite{Cao:2022ajt, Cao:2022iqh} and consider a UV theory defined by a Euclidean action, 
\begin{align}
I[\phi,\Phi]:= I_0[\phi,\Phi]+I_{\rm I}[\phi,\Phi],\label{eq:I}
\end{align}
where $\Phi$ denotes a heavy field (corresponding to a heavy particle), $\phi$ represents an external or light field (associated with a light particle), the symbol $I_{\rm I}$ signifies an interaction between $\Phi$ and $\phi$, and $I_0$ does not encompass this interaction. 
To facilitate future explanations, we will introduce a parameter $\lambda$ that characterizes the interaction in the following manner: 
\begin{align}
I_\lambda[\phi,\Phi]:=I_0[\phi,\Phi]+\lambda I_{\rm I}[\phi,\Phi].\label{eq:Ig}
\end{align}
This parameter $\lambda$ is introduced to explicitly specify the order of perturbative expansion in the interaction $I_{\rm I}$, and therefore, it is permissible to select $\lambda=1$ subsequent to the calculation of the relative entropy. 
The information from the UV theory is conveyed to an EFT through the interaction $I_{\rm I}$ by integrating $\Phi$, and the disparity between theories with and without the interaction $I_{\rm I}$ signifies the transferred information. 
In Refs~\cite{Cao:2022ajt,Cao:2022iqh}, the relative entropy between two theories is calculated for two probability distribution functions, which are defined as follows: 
\begin{align}
P_{\rm R}[\phi,\Phi]:=\frac{e^{-I_0[\phi,\Phi]}}{Z_0},~~~P_{\rm T}[\phi,\Phi]:=\frac{e^{-I_\lambda[\phi,\Phi]}}{Z_\lambda}.\label{eq:Prob}
\end{align}
The partition functions are defined as
\begin{align}
    Z_{0}:=\int d[\phi]d[\Phi]e^{-I_0[\phi,\Phi]},~~~Z_{\lambda}:=\int d[\phi]d[\Phi]e^{-I_\lambda[\phi,\Phi]},\label{eq:part}
\end{align}
where the path integral for $\phi$ is performed if $\phi$ is dynamical, but not in the case of an external field (please look at Sec.~\ref{sec:sub23} for specific details). 
In this context, if the theory complies with unitarity, {\it i.e.}, if the actions $I_0$ and $I_{\text{I}}$ are Hermitian, and if considerations of a low energy scale, such as heavy degrees of freedom not being dynamical\footnote{After integrating out those heavy fields, the effects of dynamical heavy fields can manifest as an imaginary contribution to the effective action at high energy scales, potentially violating the non-negativity condition of relative entropy.
To avoid these effects, we focus on a low energy regime where those heavy fields are no longer dynamical.
}, are taken into account, the conditions $P_{\rm R}=P_{\rm R}^{\dagger}$ and $P_{\rm T}=P_{\rm T}^{\dagger}$ are satisfied.
The non-negativity of relative entropy then follows from this premise. 
In short, the non-negativity of relative entropy under the probability distribution functions~\eqref{eq:Prob} corresponds to the concept of unitarity. 
Considering the significance of this point, we will examine this behavior from a quantum mechanical standpoint later.

According to eqs.~\eqref{eq:rel0} and \eqref{eq:Prob}, the relative entropy between $P_{\rm R}$ and $P_{\rm T}$ (representing the information transferred from the UV theory~\eqref{eq:Ig} to the EFT) is calculated as follows:
\begin{align}
    S(P_{\rm R}||P_{\rm T}):&=\int d[\phi]d[\Phi]\left[
    P_{\rm R}[\phi,\Phi]\ln P_{\rm R}[\phi,\Phi]- P_{\rm R}[\phi,\Phi]\ln P_{\rm T}[\phi,\Phi] 
    \right]\notag
    \\
    &=-\ln Z_0 +\ln Z_\lambda +\lambda \int d[\phi]d[\Phi] P_{\rm R}[\phi,\Phi] I_{\rm I}[\phi,\Phi]\notag
    \\
    &=W_0-W_\lambda+\lambda\left(\frac{\partial W_\lambda}{\partial \lambda}\right)_{\lambda=0}\geq 0,\label{eq:relFI}
\end{align}
where  $W_0:=-\ln Z_0$ and $W_\lambda:=-\ln Z_\lambda$ represent Euclidean effective actions, and we performed several straightforward calculations, $\ln P_{\rm R}=-I_0-\ln Z_0$, $\ln P_{\rm T}=-I_\lambda-\ln Z_\lambda$, $\lim_{\lambda\to 0}P_{\rm R}=P_{\rm T}$, and used the non-negativity of relative entropy~\eqref{eq:non}.
The last term in eq.~\eqref{eq:relFI} is a partial derivative concerning $\lambda$ while keeping $\phi$ and $\Phi$ constant (several examples will be presented in Sec.~\ref{sec:sub23} and Appendix~\ref{sec:delv}).
From eq.~\eqref{eq:relFI}, it was found that the relative entropy can be expressed in terms of the effective actions, and the calculation of the relative entropy, therefore, reduces to the calculation of the effective action.
In the relative entropy calculations of this work, we assume low-energy regions where heavy degrees of freedom are non-dynamical. Consequently, the effective action $W_\lambda$ indicates the EFT corresponding to $I_\lambda$ within the low-energy domain. 
In light of this observation, the inequality~\eqref{eq:relFI} means that the low-energy EFT of the theory $I_\lambda$ must maintain a specific relation (please look at Refs.~\cite{Cao:2022ajt,Cao:2022iqh} and Sec.~\ref{sec:quasi} for further details). 
In the regime of perturbative theories, each term of the EFT can be determined by, {\it e.g.}, matching the scattering amplitudes, and similarly, the relative entropy is also calculable perturbatively. 
The effective actions within this work are calculated under specific external background fields or light field solutions, as will be demonstrated later with several examples.
In Sec.~\ref{sec:quasi}, we will calculate the effective actions within an inflationary background and examine the implications of eq.~\eqref{eq:relFI} in a specific class of EFT of inflation.

Now, we attempt to depict eq.~\eqref{eq:relFI} in a quantum mechanical framework to elucidate the connection between the non-negativity of relative entropy and the unitarity more explicitly. 
Let us define a Hamiltonian of a given UV theory corresponding to eq.~\eqref{eq:Ig} as follows: 
\begin{align}
    H_\lambda:= H_0 +\lambda H_{\rm I},
\end{align}
where $H_0$ and $H_{\rm I}$ denote Hamiltonians corresponding to $I_0$ and $I_{\rm I}$, respectively.
Using partition functions $Z_0:={\rm Tr}[e^{-\beta H_0}]$ and $Z_\lambda:={\rm Tr}[e^{-\beta H_\lambda}]$, we define two probability distribution functions for systems at an inverse temperature $\beta$ as $\rho_{\rm R}:=e^{-\beta H_0}/Z_0$ and $\rho_{\rm T}:=e^{-\beta H_\lambda}/Z_\lambda$. 
Then, the Hermiticity of the probability distribution functions ($\rho_{\rm R}^{\dagger}=\rho_{\rm R}$ and $\rho_{\rm T}^{\dagger}=\rho_{\rm T}$) is ensured through the Hermiticity of two Hamiltonians ($H_0^{\dagger}=H_0$ and $H_\lambda^{\dagger}=H_\lambda$), {\it i.e.}, unitary time evolutions of the systems described by $H_0$ and $H_\lambda$. 
Upon substituting these probability distribution functions $\rho_{\rm R}$ and $\rho_{\rm T}$ into the definition of the relative entropy, the following expression is derived:
\begin{align}
    S(\rho_{\rm R}||\rho_{\rm T}):&={\rm Tr}\left[
    \rho_{\rm R}\ln \rho_{\rm R}- \rho_{\rm R}\ln \rho_{\rm T}
    \right]\notag
    \\
    &=-\ln Z_0 +\ln Z_\lambda +\lambda{\rm Tr}\left[\rho_{\rm R}\beta H_{\rm I}\right]\notag
    \\
    &=W_0 -W_\lambda +\lambda \left(\frac{\partial W_\lambda}{\partial \lambda}\right)_{\lambda=0}\geq 0.\label{eq:quanrel}
\end{align}
In the derivation above, we utilized $\rho_{\rm R}^{\dagger}=\rho_{\rm R}$, $\rho_{\rm T}^{\dagger}=\rho_{\rm T}$, and ${\rm Tr}\,[\rho_{\rm R}]={\rm Tr}\,[\rho_{\rm T}]=1$.
Although the derivation of this inequality~\eqref{eq:quanrel} is a reformulation of eq.~\eqref{eq:relFI}, it becomes more explicit that the non-negativity of relative entropy represents the unitarity. 

It is also worth mentioning that the conditions of the probability distribution function ${\rm Tr}\,[\rho_{\rm R}]={\rm Tr}\,[\rho_{\rm T}]=1$ must be maintained consistently over time due to the conservation of probability through the unitarity. 
Let us consider the quantum state of the interacting theory at time zero, denoted as $\rho_{\rm T}(0):=\rho_{\rm T}$. 
In this case, the quantum state of this interacting theory at time $t$ becomes $\rho_{\rm T}(t)=e^{-i H_{\lambda} t}\rho_{\rm T}(0)e^{+i H_{\lambda}^{\dagger} t}$. 
The time evolution of this system described by $H_{\lambda}$ is elaborated as follows: 
\begin{align}
	i\frac{d \rho_{\rm T}(t)}{dt}=[h_{\rm T}, \rho_{\rm T}(t)]+i \{ \gamma_{\rm T},\rho_{\rm T}(t)\},\label{eq:consPr}
\end{align}
where we defined $h_{\rm T}:=(H_{\lambda}+H_{\lambda}^{\dagger})/2$, and $\gamma_{\rm T}:=i(H_{\lambda}-H_{\lambda}^{\dagger})/2$. 
In the context of theories respecting the unitarity, $h_{\rm T}= H_{\lambda}$ and $\gamma_{\rm T}=0$ hold, resulting in the right-hand side of eq~\eqref{eq:consPr} becoming zero. 
The same discussions apply even to the system defined by $H_{0}$. 
In unitary theories, the condition ${\rm Tr}\,[\rho_{\rm R}]={\rm Tr}\,[\rho_{\rm T}]=1$ is thus upheld consistently over time. 
These arguments further demonstrate that the non-negativity of relative entropy reflects the unitary time evolution. 
As studied in Refs.~\cite{Cao:2022ajt,Cao:2022iqh}, the non-negativity of relative entropy, which signifies the unitarity, is linked to fundamental physics properties encompassing the second law of thermodynamics, positivity bounds on EFTs, and the mild weak gravity conjecture. Any violation of this non-negativity results in the emergence of abnormal behavior within the theory. 
In the forthcoming sections, we will explore an aspect of the non-negative relative entropy that assesses the validity of perturbative calculations utilized in calculating effective actions. 

\subsection{Weakly coupled theory}
\label{sec:weak}
Now, we elucidate assumptions in calculating the effective action as fundamental components of the relative entropy.
In subsequent sections, in the perturbative calculations of the EFT $W_\lambda$, it is assumed that the interaction $I_{\rm I}$ is small and that the following two conditions are upheld:
\begin{itembox}[c]{ Weakly coupled theory}
\begin{itemize}
    \item The impact at the tree-level is more significant than that at the loop-level. 
    \item Higher-order terms of the interaction in the effective action can be quantitatively truncated at any order. 
\end{itemize}
\end{itembox}
These two assumptions imply that the interaction between heavy and light fields is weak, and consequently, we refer to theories that satisfy these assumptions as {\it weakly coupled theories}. 
In contrast, throughout this work, theories that do not satisfy one of the two assumptions are referred to as {\it strongly coupled theories}. 
The first condition above is often assumed in the context of perturbative unitarity of $S$-matrix, and exploring parameter regions of theory satisfying its validity is of significant importance, {\it e.g.}, in predicting the SM Higgs mass~\cite{Lee:1977eg,PhysRevLett.38.883}.
%
%
It is well-known that perturbative expansions about the interaction arise during the calculation of scattering amplitudes, and the effective action can be obtained by matching these scattering amplitudes.
Based on this perspective, the tree-level approximation utilized in perturbative unitarity within the $S$-matrix is similar to the tree-level approximation applied in the EFT above calculations. 
On the other hand, the second condition represents the validity of operator expansion, meaning the validity of the EFT description in low-energy regions. 
If the operator expansion is invalid, the phenomenon cannot be adequately described without adding an infinite number of operators. 
The determination of the Wilson coefficients of an infinite number of operators requires a UV theory that corresponds to the EFT. We thus regard any theory for which the operator expansion is invalid as unsuitable for the EFT description. 
%

%
%
%
In the subsequent sections, we will not consider the loop-level contributions in evaluating the relative entropy because the emphasis is on its non-negativity in the weakly coupled theory. 
For instance, if the UV theory includes heavy fermions in addition to scalar fields, assuming the weakly coupled theory means that loop-level contributions arising from the fermion fields are negligible. 
However, our results of this study might not be applicable when the main contribution in the EFT is at the loop-level, such that the heavy degrees of freedom in the UV theory consist only of fermion fields. 
As studied in Ref.~\cite{Cao:2022ajt,Cao:2022iqh}, there are classes of UV theories to which the results of this study can be adapted at the loop-level, while there are also classes of UV theories to which they cannot be adapted\footnote{More specifically, there are theories for which the third term in the last line of eq.~\eqref{eq:relFI} is not zero at the loop-level. Such theories are not considered in this study, and the results of this study may not be available.}. 
The calculations of the relative entropy at the loop-level have been studied in Ref.~\cite{Cao:2022ajt,Cao:2022iqh}; please look at the details there. 
Assuming the weakly coupled theory also implies the validity of saddle-point approximation in the path integral, as mentioned in Ref.~\cite{Cao:2022ajt,Cao:2022iqh}. 
This is because classical-level evaluations are inadequate when performing Euclidean path integrals near maxima points of the action, and higher-order loop-level corrections are essential. 

Regarding the second condition of the weakly coupled theory, we set the condition for the validity of the operator expansion to be able to truncate higher-order operators at arbitrary orders quantitatively. 
For instance, the perturbative calculation in the following case is valid, although, at first glance, it does not meet this validity condition. 
Suppose that, up to a particular order $n$-th of the operator, the higher-order contributions of the operators are more significant than the lower-order contributions, but when the order is bigger than $n$-th, the contributions of the higher-order operators are sufficiently smaller than those of the lower-order operators. 
In this case, the higher-order operator cannot be quantitatively truncated at any order smaller than $n$-th, but the perturbative calculation can be regarded as working well for sufficiently high-order perturbative calculations. 
However, the perturbative calculation in this case would not be valid because it is essential to evaluate them to all orders, as the specific order $n$ can only be determined by considering the contributions from all perturbative orders.
In the present work, the weakly coupled theory is defined so that situations requiring infinite-order evaluation of the perturbative expansion are not included in the valid perturbation regions. 
Given these assumptions in the perturbative calculation, we now turn to the relative entropy in scalar field theories.

\subsection{Relative entropy in scalar field theory}
\label{sec:sub23}
To illustrate the process of calculating the relative entropy, it is helpful to examine scalar field theories. 
We will examine the relative entropy in three typical scenarios in the subsequent sections.

\subsubsection{Scalar field theory comprising one external background field or light field and one massive field}
\label{sec:2.2.1}

\begin{enumerate}[(i)]
    \item{\it one massive scalar field theory under an external background field ---} 
   Consider a theory defined by a Euclidean Lagrangian $\mathcal{L}\left(\phi_1,\phi_2\right)$, where $\phi_1$ represents a non-dynamical external background field, whereas $\phi_2$ denotes a dynamical scalar field (throughout this work, the field that undergoes integration with the path integral is denoted as the dynamical field, while the field that does not undergo this integration is termed the non-dynamical field) with mass.
    In this context, we are focusing on a Lagrangian parameterized as follows: 
    \begin{align}
    \mathcal{L}\left(\phi_1,\phi_2\right):=\frac{1}{2}(\partial_I\phi_2)^2+\sum_{k=0}^{\infty} a_k \left(\phi_1\right)\phi_2^k,\label{eq:lag0}
    \end{align}
where $\partial_I$ for $I=1,2,3,4$ is a derivative in Euclidean space, and $a_k$ is a function comprising $\phi_1$ and its derivative. 
Although it is straightforward to extend the following discussion to cases where higher derivative terms of the heavy fields are present in eq.~\eqref{eq:lag0} by including derivative operators in the coefficients $a_k(\phi_1)$, we neglect them since our focus is on low-energy EFTs arising from integrating out heavy fields, where the typical energy scales are smaller than the heavy field masses.
We proceed with redefining $\phi_2$ (refer to Appendix~\ref{app:red} for details) and considering it as follows: 
\begin{align}
\left(\frac{\partial \mathcal{L}(0,\phi_2)}{\partial \phi_2}\right)_{\phi_2=0}=0,\quad \left(\frac{\partial^2 \mathcal{L}(0,\phi_2)}{\partial \phi_2^2}\right)_{\phi_2=0}=m_2^2.\label{eq:par0BG}
\end{align}
Throughout this work, we examine a scenario in which $\phi_2$ is massive and $m_2^2$ is positive. 
Then, the Lagrangian can be divided as follows: 
\begin{align}
\mathcal{L}\left(\phi_1,\phi_2\right)=\mathcal{L}_0(\phi_1,\phi_2)+\mathcal{L}_{\rm I}(\phi_1,\phi_2),\label{eq:lagtwoBG}
\end{align}
where $\mathcal{L}_{\rm I}$ represents the interaction between $\phi_1$ and $\phi_2$, while $\mathcal{L}_0$ does not incorporate this interaction.
The two terms in the Lagrangian~\eqref{eq:lagtwoBG} are expressed as follows: 
\begin{align}
    \mathcal{L}_0(\phi_1,\phi_2)&=\frac{1}{2}(\partial_I\phi_2)^2+a_0(\phi_1)+\frac{m_2^2}{2}\phi_2^2+\sum_{k=3}^{\infty}a_k(0)\phi_2^k,\label{eq:L0}
    \\
    \mathcal{L}_{\rm I}(\phi_1,\phi_2)&=a_1(\phi_1)\phi_2+\left(a_2(\phi_1)-\frac{m_2^2}{2}\right)\phi_2^2+\sum_{k=3}^{\infty}\left(a_k(\phi_1)-a_k(0)\right)\phi_2^k.\label{eq:LabI}
\end{align}
To explicitly express the order of the perturbative expansion about the interaction~\eqref{eq:LabI}, we introduce the parameter $\lambda$ and consider the following Lagrangian instead of eq.~\eqref{eq:lagtwoBG}. 
\begin{align}
    \mathcal{L}_\lambda(\phi_1,\phi_2):=\mathcal{L}_0(\phi_1,\phi_2)+\lambda \mathcal{L}_{\rm I}(\phi_1,\phi_2).\label{eq:Lgscl}
\end{align}
As mentioned, $\lambda=1$ can be chosen after calculating the relative entropy. 
The Euclidean action~\eqref{eq:Ig} is defined as, 
\begin{align}
    I_\lambda [\phi_1,\phi_2]:=\int (d^4x)_{\rm E} \mathcal{L}_\lambda(\phi_1,\phi_2).\label{eq:IgsclBG}
\end{align}
Upon substituting eq.~\eqref{eq:IgsclBG} into eq.~\eqref{eq:part}, the partition functions are calculated at the tree-level as follows: 
\begin{align}
    Z_\lambda[\phi_1]&=\int d[\phi_2] e^{-I_\lambda[\phi_1,\phi_2]}=e^{-I_\lambda[\phi_1,\widetilde{\phi}_{2,\lambda}]}, \label{eq:zgBG}
    \\
        Z_0[\phi_1]&=\int d[\phi_2] e^{-I_0[\phi_1,\phi_2]}=e^{-I_0[\phi_1,\widetilde{\phi}_{2,0}]},\label{eq:zoBG}
\end{align}
where the classical solutions of $\mathcal{L}_0$ and $\mathcal{L}_\lambda$ are represented by $\widetilde{\phi}_{2,0}$ and $\widetilde{\phi}_{2,\lambda}$, respectively. 
Note that the integral for $\phi_1$ is not performed because $\phi_1$ is assumed to be the external background field. 
Since, as we will see later, the integrated degrees of freedom $\phi_2$ are replaced with the classical solution expressed by $\phi_1$, we have made $\phi_1$ dependencies explicit in the leftmost partition functions in eq.~\eqref{eq:zgBG} and \eqref{eq:zoBG}. 
The expression for the Euclidean effective action at the tree-level is derived from eqs.~\eqref{eq:zgBG} and \eqref{eq:zoBG} as follows: 
\begin{align}
    W_\lambda[\phi_{1}]=I_\lambda[\phi_{1},\widetilde{\phi}_{2,\lambda}],~~~W_0[{\phi}_{1}]=I_0[\phi_{1},\widetilde{\phi}_{2,0}].\label{eq:EucactionBG}
\end{align}
The functional derivative of $I_\lambda$ defines the classical solution $\widetilde{\phi}_{2,\lambda}$ as, 
\begin{align}
    \left(\frac{\delta I_\lambda}{\delta \phi_2}\right)_{\phi_2=\widetilde{\phi}_{2,\lambda}}&=-\partial_I^2 \widetilde{\phi}_{2,\lambda}+\lambda a_1(\phi_1)+\left\{m_2^2+\lambda\left(2 a_2(\phi_1)-m^2_2\right)\right\}\widetilde{\phi}_{2,\lambda}\notag
    \\
    &\quad\quad\quad+\sum_{k=3}^{\infty} k\left\{a_k(0)+\lambda\left(a_k(\phi_1)-a_k(0)\right)\right\}\widetilde{\phi}_{2,\lambda}^{k-1}=0.\label{eq:firsderBG}
\end{align}
The classical solution for $\widetilde{\phi}_{2,0}=0$ is calculated up to the second order of $\lambda$ as follows: 
\begin{align}
\widetilde{\phi}_{2,\lambda}
&\simeq-\lambda\frac{a_1(\phi_1)}{m_2^2\left(1-\partial_I^2/m_2^2\right)}+\lambda^2\left(\frac{a_1(\phi_1)\left(2 a_2(\phi_1)-m_2^2\right)}{m_2^4\left(1-\partial_I^2/m_2^2\right)^2}-\frac{3 \left(a_1(\phi_1)\right)^2a_3(0)}{m_2^6\left(1-\partial_I^2/m_2^2\right)^3}\right).\label{eq:PhigBG}
\end{align}
Regarding the classical solution at $\lambda$ higher than the second order, please refer to Appendix~\ref{app:B}. 
After substituting eq.~\eqref{eq:PhigBG} into eq.~\eqref{eq:EucactionBG} and expanding up to the third order of $\lambda$, the resulting expression is as follows:
\begin{align}
    W_\lambda[\phi_1]&\simeq \int (d^4x)_{\rm E}\Bigg[
    a_0(\phi_{1})-\lambda^2 \frac{1}{2m^2_2\left(1-\partial_I^2/m_2^2\right)}\left(a_1({\phi}_{1})\right)^2+\lambda^3 \frac{1}{m_2^6\left(1-\partial_I^2/m_2^2\right)^3}\notag
    \\
    &\times\Bigg\{
    -a_3(0)\left(a_1({\phi}_{1})\right)^3
    +m^2_2 \left(1-\frac{\partial_I^2}{m_2^2}\right)\left(a_1({\phi}_{1})\right)^2
    \left(a_2({\phi}_{1})-\frac{m^2_2}{2}\right)
    \Bigg\}
    \Bigg],\label{eq:WgsclBG}
    \\
     W_0[\phi_{1}]&=\int (d^4x)_{\rm E}
    a_0(\phi_{1}).   \label{eq:W0sclBG}
\end{align}
Note that the main perturbative correction to $W_\lambda$ is quadratic in $\lambda$ because of eq.~\eqref{eq:par0BG}. 
Regarding the effective action up to the higher order of $\lambda$ than the third order, look at Appendix~\ref{app:B}.  
The effective action $W_{\lambda}$ corresponds to the low-energy EFT of $I_{\lambda}$, establishing a connection between the relative entropy and the EFT through eq.~\eqref{eq:relFI}. 
From eq.~\eqref{eq:EucactionBG}, the third term of eq.~\eqref{eq:relFI} becomes, 
\begin{align}
\left(\frac{\partial W_\lambda[\phi_1]}{\partial \lambda}\right)_{\lambda=0}=\frac{\partial I_\lambda[\phi_1,\widetilde{\phi}_{2,0}]}{\partial \lambda}=0,\label{eq:dWgBG}
\end{align}
where $\widetilde{\phi}_{2,0}=0$ was utilized. 
Note here that eq.~\eqref{eq:dWgBG} is derived under the tree-level approximation without relying on operator expansions.

Given the considerations above, we will calculate the relative entropy, assuming that $I_\lambda$ represents the weakly coupled theory. 
Upon substituting eqs.~\eqref{eq:WgsclBG}, \eqref{eq:W0sclBG}, and \eqref{eq:dWgBG} into eq.~\eqref{eq:relFI}, the relative entropy is calculated up to the third order of $\lambda$ as, 
\begin{align}
    S(P_{\rm R}||P_{\rm T})&=W_0[\phi_{1}]-W_\lambda[\phi_{1}],\label{eq:SscalBG0}
    \\
    &\simeq \int (d^4 x)_{\rm E}\Bigg[\lambda^2 \frac{1}{2m_2^2\left(1-\partial_I^2/m_2^2\right)} \left(a_1({\phi}_{1})\right)^2-\lambda^3\frac{1}{m_2^6\left(1-\partial_I^2/m_2^2\right)^3}\notag
    \\
    &\times\Bigg\{
     -a_3(0)\left(a_1({\phi}_{1})\right)^3+m^2_2\left(1-\frac{\partial_I^2}{m_2^2}\right) \left(a_1({\phi}_{1})\right)^2
    \left(a_2({\phi}_{1})-\frac{m^2_2}{2}\right)
    \Bigg\}\Bigg],\label{eq:SscalBG}
\end{align}
where in the first line, we applied $(\partial  W_\lambda/\partial \lambda)_{\lambda=0}=0$ following eq.~\eqref{eq:dWgBG}. 
From eq.~\eqref{eq:SscalBG}, the second order of $\lambda$ in the relative entropy, {\it i.e.}, the main order of the perturbative expansion, is generally non-negative in low-energy domains. 
In contrast, the third order of $\lambda$ in the relative entropy can attain a negative value, raising the possibility of violating the non-negativity property in the perturbative calculation. 
Even for tertiary and subsequent $\lambda$ contributions, the non-negativity of relative entropy may be disrupted, and the sign of relative entropy may vary based on the perturbation order (see Sec.~\ref{sec:quasi}). 
Since the relative entropy calculated without approximation satisfies the non-negativity~\eqref{eq:non}, the right-hand side of eq.~\eqref{eq:SscalBG} should be non-negative if the perturbative calculation is valid. 
Based on the facts above, it can be inferred that the sign of the perturbatively calculated relative entropy serves as a potential indicator of the effectiveness of the perturbative expansion of the interaction $\lambda I_{\rm I}$. 
In short, theories that violate the non-negativity of relative entropy are strongly coupled, rendering the perturbative description of physics invalid. 
In Sec.~\ref{sec:quasi}, two scalar field theories are analyzed as examples, and the validity of perturbation theory is explored in more detail through the perturbative calculation of the relative entropy, followed by a comparison with the non-perturbative calculation results.

    \item {\it Two scalar fields theory comprising one light field and one massive field ---} Consider a theory described by a Euclidean Lagrangian $\mathcal{L}\left(\phi_1,\phi_2\right)$, where $\phi_2$ is a massive scalar field, and the mass of $\phi_1$ is smaller than that of $\phi_2$.  
Note here that $\phi_1$ is regarded as a dynamical field rather than an external one. 
From eqs.~\eqref{eq:IgsclBG} and \eqref{eq:part}, the partition functions at the tree-level are given as,
\begin{align}
    Z_\lambda[\widetilde{\phi}_{1,\lambda}]&=\int d[\phi_1]d[\phi_2]e^{-I_\lambda[\phi_1,\phi_2]}=e^{-I_\lambda[\widetilde{\phi}_{1,\lambda},\widetilde{\phi}_{2,\lambda}]},\label{eq:Zgdy}
    \\
    Z_0[\widetilde{\phi}_{1,0}]&=\int d[\phi_1]d[\phi_2]e^{-I_0[\phi_1,\phi_2]}=e^{-I_0[\widetilde{\phi}_{1,0},\widetilde{\phi}_{2,0}]},   \label{eq:Z0dy}
\end{align}
where $\widetilde{\phi}_{1,\lambda}$ and $\widetilde{\phi}_{2,\lambda}$ represent classical solutions of $\mathcal{L}_\lambda$. 
In eqs.~\eqref{eq:Zgdy} and \eqref{eq:Z0dy}, we conducted the path integral over the dynamical field $\phi_1$, in contrast to the scenario in which $\phi_1$ serves as the external field. 
From eqs.~\eqref{eq:Zgdy} and \eqref{eq:Z0dy}, the Euclidean effective actions at the tree-level are obtained as, 
\begin{align}
    W_\lambda[\widetilde{\phi}_{1,\lambda}]=I_\lambda[\widetilde{\phi}_{1,\lambda},\widetilde{\phi}_{2,\lambda}],~~~W_0[\widetilde{\phi}_{1,0}]=I_0[\widetilde{\phi}_{1,0},\widetilde{\phi}_{2,0}],\label{eq:Eucaction}
\end{align}
where $\widetilde{\phi}_{2,\lambda}$ is identical to eq.~\eqref{eq:PhigBG} except for replacing the external field $\phi_1$ with the classical solution $\widetilde{\phi}_{1,\lambda}$. 
Upon substituting eq.~\eqref{eq:PhigBG} into eq.~\eqref{eq:Eucaction}, the effective action in the weakly coupled theory is evaluated up to the third order of $\lambda$ as follows: 
\begin{align}
    W_\lambda[\widetilde{\phi}_{1,\lambda}]&\simeq\int (d^4x)_{\rm E}\Bigg[
    a_0(\widetilde{\phi}_{1,0})-\lambda^2 \frac{1}{2m^2_2\left(1-\partial_I^2/m_2^2\right)}\left(a_1(\widetilde{\phi}_{1,0})\right)^2+\lambda^3 \frac{1}{m_2^6\left(1-\partial_I^2/m_2^2\right)^3}\notag
    \\
    &\times\Bigg\{
    -a_3(0)\left(a_1(\widetilde{\phi}_{1,0})\right)^3
    +m^2_2 \left(1-\frac{\partial_I^2}{m_2^2}\right)\left(a_1(\widetilde{\phi}_{1,0})\right)^2
    \left(a_2(\widetilde{\phi}_{1,0})-\frac{m^2_2}{2}\right)
    \Bigg\}
    \Bigg],\label{eq:Wgscl}
    \\
     W_0[\widetilde{\phi}_{1,0}]&=\int (d^4x)_{\rm E}
    a_0(\widetilde{\phi}_{1,0}),   \label{eq:W0scl}
\end{align}
where $W_\lambda[\widetilde{\phi}_{1,\lambda}]=W_\lambda[\widetilde{\phi}_{1,0}]+\mathcal{O}(\lambda^4) $ was used in conjunction with $\widetilde{\phi}_{1,\lambda}=\widetilde{\phi}_{1,0}+\mathcal{O}(\lambda^2)$ (see Appendix~\ref{sec:phi1}). 
Note that, according to eq.~\eqref{eq:par0BG}\footnote{Here, we consider the Lagrangian~\eqref{eq:lag0}, which satisfies eq.~\eqref{eq:par0BG}.
Then, the classical solution for 
$\phi_2$ is zero at zeroth order in 
$\lambda$, and the tadpole diagrams associated with 
$\phi_2$ also vanish at zeroth order in 
$\lambda$.
From this, Feynman diagrams contributing to the effective action $W_{\lambda}$ when integrating the heavy field $\phi_2$ out cannot appear at the first order of $\lambda$ because the interaction $\lambda I_I$ between $\phi_2$ and the light fields is the first order of $\lambda$.
}, the main perturbative correction to $W_\lambda$ is the second order of $\lambda$.
The third term of eq.~\eqref{eq:relFI} can be obtained from eq.~\eqref{eq:Eucaction} and $\widetilde{\phi}_{2,0}=0$ as, 
\begin{align}
    \left(\frac{\partial W_\lambda[\widetilde{\phi}_{1,0}]}{\partial \lambda}\right)_{\lambda=0}=\frac{\partial I_\lambda [\widetilde{\phi}_{1,0},\widetilde{\phi}_{2,0}]}{\partial \lambda}=0.\label{eq:dWgdy}
\end{align}
Upon substituting eqs.~\eqref{eq:Wgscl}, \eqref{eq:W0scl}, and \eqref{eq:dWgdy} into eq.~\eqref{eq:relFI}, the relative entropy within the context of the weakly coupled theory is expressed up to the third order of $\lambda$ as follows: 
\begin{align}
    S(P_{\rm R}||P_{\rm T})&=W_0[\widetilde{\phi}_{1,0}]-W_\lambda[\widetilde{\phi}_{1,\lambda}],\label{eq:rel00}
    \\
    &\simeq \int (d^4 x)_{\rm E}\Bigg[
    \lambda^2 \frac{1}{2m_2^2\left(1-\partial_I^2/m_2^2\right)} \left(a_1(\widetilde{\phi}_{1,0})\right)^2-\lambda^3\frac{1}{m_2^6\left(1-\partial_I^2/m_2^2\right)^3}\notag
    \\
    &\times\Bigg\{
     -a_3(0)\left(a_1(\widetilde{\phi}_{1,0})\right)^3+m^2_2\left(1-\frac{\partial_I^2}{m_2^2}\right) \left(a_1(\widetilde{\phi}_{1,0})\right)^2
    \left(a_2(\widetilde{\phi}_{1,0})-\frac{m^2_2}{2}\right)
    \Bigg\}\Bigg],\label{eq:Sscal}
\end{align}
where, in the first line, $(\partial W_\lambda/ \partial \lambda)_{\lambda=0}=0$ was used following eq.~\eqref{eq:dWgdy}. 
Equation~\eqref{eq:Sscal} is equivalent to eq.~\eqref{eq:SscalBG} up to the third order of $\lambda$ by substituting $\phi_1$ with $\widetilde{\phi}_{1,0}$.
Thus, it becomes evident that up to the low orders of perturbative expansion, regardless of whether $\phi_1$ is the dynamical or background field, the method and results of the calculation remain the same, with the only difference being the replacement of $\phi_1\to \widetilde{\phi}_{1,0}$. 
It should be mentioned that the $\lambda$ dependence of $\widetilde{\phi}_{1,\lambda}$ can be crucial in the higher-order perturbative calculations.

\end{enumerate}

\subsubsection{$N$ scalar fields theory with one heavy field}
\label{sec:2.2.2}
Generalizing the discussions from Sec.~\ref{sec:2.2.1} to $N$ scalar fields theory is straightforward. 
The calculations in Sec.~\ref{sec:2.2.1} remain mostly unchanged, regardless of whether the light degrees of freedom are considered as an external or a dynamical field, as demonstrated in (i) and (ii). Our discussion will go ahead assuming that all $N$ scalar fields are dynamical. 
    Let us consider a theory defined by a Euclidean Lagrangian $\mathcal{L}^{(N)}\left(\phi_1,\cdots,\phi_N\right)$, where $\phi_N$ is a massive scalar field, and the mass of $(\phi_1,\cdots,\phi_{N-1})$ is smaller than that of $\phi_N$.  
    Similar to the approach taken in the two scalar fields theory, the Lagrangian is parametrized as follows: 
    \begin{align}
    \mathcal{L}^{(N)}\left(\phi_1,\cdots,\phi_N\right):=
    \frac{1}{2}(\partial_I\phi_N)^2+\sum_{k=0}^{\infty} a_k^{(N)}\left(\phi_1,\cdots,\phi_{N-1}\right)\phi_N^k,\label{eq:LN}
    \end{align}
    where $a_k^{(N)}$ represents a function dependent on the fields $(\phi_1,\cdots,\phi_{N-1})$ and its derivative. 
    After appropriately redefining the field $\phi_N$ (see Appendix~\ref{app:red}), the following relations hold: 
    \begin{align}
    \left(\frac{\partial \mathcal{L}^{(N)}\left(0,\cdots,0,\phi_N\right)}{\partial \phi_N}\right)_{\phi_N=0}=0,~~~
    \left(\frac{\partial^2 \mathcal{L}^{(N)}\left(0,\cdots,0,\phi_N\right)}{\partial \phi_N^2}\right)_{\phi_N=0}=m_N^2,\label{eq:redN}
    \end{align}
    where $m_N^2$ is positive, similar to (i) and (ii). 
    Here, we divide the Lagrangian into the following: 
    \begin{align}
        \mathcal{L}^{(N)}\left(\phi_1,\cdots,\phi_N\right)=\mathcal{L}^{(N)}_0\left(\phi_1,\cdots,\phi_N\right)+\mathcal{L}^{(N)}_{\rm I}\left(\phi_1,\cdots,\phi_N\right),
    \end{align}
    where $\mathcal{L}_{\rm I}^{(N)}$ signifies the interaction between $(\phi_1,\cdots,\phi_{N-1})$ and $\phi_N$, while $\mathcal{L}_{0}^{(N)}$ does not encompass this interaction.
    Take note here that $\mathcal{L}_{0}^{(N)}$ generally encompasses interactions between fields $(\phi_1,\cdots,\phi_{N-1})$. 
    These two terms are then parameterized as indicated below: 
    \begin{align}
    \mathcal{L}_{0}^{(N)}\left(\phi_1,\cdots,\phi_N\right)&=\frac{1}{2}(\partial_I\phi_N)^2+
    a_0^{(N)}\left(\phi_1,\cdots,\phi_{N-1}\right)+\frac{m^2_N}{2}\phi_N^2 +\sum_{k=3}^{\infty}a_k^{(N)}\left(0,\cdots,0\right)\phi_N^k,
    \\
    \mathcal{L}_{\rm I}^{(N)}\left(\phi_1,\cdots,\phi_N\right)&=a_1^{(N)}\left(\phi_1,\cdots,\phi_{N-1}\right)\phi_N +
    \left(
    a_2^{(N)}\left(\phi_1,\cdots,\phi_{N-1}\right)-\frac{m^2_N}{2}
    \right)\phi_N^2\notag
    \\
    &\quad\quad\quad\quad+\sum_{k=3}^{\infty}\left(
    a_k^{(N)}\left(\phi_1,\cdots,\phi_{N-1}\right)-a_k^{(N)}\left(0,\cdots,0\right)
    \right)\phi_N^k.\label{eq:LIN}
    \end{align}
    To make the order of the perturbative expansion explicit, we introduce the parameter $\lambda$ and define the following Lagrangian: 
    \begin{align}
    \mathcal{L}_{\lambda}^{(N)}\left(\phi_1,\cdots,\phi_N\right):=\mathcal{L}_{0}^{(N)}\left(\phi_1,\cdots,\phi_N\right)+ \lambda \mathcal{L}_{\rm I}^{(N)}\left(\phi_1,\cdots,\phi_N\right).\label{eq:NLG}
    \end{align}
    Upon completion of the relative entropy calculation, $\lambda=1$ can be set.
    From eq.~\eqref{eq:NLG}, the Euclidean action~\eqref{eq:Ig} is defined as, 
\begin{align}
    I_{\lambda}^{(N)}[\phi_1,\cdots,\phi_N]&:=\int (d^4x)_{\rm E} \mathcal{L}_{\lambda}^{(N)}\left(\phi_1,\cdots,\phi_N\right).\label{eq:NIgscl}
\end{align}
Upon substituting eq.~\eqref{eq:NIgscl} into eq.~\eqref{eq:part}, the Euclidean actions are calculated as follows: 
\begin{align}
    &W_{\lambda}^{(N)}[\widetilde{\phi}_{1,\lambda}^{(N)},\cdots,\widetilde{\phi}_{N-1,\lambda}^{(N)}]=I_{\lambda}^{(N)}[\widetilde{\phi}_{1,\lambda}^{(N)},\cdots,\widetilde{\phi}_{N-1,\lambda}^{(N)},\widetilde{\phi}_{N,\lambda}^{(N)}],\label{eq:WgN}
    \\
    &W_{0}^{(N)}[\widetilde{\phi}_{1,0}^{(N)},\cdots,\widetilde{\phi}_{N-1,0}^{(N)}]=I_{0}^{(N)}[\widetilde{\phi}_{1,0}^{(N)},\cdots,\widetilde{\phi}_{N-1,0}^{(N)},\widetilde{\phi}_{N,0}^{(N)}],\label{eq:W0N}
\end{align}
where the classical solutions of $\mathcal{L}_{\lambda}^{(N)}$ are denoted as $(\widetilde{\phi}_{1,\lambda}^{(N)},\cdots,\widetilde{\phi}_{N,\lambda}^{(N)})$, while $\widetilde{\phi}_{N,\lambda}^{(N)}$ up to the second order of $\lambda$ is expressly provided as follows: 
\begin{align}
    \widetilde{\phi}_{N,\lambda}^{(N)}&\simeq
    -\lambda \frac{a_1^{(N)}\left(\widetilde{\phi}_{1,\lambda}^{(N)},\cdots,\widetilde{\phi}_{N-1,\lambda}^{(N)}\right)}{m^2_N\left(1-\partial_I^2/m^2_N\right)}\notag
    \\
    &+\lambda^2\Bigg(
\frac{a_1^{(N)}\left(\widetilde{\phi}_{1,\lambda}^{(N)},\cdots,\widetilde{\phi}_{N-1,\lambda}^{(N)}\right) \left(2 a_2^{(N)}\left(\widetilde{\phi}_{1,\lambda}^{(N)},\cdots,\widetilde{\phi}_{N-1,\lambda}^{(N)}\right)-m^2_N\right)}{m^4_N\left(1-\partial_I^2/m^2_N\right)^2}\notag
\\
&\quad\quad\quad\quad\quad\quad\quad\quad\quad\quad\quad\quad-\frac{3 \left(a_1^{(N)} \left(\widetilde{\phi}_{1,\lambda}^{(N)},\cdots,\widetilde{\phi}_{N-1,\lambda}^{(N)}\right)\right)^2 a_3^{(N)}\left(0,\cdots,0\right)}{m^6_N\left(1-\partial_I^2/m^2_N\right)^3}
    \Bigg).\label{eq:PHGN}
\end{align}
Upon substituting eq.~\eqref{eq:PHGN} into eq.~\eqref{eq:WgN}, the effective action of the weakly coupled theory, up to the third order of $\lambda$, is obtained as,
\begin{align}
    W_{\lambda}^{(N)}&\simeq \int (d^4x)_{\rm E}\Bigg[
    a_0^{(N)}\left(\widetilde{\phi}_{1,\lambda}^{(N)},\cdots,\widetilde{\phi}_{N-1,\lambda}^{(N)}\right)\notag
    \\
    &-\lambda^2 \frac{1}{2m_N^2\left(1-\partial^2_I/m^2_N\right)}\left(a_1^{(N)}\left(\widetilde{\phi}_{1,\lambda}^{(N)},\cdots,\widetilde{\phi}_{N-1,\lambda}^{(N)}\right)\right)^2\notag
    \\
    &+\lambda^3 \frac{1}{m^6_N\left(1-\partial^2_I/m^2_N\right)^3}\bigg\{
    -a_3^{(N)}\left(0,\cdots,0\right)\left(a_1^{(N)}\left(\widetilde{\phi}_{1,\lambda}^{(N)},\cdots,\widetilde{\phi}_{N-1,\lambda}^{(N)}\right)\right)^3\notag
    \\
    &+
    m^2_N\left(1-\frac{\partial^2_I}{m^2_N}\right) \left(a_1^{(N)}\left(\widetilde{\phi}_{1,\lambda}^{(N)},\cdots,\widetilde{\phi}_{N-1,\lambda}^{(N)}\right)\right)^2
    \left(a_2^{(N)}\left(\widetilde{\phi}_{1,\lambda}^{(N)},\cdots,\widetilde{\phi}_{N-1,\lambda}^{(N)}\right)-\frac{m^2_N}{2}\right)
    \bigg\}
    \Bigg],\label{eq:NWg}
    \\
    W_{0}^{(N)}&=\int (d^4x)_{\rm E}
    a_0^{(N)}\left(\widetilde{\phi}_{1,0}^{(N)},\cdots,\widetilde{\phi}_{N-1,0}^{(N)}\right).\label{eq:NW0}
\end{align}
From eq.~\eqref{eq:WgN} and $\widetilde{\phi}_{N,0}^{(N)}=0$, we also obtain
\begin{align}
    \left(\frac{\partial W_\lambda^{(N)}}{\partial \lambda}\right)_{\lambda=0}=0.\label{eq:delWgN}
\end{align}
Note that the partial derivative is performed at $\lambda=0$ (also see Appendix~\ref{sec:delv} for the derivation of eq.~\eqref{eq:delWgN}).
From eqs.~\eqref{eq:relFI}, \eqref{eq:NWg}, \eqref{eq:NW0}, and \eqref{eq:delWgN}, the relative entropy of the weakly coupled theory, up to the third order of $\lambda$, is obtained as,
\begin{align}
    S\left(P_{\rm R}||P_{\rm T}\right)^{(N)}&=
    W_0^{(N)}[\widetilde{\phi}_{1,0}^{(N)},\cdots,\widetilde{\phi}_{N-1,0}^{(N)}]-W_{\lambda}^{(N)}[\widetilde{\phi}_{1,\lambda}^{(N)},\cdots,\widetilde{\phi}_{N-1,\lambda}^{(N)}],\label{eq:relN}
    \\
    &\simeq \int (d^4x)_{\rm E}\Bigg[\lambda^2 \frac{1}{2m_N^2\left(1-\partial_I^2/m^2_N\right)}\left(a_1^{(N)}\left(\widetilde{\phi}_{1,\lambda}^{(N)},\cdots,\widetilde{\phi}_{N-1,\lambda}^{(N)}\right)\right)^2\notag
    \\
    &-\lambda^3 \frac{1}{m^6_N\left(1-\partial_I^2/m^2_N\right)^3}\bigg\{
    -a_3^{(N)}\left(0,\cdots,0\right)\left(a_1^{(N)}\left(\widetilde{\phi}_{1,\lambda}^{(N)},\cdots,\widetilde{\phi}_{N-1,\lambda}^{(N)}\right)\right)^3\notag
    \\
    &+
    m^2_N \left(1-\frac{\partial_I^2}{m^2_N}\right)\left(a_1^{(N)}\left(\widetilde{\phi}_{1,\lambda}^{(N)},\cdots,\widetilde{\phi}_{N-1,\lambda}^{(N)}\right)\right)^2\notag
    \\
    &\times
    \left(a_2^{(N)}\left(\widetilde{\phi}_{1,\lambda}^{(N)},\cdots,\widetilde{\phi}_{N-1,\lambda}^{(N)}\right)-\frac{m^2_N}{2}\right)
    \bigg\}\Bigg],
\end{align}
where $\left(\partial W^{(N)}_{\lambda}/\partial \lambda\right)_{\lambda=0}=0$ was used, following eq.~\eqref{eq:delWgN}. 
Equation~\eqref{eq:relN} depicts an extended form of eq.~\eqref{eq:rel00} within the framework of $N$ scalar fields theory. 
Similarly to (i) and (ii), the second-order term of $\lambda$ in the relative entropy consistently remains non-negative. In contrast, the third-order term of $\lambda$ in the relative entropy can breach the non-negativity constraint.

\subsubsection{$N$ scalar fields theory with $N-L$ heavy fields}
\label{app:Nsc}
Given the two Secs.~\ref{sec:2.2.1} and \ref{sec:2.2.2} above, we will analyze the more general case.
In the concrete systems of the following sections, the formulas in this case will not be applied, but it may be worthwhile to consider this case to derive bounds for more general theories. 
We are examining the identical Lagrangian as outlined in eq.~\eqref{eq:LN}. However, we are assuming that $\vec{\Phi}_{N-L}:=(\phi_{L+1},\cdots,\phi_{N})$ represents a set of heavy real scalar fields, while $\vec{\phi}_L:=(\phi_1,\cdots,\phi_{L})$ corresponds to a set of real scalar fields associated with particles having masses smaller than $\vec{\Phi}_{N-L}$'s mass.
Under these assumptions, the Lagrangian in eq.~\eqref{eq:LN} can be expressed as follows: 
\begin{align}
\mathcal{L}^{(N)}\left(\phi_1,\cdots,\phi_N\right):&=\frac{1}{2}\sum_{i=L+1}^N(\partial_I\phi_i)^2\notag
\\
&+
    \sum_{n_{L+1},\cdots,n_{N}=0}^{\infty} a_{n_{L+1},\cdots,n_N}^{(N)}\left(\phi_1,\cdots,\phi_{L}\right)\phi_{L+1}^{n_{L+1}}\cdots \phi_{N}^{n_{N}}.
\end{align}
For simplicity in subsequent calculations, the Lagrangian of a system encompassing $N-1$ real scalar fields, with $N-1-L$ of them identified as heavy real scalar fields, is denoted as $\mathcal{L}^{(N-1)}\left(\phi_1,\cdots,\phi_{N-1}\right):=\mathcal{L}^{(N)}\left(\phi_1,\cdots,\phi_{N-1},0\right)$.
By diagonalizing the mass matrix of the heavy fields and redefining $\vec{\Phi}_{N-L}$ appropriately, we obtain the following relations: 
\begin{align}
    &\left(\frac{\partial \mathcal{L}^{(N)}\left(\vec{0},\vec{\Phi}_{N-L}\right)}{\partial \phi_n}\right)_{\vec{\Phi}_{N-L}=\vec{0}}=0,~
    \left(\frac{\partial^2 \mathcal{L}^{(N)}\left(\vec{0},\vec{\Phi}_{N-L}\right)}{\partial \phi_n^2}\right)_{\vec{\Phi}_{N-L}=\vec{0}}=m_n^2,\notag
    \\
    &\left(\frac{\partial^2 \mathcal{L}^{(N)}\left(\vec{0},\vec{\Phi}_{N-L}\right)}{\partial \phi_n \partial \phi_m}\right)_{\vec{\Phi}_{N-L}=0}=0~~{\rm for}~n\neq m,\label{eq:LNlin}
    \end{align}
    where $n,m\in \{L+1,L+2,\cdots,N\}$, and $m_n^2$ is positive. 
    The Lagrangian $\mathcal{L}^{(N)}$ is then divided as follows: 
    \begin{align}
        \mathcal{L}^{(N)}\left(\phi_1,\cdots,\phi_N\right)=\widetilde{\mathcal{L}}^{(N)}_0\left(\phi_1,\cdots,\phi_N\right)+\widetilde{\mathcal{L}}^{(N)}_{\rm I}\left(\phi_1,\cdots,\phi_N\right),
    \end{align}
    where $\widetilde{\mathcal{L}}^{(N)}_{\rm I}$ corresponds to the interaction between $\vec{\phi}_L$ and $\vec{\Phi}_{N-L}$, and $\widetilde{\mathcal{L}}^{(N)}_{0}$ is defined in a manner that excludes this interaction. 
    Note here that interactions between the fields $\vec{\phi}_{L}$ and between the fields $\vec{\Phi}_{N-L}$ are generally included in $\widetilde{\mathcal{L}}^{(N)}_{0}$.
    Similarly, $\widetilde{\mathcal{L}}^{(N-1)}_{\rm I}$ and $\widetilde{\mathcal{L}}^{(N-1)}_{0}$ are defined as the interaction between $\vec{\phi}_L$ and $\vec{\Phi}_{N-1-L}$ and a term that does not include this interaction, respectively. 
    The Lagrangians $\widetilde{\mathcal{L}}^{(N)}_{0}$ and $\widetilde{\mathcal{L}}^{(N)}_{\rm I}$ are expressed as follows: 
    \begin{align}
    \widetilde{\mathcal{L}}^{(N)}_{0}&=\frac{1}{2}\sum_{i=L+1}^N (\partial_I\phi_i)^2+a^{(N)}_{0,\cdots,0}\left(\phi_1,\cdots,\phi_L\right)
    +\sum_{i=L+1}^N\frac{m^2_i}{2}\phi^2_i\notag
    \\
 &+\sum_{n_{L+1},\cdots,n_N=0}^{\infty}a_{n_{L+1},\cdots,n_N}^{(N)}\left(0,\cdots,0\right) H\left[\sum_{i=L+1}^N n_i-3\right]\phi_{L+1}^{n_{L+1}}\cdots \phi_{N}^{n_{N}},
    \\
    \widetilde{\mathcal{L}}^{(N)}_{\rm I}&=\sum_{i=L+1}^{N}a^{(N)}_{0,\cdots,0,n_i=1,0,\cdots,0}\left(\phi_1,\cdots,\phi_L\right)\phi_i+\sum_{i,j=L+1,i\neq j}^{N}a_{0,\cdots,0,n_i=1,0,\cdots,0,n_j=1,0,\cdots,0}^{(N)}\left(\phi_1,\cdots,\phi_L\right)\phi_i \phi_j\notag
    \\
    &+\sum_{i=L+1}^N \left(a_{0,\cdots,0,n_i=2,0,\cdots,0}^{(N)}\left(\phi_1,\cdots,\phi_L\right)-\frac{m_i^2}{2}\right)\phi_i^2\notag
    \\
    & +\sum_{n_{L+1},\cdots,n_N=0}^{\infty}\left(a_{n_{L+1},\cdots,n_N}^{(N)}\left(\phi_1,\cdots,\phi_L\right)-a_{n_{L+1},\cdots,n_N}^{(N)}\left(0,\cdots,0\right) \right) H\left[\sum_{i=L+1}^N n_i-3\right]\phi_{L+1}^{n_{L+1}}\cdots \phi_{N}^{n_{N}},
    \end{align}
    where $H[n]$ is the Heaviside step function defined as,
    \begin{align}
    H[n]:=
    \begin{cases}
    0,~n< 0,
    \\
    1,~n\geq 0.
    \end{cases}
    \end{align}
    We are introducing the parameter $\lambda$ and defining the following Lagrangian:
    \begin{align}
    \widetilde{\mathcal{L}}_\lambda ^{(N)}\left(\phi_1,\cdots,\phi_N\right):=\widetilde{\mathcal{L}}_0^{(N)}\left(\phi_1,\cdots,\phi_N\right)+\lambda\widetilde{\mathcal{L}}_{\rm I}^{(N)}\left(\phi_1,\cdots,\phi_N\right).\label{eq:LgLN}
    \end{align}
    From eq.~\eqref{eq:LgLN}, a Euclidean action is defined as,
    \begin{align}
    \widetilde{I}^{(N)}_\lambda[\phi_1,\cdots,\phi_N]:=\int (d^4x)_{\rm E}\widetilde{\mathcal{L}}_\lambda^{(N)}\left(\phi_1,\cdots,\phi_N\right).\label{eq:wideIg}
    \end{align}
    Upon substituting eq.~\eqref{eq:wideIg} into eq.~\eqref{eq:part}, the Euclidean effective actions are given as, 
    \begin{align}
    &\widetilde{W}_\lambda^{(N)}[
\hat{\phi}^{(N)}_{1,\lambda},\cdots,
    \hat{\phi}^{(N)}_{L,\lambda}]=\widetilde{I}^{(N)}_\lambda[
\hat{\phi}^{(N)}_{1,\lambda},\cdots,
    \hat{\phi}^{(N)}_{N,\lambda}],
    \\
    &\widetilde{W}_0^{(N)}[
\hat{\phi}^{(N)}_{1,0},\cdots,
    \hat{\phi}^{(N)}_{L,0}]=\widetilde{I}^{(N)}_0[
\hat{\phi}^{(N)}_{1,0},\cdots,
    \hat{\phi}^{(N)}_{N,0}],
    \end{align}
    where $(\hat{\phi}^{(N)}_{1,\lambda},\cdots,
    \hat{\phi}^{(N)}_{N,\lambda})$ denote classical solutions of $\widetilde{\mathcal{L}}_\lambda^{(N)}$. 
    From eq.~\eqref{eq:relFI}, the relative entropy between the two theories $\widetilde{I}_\lambda^{(N)}$ and $\widetilde{I}_0^{(N)}$ is obtained as,
\begin{align}
    \widetilde{S}\left(P_{\rm R}||P_{\rm T}\right)^{(L,N)}=
    \widetilde{W}_0^{(N)}[\hat{\phi}^{(N)}_{1,0},\cdots,\hat{\phi}^{(N)}_{L,0}]-\widetilde{W}_{\lambda}^{(N)}[\hat{\phi}^{(N)}_{1,\lambda},\cdots,\hat{\phi}^{(N)}_{L,\lambda}]\geq 0,\label{eq:reltilLN}
\end{align}
where, from eq.~\eqref{eq:LNlin}, it was used that there is no linear term of $\phi_n$ in $\widetilde{\mathcal{L}}^{(N)}_0$ and that $\left(\partial\widetilde{W}_{\lambda}^{(N)}/\partial \lambda\right)_{\lambda=0}=0$ holds.
Regarding $\left(\partial\widetilde{W}_{\lambda}^{(N)}/\partial \lambda\right)_{\lambda=0}$, it is important to note that the partial derivative is evaluated at $\lambda=0$.
Also, this inequality is a generalization of eq.~\eqref{eq:rel00}.

Now, consider the relation between the relative entropies $\widetilde{S}^{(L,N)}$ and $\widetilde{S}^{(L,N-1)}$. 
Considering eq.~\eqref{eq:LNlin}, we will focus on classical solutions of $\widetilde{\mathcal{L}}_0^{(N)}$ that meet $\hat{\phi}^{(N)}_{L+1,0}=\hat{\phi}^{(N)}_{L+2,0}=\cdots=\hat{\phi}^{(N)}_{N,0}=0$. 
We then obtain the following relation: 
\begin{align}
    &\widetilde{W}_0^{(N)}[\hat{\phi}^{(N)}_{1,0},\cdots,\hat{\phi}^{(N)}_{L,0}]=\widetilde{W}_0^{(N-1)}[\hat{\phi}^{(N-1)}_{1,0},\cdots,\hat{\phi}^{(N-1)}_{L,0}],\label{eq:rell1}
\end{align}
where we used $\widetilde{I}^{(N)}_0[
\hat{\phi}^{(N)}_{1,0},\cdots,
    \hat{\phi}^{(N)}_{N,0}]=\widetilde{I}^{(N)}_0[
\hat{\phi}^{(N)}_{1,0},\cdots,\hat{\phi}^{(N)}_{L,0},0,\cdots,0]=\widetilde{I}^{(N-1)}_0[
\hat{\phi}^{(N)}_{1,0},\cdots,
    \hat{\phi}^{(N)}_{N-1,0}]$, and $\hat{\phi}^{(N)}_{i,0}=\hat{\phi}^{(N-1)}_{i,0}$ for $i=1,\cdots,L$. 
Using eqs.~\eqref{eq:WgN} and \eqref{eq:W0N}, we also find relations for $L\leq N-1$ to be: 
\begin{align}
    &W_0^{(N)}[\widetilde{\phi}_{1,0}^{(N)},\cdots,\widetilde{\phi}_{N-1,0}^{(N)}]=W_{\lambda=1}^{(N-1)}[\widetilde{\phi}_{1,1}^{(N-1)},\cdots,\widetilde{\phi}_{N-2,1}^{(N-1)}]=\widetilde{W}_{\lambda=1}^{(N-1)}[\hat{\phi}_{1,1}^{(N-1)},\cdots,\hat{\phi}_{L,1}^{(N-1)}],\label{eq:rell2}
    \\
    &\widetilde{W}_{\lambda=1}^{(N)}[\hat{\phi}_{1,1}^{(N)},\cdots,\hat{\phi}_{L,1}^{(N)}]={W}_{\lambda=1}^{(N)}[\widetilde{\phi}_{1,1}^{(N)},\cdots,\widetilde{\phi}_{N-1,1}^{(N)}],\label{eq:rell3}
\end{align}
where the utilization of $\widetilde{\phi}_{i,0}^{(N)}=\widetilde{\phi}_{i,1}^{(N-1)}$ for $i=1,\cdots,N-1$ was incorporated alongside $\widetilde{\phi}_{N,0}^{(N)}=0$ and $\mathcal{L}^{(N-1)}\left(\phi_1,\cdots,\phi_{N-1}\right)=\mathcal{L}^{(N)}\left(\phi_1,\cdots,\phi_{N-1},0\right)$. 
Upon combining eq.~\eqref{eq:relN} with eqs.~\eqref{eq:reltilLN}\text{-}\eqref{eq:rell3}, for $L\leq N-1$ and $\lambda=1$, we obtain the following: 
\begin{align}
    \widetilde{S}\left(P_{\rm R}||P_{\rm T}\right)^{(L,N-1)}&=\widetilde{W}_0^{(N-1)}[\hat{\phi}_{1,0}^{(N-1)},\cdots,\hat{\phi}_{L,0}^{(N-1)}]-\widetilde{W}_{\lambda=1}^{(N-1)}[\hat{\phi}^{(N-1)}_{1,1},\cdots,\hat{\phi}^{(N-1)}_{L,1}]\notag
    \\
    &=\widetilde{W}_0^{(N)}[\hat{\phi}^{(N)}_{1,0},\cdots,\hat{\phi}^{(N)}_{L,0}]-{W}_{0}^{(N)}[\widetilde{\phi}_{1,0}^{(N)},\cdots,\widetilde{\phi}_{N-1,0}^{(N)}]\notag
    \\
    &\leq \widetilde{W}_0^{(N)}[\hat{\phi}^{(N)}_{1,0},\cdots,\hat{\phi}^{(N)}_{L,0}]-{W}_{\lambda=1}^{(N)}[\widetilde{\phi}_{1,1}^{(N)},\cdots,\widetilde{\phi}_{N-1,1}^{(N)}]\notag
    \\
    &=\widetilde{W}_0^{(N)}[\hat{\phi}^{(N)}_{1,0},\cdots,\hat{\phi}^{(N)}_{L,0}]-\widetilde{W}_{\lambda=1}^{(N)}[\hat{\phi}^{(N)}_{1,1},\cdots,\hat{\phi}^{(N)}_{L,1}]\notag
    \\
    &=\widetilde{S}\left(P_{\rm R}||P_{\rm T}\right)^{(L,N)}.\label{eq:lasine}
\end{align}
This inequality implies that the relative entropy increases as each heavy field is added to a given theory. 
In the derivation of eq.~\eqref{eq:lasine}, classical-level relative entropy calculations were employed, and the assumption of non-negativity of the relative entropy was made. 
When the relative entropy is calculated perturbatively for the strongly coupled theories, the non-negativity of relative entropy may not hold, and therefore, inequality~\eqref{eq:lasine} can be violated.

\section{Relative entropy in an effective field theory of inflation}
\label{sec:quasi}
%
 Utilizing the formulae outlined in Sec.~\ref{sec:entropy}, we revisit an EFT of single-field inflation~\cite{Garriga:1999vw,Chen:2006nt}, as described by the following action:
\begin{align}
I_{X}=\int d^4x \sqrt{-g}\left[
\frac{1}{2}M^2_{\rm Pl} R+
X+\sum_{j=1}^nc_{j+1} X^{j+1}-V_{\rm soft}(\phi)
\right]
,\label{eq:infEFT0}
\end{align}
where $M_{\rm Pl}$ represents the reduced Planck mass, $R$ is the Ricci scalar, $\phi$ is an inflaton field, $X:=-\frac{1}{2}(\partial_{\mu}\phi)^2$, $c_j$ is the Wilson coefficient of the operator $X^j$, and $V_{\rm soft}(\phi)$ is assumed to softly break a shift symmetry of $\phi$.\footnote{
As discussed later, we assume that the soft-breaking term $V_{\rm soft}$ does not affect hard processes associated with heavy fields.
We thus refer to $V_{\rm soft}$ as the soft-breaking term.
}.
In general, EFT-based analyses of physical phenomena offer the advantage of making quantitative predictions without requiring associated UV theories that generate the EFT. 
It is, therefore, essential to identify parameter regions where the EFT loses its ability to predict physical phenomena quantitatively (where the UV theory generating the EFT becomes indispensable) to ensure the reliability of EFT-based predictions of phenomena. 
Furthermore, the existence of a UV theory corresponding to the EFT is essential for attributing significance to the EFT predictions. 
We assume the validity of the EFT description in eq.~\eqref{eq:infEFT0}, predicated on the validity of the operator $X$ expansion, which refers to the quantitative truncation of higher-dimensional operators at any finite order $n\geq 1$ (also see Sec.~\ref{sec:weak} for discussions of this assumption). 
If that assumption is violated, an infinite number of higher-dimensional operators could be associated with the correct physical description. This implies that the UV theory generating eq.~\eqref{eq:infEFT0} is essential for describing physical phenomena. 

From eq.~\eqref{eq:infEFT0}, the inflationary background is described as,
\begin{align}
    \phi=\phi_0(t),~~~~ds^2 =-dt^2+a(t)^2 d\vec{x}^2.\label{eq:inBG}
\end{align}
The equation of motion for $\phi$ is given as,
\begin{align}
    \ddot{\phi}_0+ 3H \dot{\phi}_0+V'_{\rm soft}(\phi_0)+\sum_{j=1}^nc_{j+1}(j+1) \frac{(\dot{\phi}_0)^{2j}}{2^j} \left(
    (1+2j)\ddot{\phi}_0+3 H\dot{\phi}_0 
    \right)=0,
\end{align}
where $H:=\dot{a}/a$ is the Hubble parameter.
The Friedman equation and the continuity equation are then as follows:
\begin{align}
    &3 M^2_{\rm Pl} H^2 = \rho,~~~-2 M^2_{\rm Pl} \dot{H}=\rho+P,
\end{align}
where the energy density $\rho$ and the pressure $P$ are given by,
\begin{align}
    \rho=\frac{1}{2}\dot{\phi}_0^2+\sum_{j=1}^nc_{j+1} (2j+1) \frac{(\dot{\phi}_0)^{2(j+1)}}{2^{j+1}}+V_{\rm soft}(\phi_0),~~~P=\frac{1}{2}\dot{\phi}_0^2+\sum_{j=1}^n c_{j+1} \frac{(\dot{\phi}_0)^{2(j+1)}}{2^{j+1}}-V_{\rm soft}(\phi_0).
\end{align}
The slow-roll parameters are defined as follows: 
\begin{align}
    \epsilon:&= -\frac{\dot{H}}{H^2}=\frac{3}{\rho}\frac{\dot{\phi}_0^2}{2}\left[
    1+\sum_{j=1}^nc_{j+1}(j+1)\frac{(\dot{\phi}_0)^{2j}}{2^j}
    \right],
    \\
    \eta:&=\frac{\dot{\epsilon}}{H\epsilon}=2\left[\epsilon+\frac{\ddot{\phi}_0}{H\dot{\phi}_0}\times\frac{1+\sum_{j=1}^nc_{j+1}(j+1)^2 (\dot{\phi}_0)^{2j}/2^j}{1+\sum_{j=1}^nc_{j+1}(j+1) (\dot{\phi}_0)^{2j}/2^j}\right].
\end{align}
Under the slow-roll approximation $\dot{\phi}_0={\rm const}.$, conditions $\epsilon, |\eta|\ll 1$ can be satisfied, leading to a persistent inflation. 
In the following, we will examine UV scalar field theories that produce the above EFT~\eqref{eq:infEFT0}: two and three scalar fields theories. We will then analyze the parameter ranges in which the EFT~\eqref{eq:infEFT0} description is valid (in other words, there is a UV theory that corresponds to the EFT, and it is a weakly coupled theory) in more general scalar field theories. 

\subsection{Two scalar fields theory}
Let us consider a UV theory consisting of two scalar fields: an inflaton (a light or background field) and a heavy scalar field. 
In this subsection, we will analyze an inflation model that encompasses two scalar fields, presented as follows: 
\begin{align}
     I^{(2)}=\int d^4x\sqrt{-g}\left[\frac{1}{2}M_{\rm{Pl}}^2R-\sum_{i=1}^2\frac{1}{2}(\partial_{\mu}\phi_{i})^2-V(\phi_{i})\right],\label{eq:3:1}
\end{align}
where $\phi_1$ and $\phi_2$ refer to real scalar fields, while $V(\phi_i)$ denotes a potential. 
To maintain the flatness of the inflaton potential, we postulate an approximate U(1) symmetry under a phase shift of $\phi_1+i\phi_2:=r e^{i\theta}$ for the potential $V(\phi_i)$.
When considering the potential $V(\phi_i)$ at the renormalizable level, the action~\eqref{eq:3:1} is reformulated by incorporating the fields $r$ and $\theta$ in the following manner: 
\begin{align}
    I^{(2)}=\int d^4x\sqrt{-g}\left[\frac{1}{2}M_{\rm{Pl}}^2R-\frac{1}{2}r^2(\partial_{\mu}\theta)^2-\frac{1}{2}(\partial_{\mu}r)^2-V_{r}(r)-V_{\text{soft}}(\theta)\right],\label{eq:3:3}
\end{align}
where $V_{r}(r)$ remains invariant under a constant shift of $\theta$, while $V_{\text{soft}}(\theta)$ introduces a soft break in the shift symmetry and does not impact hard processes involving the heavy field $r$.
The following analysis examines the EFT of inflation, originating from theories featuring two typical potentials: a U(1) symmetric potential and a modified potential.
The focus of the analysis is to evaluate the conditions under which the EFT description remains valid.

\subsubsection{U(1) symmetric potential}
\label{sec:U(1)}
Consider the following U(1) symmetric renormalizable potentials: 
\begin{align}
    V_r(r)&=\frac{m^2_2 }{8r^2_{\rm min}}\left(r^2-r_{\rm min}^2\right)^2\notag
    \\
    &=\frac{m^2_{2}r^2_{\rm min}}{2}\left[
    \left(\frac{\tilde{\sigma}}{r_{\rm min}}\right)^2
    +\left(\frac{\tilde{\sigma}}{r_{\rm min}}\right)^3
    +\frac{1}{4}\left(\frac{\tilde{\sigma}}{r_{\rm min}}\right)^4
    \right]~\label{eq:U1sym},
\end{align}
where we assume the potential $V_r$ has a local minimum at a point $r_{\rm min}$, which is not the origin, and define $\tilde{\sigma}:=r-r_{\rm min}$ as well as a positive dimensionful parameter $m^2_2$. 
After substituting eq.~\eqref{eq:U1sym} into the action~\eqref{eq:3:3}, the following action is obtained:
\begin{align}
    I_{\rm U(1)}
    &:=\int d^4x\sqrt{-g}\Bigg[
    -\frac{1}{2}\left(1+\frac{\tilde{\sigma}}{r_{\rm min}}\right)^2 (\partial_{\mu}\phi)^2
    -\frac{1}{2}(\partial_{\mu}\tilde{\sigma})^2\notag
    \\
    &\quad\quad\quad-\frac{m^2_{2}r^2_{\rm min}}{2}\left[
    \left(\frac{\tilde{\sigma}}{r_{\rm min}}\right)^2
    +\left(\frac{\tilde{\sigma}}{r_{\rm min}}\right)^3
    +\frac{1}{4}\left(\frac{\tilde{\sigma}}{r_{\rm min}}\right)^4
    \right]-V_{\rm soft}(\phi)
    \Bigg],\label{eq:IU(1)}
\end{align}
where we defined $\phi:=r_{\rm min}\theta$ and omitted the term proportional to the Ricci scalar for brevity. 
We direct our attention toward the dynamical field $\tilde{\sigma}$ under the inflationary background represented by eq.~\eqref{eq:inBG} to assess the relative entropy. 
Following the setup described above, we can determine the coefficients in eqs.~\eqref{eq:L0} and \eqref{eq:LabI} as follows: 
\begin{align}
    &a_0(\phi)=\frac{1}{2} \left(\partial_{\mu}\phi\right)^2+V_{\rm soft}(\phi),~~~a_1(\phi)=\frac{1}{r_{\rm min}} \left(\partial_{\mu}\phi\right)^2,~~~a_2(\phi)=
    \frac{m^2_{2}}{2}+\frac{1}{2r_{\rm min}^2}\left(\partial_{\mu}\phi\right)^2,\notag
    \\
    &a_3(\phi)=\frac{m^2_{2}}{2r_{\rm min}},~~~a_4(\phi)=\frac{m^2_{2}}{8 r_{\rm min}^2},~~~a_k(\phi)=0~{\rm for}~k\geq 5.\label{eq:coU(1)}
\end{align}
Upon incorporating eq.~\eqref{eq:coU(1)} into eqs.~\eqref{eq:WgsclBG} and \eqref{eq:W0sclBG}, the Euclidean effective actions are calculated up to the sixth order of $\lambda$ as, 
\begin{align}
   W_\lambda^{\rm U(1)}[\phi]&= \int (d^4x)_{\rm E}\sqrt{g}\bigg[
    \frac{1}{2} \left(\partial_{\mu}\phi\right)^2+V_{\rm soft}(\phi)-\lambda^2\frac{1}{2m^2_{2}r_{\rm min}^2}\left(\partial_{\mu}\phi\right)^4+\mathcal{O}(\lambda^7)
    \bigg],\label{eq:WgU1}
    \\
    W_0^{\rm U(1)}[\phi]&=\int (d^4x)_{\rm E}\sqrt{g}\bigg[
     \frac{1}{2} \left(\partial_{\mu}\phi\right)^2+V_{\rm soft}(\phi)\bigg],\label{eq:W0U1}
\end{align}
where, in eq.~\eqref{eq:WgU1}, higher derivative terms of $\phi$ vanish for the inflationary background $\phi=\phi_0$ due to the slow-roll approximation. 
Please refer to Appendix~\ref{app:B} for instructions on calculating eq.~\eqref{eq:WgU1}.
In particular, the effective action~\eqref{eq:WgU1} corresponds to the EFT~\eqref{eq:infEFT0}, and the Wilson coefficients in eq.~\eqref{eq:infEFT0} are provided as,
\begin{align}
    c_2=\lambda^2 \frac{2}{m_2^2 r^2_{\rm min}},~~~c_j=0~{\rm for}~j>3.
\end{align}
The relative entropy up to the sixth order of $\lambda$ is determined from eqs.~\eqref{eq:WgU1}, \eqref{eq:W0U1}, and \eqref{eq:SscalBG} as follows: 
\begin{align}
    S(P_{\rm R}||P_{\rm T})_{\rm U(1)}=\lambda^2 \frac{1}{2m^2_{2}r_{\rm min}^2}\int (d^4 x)_{\rm E}\sqrt{g} (\partial_{\mu}\phi)^4+\mathcal{O}(\lambda^7).\label{eq:relU1}
\end{align}
It should be noted here that the relative entropy equals a linear combination of the higher-order operator appearing in the eqs.~\eqref{eq:WgU1} or \eqref{eq:infEFT0}.
From eq.~\eqref{eq:relU1}, we observe that the theory~\eqref{eq:IU(1)} fully upholds the non-negativity of relative entropy.
In contrast, as we will see later, the non-negativity of the relative entropy can break down in the modified case of the U(1)-symmetric potential~\eqref{eq:U1sym}, where perturbative calculations become invalid.

Now, we reconsider the relative entropy for the U(1) symmetric potential~\eqref{eq:U1sym} up to the full order of $\lambda$ and reconfirm its non-negativity. 
From eqs.~\eqref{eq:coU(1)} and \eqref{eq:firsderBG}, the classical solution is obtained as, 
\begin{align}
    \widetilde{\sigma}_\lambda=-r_{\rm min}\pm \sqrt{r_{\rm min}^2-\lambda \frac{2}{m_2^2} (\partial_{\mu}\phi)^2}.\label{eq:solU1}
\end{align}
Upon substituting eq.~\eqref{eq:solU1} into the action~\eqref{eq:IU(1)}, the Euclidean action is calculated up to the full order of $\lambda$ as, 
\begin{align}
    W_\lambda^{\rm U(1)}[\phi]&=\int (d^4x)_{\rm E}\sqrt{g}\bigg[
    \frac{1}{2} \left(\partial_{\mu}\phi\right)^2
    +V_{\rm soft}(\phi)-\lambda^2\frac{1}{2m^2_2r_{\rm min}^2} (\partial_{\mu}\phi)^4
    \bigg].\label{eq:WgU2}
\end{align}
Upon comparing this obtained result with eq.~\eqref{eq:WgU1}, it is apparent that the term $\mathcal{O}(\lambda^7)$ in eq.~\eqref{eq:WgU1} is zero. 
According to eqs.~\eqref{eq:SscalBG} and \eqref{eq:WgU2}, the relative entropy up to the full order of $\lambda$ is expressed as follows: 
\begin{align}
    S(P_{\rm R}||P_{\rm T})_{\rm U(1)}=\lambda^2 \frac{1}{2m_2^2r_{\rm min}^2}\int (d^4 x)_{\rm E}\sqrt{g} (\partial_{\mu}\phi)^4\geq 0.\label{eq:relfull}
\end{align}
Therefore, we have observed that the relative entropy remains non-negative for any perturbation order. 
In this sense, the perturbative calculations of this theory denote a faithful representation of physics, and its U(1) symmetric case~\eqref{eq:U1sym} would be a weakly coupled theory. 
In Ref.~\cite{Kim:2021pbr} (also see Sec.~\ref{sec:non-G} of this paper), it was discussed that the perturbative unitarity of this model is satisfied even in high-energy scattering amplitudes. This is qualitatively consistent with the considerations of relative entropy mentioned above, which implicitly encodes the unitarity.

\subsubsection{A modified version of potential in eq.~\eqref{eq:U1sym}}
\label{sec:3.1.2}
Next, we consider a modified version of the potential in eq.~\eqref{eq:U1sym} at the renormalizable level defined as follows: 
\begin{align}
V_r(r)=\frac{m^2_{2} r^2_{\rm min}}{2}\left[
\left(\frac{\tilde{\sigma}}{r_{\rm min}}\right)^2+\tilde{\lambda}_3 \left(\frac{\tilde{\sigma}}{r_{\rm min}}\right)^3
+\frac{\tilde{\lambda}_4}{4} \left(\frac{\tilde{\sigma}}{r_{\rm min}}\right)^4
\right],\label{eq:Ubpo}
\end{align}
where $\tilde{\lambda}_3$ and $\tilde{\lambda}_4$ are dimensionless parameters characterizing the deviations from eq.~\eqref{eq:U1sym}, and eq.~\eqref{eq:Ubpo} is the same as eq.~\eqref{eq:U1sym} for $\tilde{\lambda}_3=\tilde{\lambda}_4=1$. 
This work focuses on the renormalizable potential in eq.~\eqref{eq:3:1}, where $\phi_1+i\phi_2=r e^{i\theta}$.
Regarding the parameterization of eq.~\eqref{eq:Ubpo}, please also refer to Ref.~\cite{Kim:2021pbr}.
From eq.~\eqref{eq:Ubpo}, the action~\eqref{eq:3:3} is expressed as,
\begin{align}
I_{\cancel{\rm U(1)}}:&= \int d^4x \sqrt{-g}\Bigg[
-\frac{1}{2}\left(1+\frac{\tilde{\sigma}}{r_{\rm min}}\right)^2 (\partial_{\mu}\phi)^2-\frac{1}{2}(\partial_{\mu}\tilde{\sigma})^2\notag
\\
&\quad\quad-\frac{m^2_{2} r_{\rm min}^2}{2}\left[
\left(\frac{\tilde{\sigma}}{r_{\rm min}}\right)^2
+\tilde{\lambda}_3\left(\frac{\tilde{\sigma}}{r_{\rm min}}\right)^3
+\frac{\tilde{\lambda}_4}{4}\left(\frac{\tilde{\sigma}}{r_{\rm min}}\right)^4
\right]-V_{\rm soft}(\phi)
\Bigg],\label{eq:acBR}
\end{align}
where $\phi=r_{\rm min}\theta$ was used. 
In the subsequent Sec.~\ref{sec:three}, we will be discussing a specific UV model in which the displacements of $\tilde{\lambda}_3$ and $\tilde{\lambda}_4$ from unity are induced through integrating an additional heavy field. 
According to eq.~\eqref{eq:acBR}, the coefficients in eqs.~\eqref{eq:L0} and \eqref{eq:LabI} are as follows: 
\begin{align}
 &a_0(\phi)=\frac{1}{2} \left(\partial_{\mu}\phi\right)^2+V_{\rm soft}(\phi),~~~a_1(\phi)=\frac{1}{r_{\rm min}}  \left(\partial_{\mu}\phi\right)^2
 ,~~~a_2(\phi)=\frac{m^2_{2}}{2}+\frac{1}{2 r_{\rm min}^2}\left(\partial_{\mu}\phi\right)^2,\notag
\\
 &a_3(\phi)=\tilde{\lambda}_3\frac{m^2_{2}}{2r_{\rm min}},~~~a_4(\phi)=\tilde{\lambda}_4 \frac{m^2_{2}}{8 r_{\rm min}^2},~~~a_k(\phi)=0~{\rm for}~k\geq 5.\label{eq:co2}
\end{align}
Let us concentrate on the inflationary background and calculate the relative entropy, similar to the U(1) symmetric potential in subsubsection~\ref{sec:U(1)}. 
From eqs.~\eqref{eq:WgsclBG}, \eqref{eq:W0sclBG}, and \eqref{eq:co2}, the Euclidean effective action is calculated, {\it e.g.}, up to the third order of $\lambda$, as follows: 
\begin{align}
    W_\lambda^{\cancel{\rm U(1)}}[\phi]&=\int (d^4 x)_{\rm E} \sqrt{g} \Bigg[\frac{1}{2} \left(\partial_{\mu}\phi\right)^2+V_{\rm soft}(\phi)-\lambda^2 \frac{1}{2m_2^2 r_{\rm min}^2}\left(\partial_{\mu}\phi\right)^4+\lambda^3 \frac{1}{2m_2^4 r_{\rm min}^4}
     \left(1-
    \tilde{\lambda}_3
    \right)
    \left(\partial_{\mu}\phi\right)^6
    +\mathcal{O}(\lambda^4)
    \Bigg],\label{eq:WgU1br}
    \\
    W_0^{\cancel{\rm U(1)}}[\phi]&=\int (d^4 x)_{\rm E} \sqrt{g} \bigg[\frac{1}{2} \left(\partial_{\mu}\phi\right)^2
    +V_{\rm soft}(\phi)\bigg],\label{eq:W0U1br}
\end{align}
where the higher derivative term of $\phi$ does not manifest due to the slow-roll approximation. 
The effective action~\eqref{eq:WgU1br} represents the EFT generated from the UV theory~\eqref{eq:acBR}. 
Upon comparing eq.~\eqref{eq:WgU1br} with eq.~\eqref{eq:infEFT0}, the Wilson coefficients can be derived as, 
\begin{align}
    c_2=\lambda^2 \frac{2}{m^2_2 r^2_{\rm min}},~~~c_3=\lambda^3\frac{4}{m_2^4 r^4_{\rm min}}(1-\tilde{\lambda}_3),\cdots\label{eq:twoscl_c2_c3}
\end{align}
In Appendix~\ref{app:B}, we present the effective action up to the sixth order of $\lambda$. 
Upon combining eqs.~\eqref{eq:SscalBG}, \eqref{eq:WgU1br}, and \eqref{eq:W0U1br}, the relative entropy can be derived as, 
\begin{align}
    S(P_{\rm R}||P_{\rm T})_{\cancel{\rm U(1)}}=\sum_{i=2}^{\infty}\lambda^i \int (d^4 x)_{\rm E} \sqrt{g} S^{(i)}.\label{eq:relBR}
\end{align}
Here, the relative entropy equals a linear combination of higher dimensional operators appearing in the EFT~\eqref{eq:WgU1br}. 
This is because the difference between theories with and without higher dimensional operators is quantified in terms of the relative entropy.
The relative entropy up to the sixth order of $\lambda$ is calculated in Appendix~\ref{app:B}.
The following are the listed results:
\begin{align}
    &S^{(2)}:=\frac{ \left(\partial_{\mu}\phi\right)^4}{2m_2^2r^2_{\rm min}},\label{eq:s2}
    \\
    &S^{(3)}:=- 
     \left(1-
    \tilde{\lambda}_3
    \right)\frac{\left(\partial_{\mu}\phi\right)^6}{2m_2^4r_{\rm min}^4} ,
    \\
    &S^{(4)}:=\left(
    4-12 \tilde{\lambda}_3 +\left(9 \tilde{\lambda}_3^2-\tilde{\lambda}_4\right) 
    \right)\frac{ \left(\partial_{\mu}\phi\right)^8}{8m_2^6 r_{\rm min}^6} ,
    \\
    &S^{(5)}:=-\left(4-4 \tilde{\lambda}_4 -6\tilde{\lambda}_3 \left(4-\tilde{\lambda}_4  \right)+45 \tilde{\lambda}_3^2 -27\tilde{\lambda}_3^3  
    \right)\frac{ \left(\partial_{\mu}\phi\right)^{10}}{8 m_2^8 r_{\rm min}^8} ,\label{eq:S5}
    \\
    &S^{(6)}:=\Bigg[
    8+ \left( -20 \tilde{\lambda}_4+2 \tilde{\lambda}_4^2 \right)+\tilde{\lambda}_3 \left(
    -80 +72  \tilde{\lambda}_4 
    \right)
    +\tilde{\lambda}_3^2 \left(
    270-63 \tilde{\lambda}_4  
    \right)
    -378 \tilde{\lambda}_3^3  +189 \tilde{\lambda}_3^4 
    \Bigg]\frac{ \left(\partial_{\mu}\phi\right)^{12}}{16 m_2^{10} r_{\rm min}^{10}}.\label{eq:S6}
\end{align}
From eq.~\eqref{eq:s2}, it is evident that the second order of $\lambda$ in eq.~\eqref{eq:relBR} is non-negative, and the main order of the operator expansion adheres to the non-negativity of relative entropy. 
In contrast, according to eqs.~\eqref{eq:s2}-\eqref{eq:S6}, the third and higher order terms of $\lambda$ may take negative values. 
That is, when the operator expansions in the EFT are truncated at finite order, the non-negativity of relative entropy can be compromised, particularly for large values of $\tilde{\lambda}_3$ and $(\partial_{\mu}\phi)^2/m_2^2 r^2_{\rm min}$. 
Hence, it can be inferred from our analysis that the non-negativity of relative entropy, serving as an implicit representation of the unitarity of the theory, delineates the parameter region wherein the operator $(\partial_{\mu}\phi)^2/m_2^2 r^2_{\rm min}$ expansion, specifically the finite truncation of the $(\partial_{\mu}\phi)^2/m_2^2 r^2_{\rm min}$ power series, holds quantitative validity. 
As demonstrated in Sec.~\ref{sec:non-G}, the non-linear parameter $f_{\rm NL}$ in the theory~\eqref{eq:acBR} exhibits a proportional increase relative to $\tilde{\lambda}_3$ and $(\partial_{\mu}\phi)^2/m_2^2 r^2_{\rm min}$. 
From these, in Sec.~\ref{sec:non-G}, it is revealed that the non-negativity of relative entropy imposes constraints on the parameter space within which observable non-Gaussianity with $|f_{\rm NL}|\gtrsim 1$ can be realized while also ensuring the validity of the EFT description.

\begin{figure}
\centering
\includegraphics[width=0.9\textwidth]{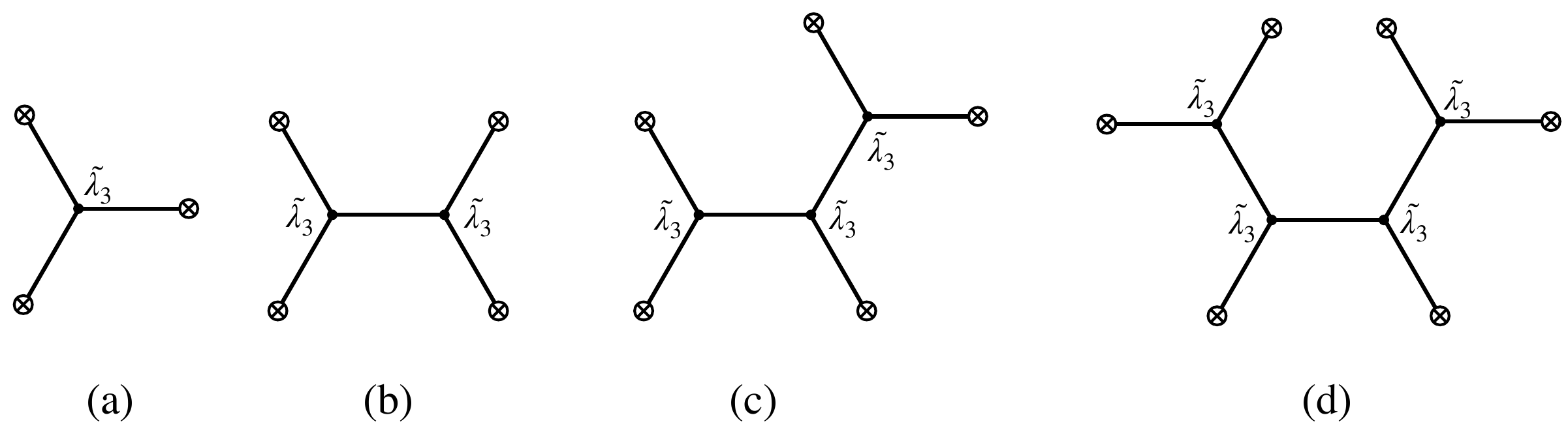}
\caption{
The dominant Feynman diagrams contributing to $S^{(j)}$ in $1\ll|\tilde{\lambda}_3|$ regimes.
The diagrams (a), (b), (c), and (d) represent the contributions to $S^{(3)}$, $S^{(4)}$, $S^{(5)}$, and $S^{(6)}$, respectively. 
The black and the circled cross vertices indicate the interaction proportional to $\tilde{\lambda}_3$ and $a_1(\phi)$, respectively.
For $\tilde{\lambda}_3>0$, the odd order of the perturbative expansion in the relative entropy takes negative values while the even ones are positive because of $S^{(j)}\propto (\tilde{\lambda}_3)^{j-2}(-\dot{\phi}_0^2)^{j}$, where $j>3$.
In contrast, for $\tilde{\lambda}_3<0$, the non-negativity of relative entropy holds.
}
\label{fig:eff}
\end{figure}

Before conducting numerical evaluations of the relative entropy~\eqref{eq:relBR}, we present an analysis of the behavior of the relative entropy in the large $\tilde{\lambda}_3$ region.
Let us consider the cases $ 1 \ll |\tilde{\lambda}_3|$ and $|\tilde{\lambda}_4|\sim 1$. 
The dominant contribution to $S^{(j)}$ for $j\geq 3$ is then proportional to $(\tilde{\lambda}_3)^{j-2}$, as illustrated in the Feynman diagrams in Fig.~\ref{fig:eff}. 
From eqs.~\eqref{eq:s2}\text{-}\eqref{eq:S6}, we can further comprehend this behavior. 
After substituting the background field $(\partial_{\mu}\phi)^2=-\dot{\phi}_0^2<0$, it is important to note that for $0<\tilde{\lambda}_3$ and $1\ll|\tilde{\lambda}_3|$, the relative entropy is positive at even orders of the perturbative expansions and negative at odd orders of the perturbative expansions. 
In short, this denotes that the relative entropy lacks a clear sign in the perturbative calculations for the strongly coupled theory. 

We will now proceed to the numerical assessment of the relative entropy. 
Let us normalize the relative entropy of \eqref{eq:relBR} for $S^{(2)}$ and define the normalized relative entropy up to the $j$-th order of the perturbative expansion as,
\begin{align}
       s^{(j)}:=\sum_{i=2}^{j}
       \frac{S^{(i)}}{S^{(2)}},\label{eq:nomrelJ}
\end{align}
where $\lambda=1$ has been selected.
In the weakly coupled theory, the perturbative calculations faithfully describe physics, and due to the non-negativity of relative entropy, $s^{(j)}$ should be non-negative regardless of $j$. 
Due to the non-negativity of relative entropy, eq.~\eqref{eq:nomrelJ} should satisfy the following inequalities for the weakly coupled theory, {\it e.g.}, for $j=3$ and 5, 
    \begin{align}
    s^{(3)}&=1+\left(1-\tilde{\lambda}_3 \right)\left(\frac{\dot{\phi}_0}{m_2 r_{\rm min}}\right)^2\gtrsim 0,\label{eq:nomrel3}
    \\
    s^{(5)}&=1+\left(1-\tilde{\lambda}_3 \right)\left(\frac{\dot{\phi}_0}{m_2 r_{\rm min}}\right)^2+\frac{1}{4}\left(
    4 -12 \tilde{\lambda}_3 
    +9\tilde{\lambda}_3^2 -\tilde{\lambda}_4
    \right)\left(\frac{\dot{\phi}_0}{m_2 r_{\rm min}}\right)^4\notag
    \\
    &+\frac{1}{4}\left[
    4-4\tilde{\lambda}_4 
    -6 \tilde{\lambda}_3 \left(
    4 -\tilde{\lambda}_4 
    \right)
    +45\tilde{\lambda}_3^2 -27 \tilde{\lambda}_3^3 
    \right]\left(\frac{\dot{\phi}_0}{m_2 r_{\rm min}}\right)^6\gtrsim 0,\label{eq:nomrel5}
    \end{align}
    where we assume that the symbol $\gtrsim$ in this paper indicates that the relative entropy is calculated approximately ({\it i.e.}, perturbatively) and approximately satisfies non-negativity.
In the following, we will analyze $s^{(j)}$ numerically and present the results for two typical parameter sets.

\begin{itemize}
 \item $\tilde{\lambda}_4=1$ and $\tilde{\lambda}_3=10^2$  ---  We select a positive value for $\tilde{\lambda}_3$ to examine the non-negativity of relative entropy and its violation in the perturbative calculations. 
    From eqs.~\eqref{eq:s2}\text{--}\eqref{eq:S6}, the $j$-th order effect of the perturbative expansion to the normalized relative entropy~\eqref{eq:nomrelJ} increases linearly with $(\dot{\phi}_0/m_2 r_{\rm min})^{2j-4}$. 
    In the left panel of Fig.~\ref{fig:lam31p5}, we show the normalized relative entropy up to the $j$-th order as a function of $\dot{\phi}_0/m_2 r_{\rm min}$.
    The grey line shows $s^{(2)}$, {\it i.e.,} the main contribution in the perturbative calculation. 
    The red, blue, green, and purple curves represent $s^{(3)}$, $s^{(4)}$, $s^{(5)}$, and $s^{(6)}$, respectively. 
    We also display the numerically evaluated $s^{(\infty)}$ (full order result at the tree-level) as a black dotted curve. 
    In the small $\dot{\phi}_0/m_2 r_{\rm min}$ regions, the colored curves almost coincide with the grey line and black dotted curve, and it is clear that the perturbative calculation is working well. 
    As $\dot{\phi}_0/m_2 r_{\rm min}$ increases, the colored curves deviate from the grey line and the black dotted curve, and the perturbative calculations become progressively invalid. 
    In particular, $s^{(3)}$ (red) almost reaches 10\%-100 \% of $s^{(2)}$ (grey) and $s^{(\infty)}$(black dotted) around $\dot{\phi}_0/m_2 r_{\rm min}\sim 0.1$, and the non-negativity of relative entropy is broken in the third order of the perturbative expansion.
     In regions where the value of $\dot{\phi}_0/m_2 r_{\rm min}$ exceeds $\sim 0.1$, the sign of relative entropy changes with each higher order of $\lambda$, rendering the perturbative calculation an inaccurate representation of physics. 
    The variation in the sign of the relative entropy, depending on the order of the perturbative expansion, can also be understood from Fig.~\ref{fig:eff}. 
    The non-negativity of relative entropy is linked to fundamental physics properties, notably unitarity. When the sign of the relative entropy changes, it indicates pathological behavior in the perturbative calculations, and the EFT requires evaluation incorporating all orders of the operator expansion.

    In the right panel of Fig.~\ref{fig:lam31p5}, we display the normalized relative entropy of the $j$-th order of $\lambda$ compared to that of the second order of $\lambda$ in percentage, {\it i.e.}, $100\times|s^{(j)}-s^{(2)}|/|s^{(2)}|$, as a function of $\dot{\phi}_0/m_2 r_{\rm min}$. 
    Here, the $s^{(2)}=1$ from eq.~\eqref{eq:nomrelJ}.
    The grey line indicates that its size is 100 \%. 
    The curves represent the third, fourth, fifth, and sixth orders of $\lambda$ of the normalized relative entropy compared to their second order of $\lambda$, with red, blue, green, and purple, respectively. 
    The black dotted curve displays its numerically evaluated magnitude at an infinite order of $\lambda$. 
    In regions where $\dot{\phi}_0/m_2 r_{\rm min}$ is considerably smaller than $\sim 0.1$, each order effect is less than 100 \% compared to $s^{(2)}$, indicating the validity of the perturbative calculation. 
    For instance, the sign of the relative entropy up to the third and fifth orders of $\lambda$ changes when the red and green curves intersect the grey line, respectively. 
    In the regions with $\dot{\phi}_0 /m_2 r_{\rm min}$ larger than $\sim 0.1$, the effect of each finite order is more than 100 \% compared to $s^{(2)}$, and the perturbation theory is invalid. 
    Based on the analysis of these panels, it is evident that the points where the sign flips of the relative entropy occur represent the boundaries at which the operator expansion is either valid or invalid and quantify the validity of the EFT description. 
    This fact can be understood analytically through the order-of-magnitude evaluation for $j>2$, $|S^{(j)}/S^{(j-1)}|\sim |S^{(3)}/S^{(2)}|$. 
    According to eq.~\eqref{eq:s2}, $S^{(2)}$ is non-negative, and the non-negativity breaking of $s^{(3)}$ (representing $S^{(2)}+S^{(3)}\sim 0$) indicates $|S^{(j)}/S^{(j-1)}|\sim 1$ for $j>2$, as confirmed by numerical calculation. 
    At this juncture, it becomes unfeasible to quantitatively truncate the operator expansion beyond the third order, causing a breakdown in the operator expansion (or the EFT description).

    \begin{figure}
\centering
\includegraphics[width=0.48\textwidth]{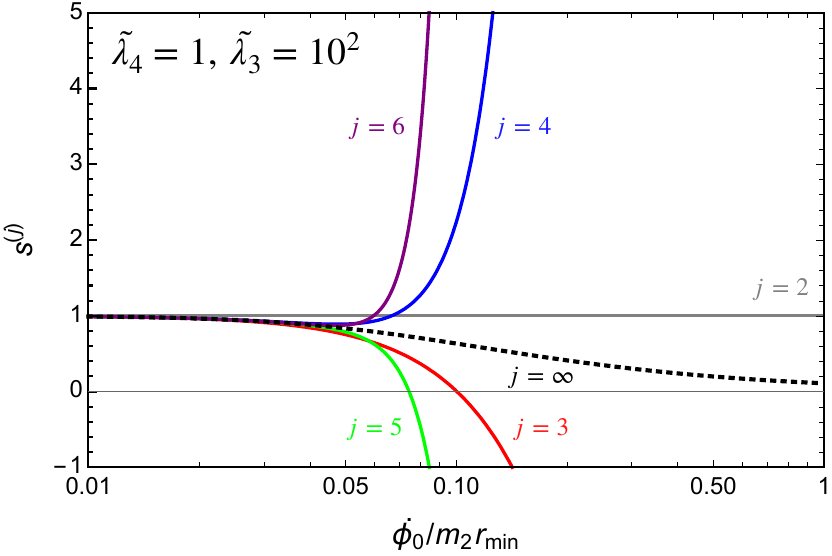}
\includegraphics[width=0.49\textwidth]{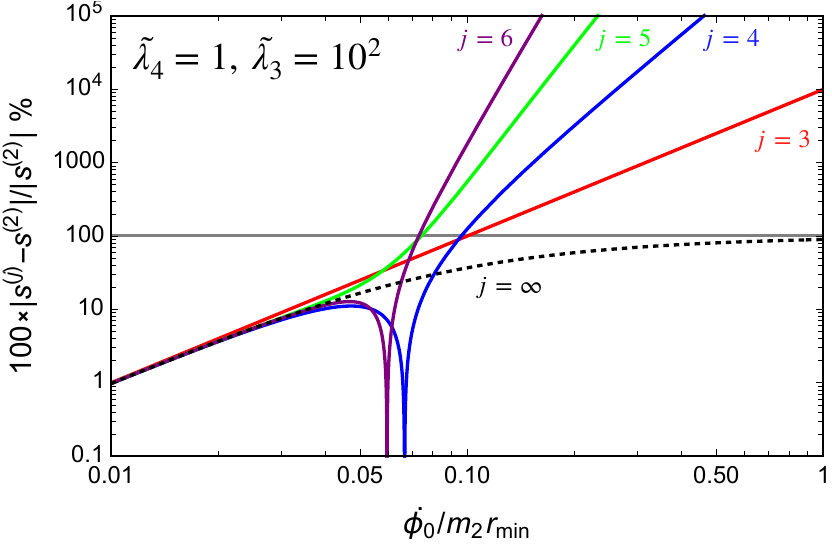}
\caption{
  Left panel: Normalized relative entropies, $s^{(2)}$ (gray), $s^{(3)}$ (red), $s^{(4)}$ (blue), $s^{(5)}$ (green), $s^{(6)}$ (purple), and $s^{\infty}$ (black dotted) as a function of $\dot{\phi}_0/m_2 r_{\rm min}$.
  Right panel: The size of normalized relative entropies, $s^{(3)}$ (red), $s^{(4)}$ (blue), $s^{(5)}$ (green), $s^{(6)}$ (purple), and $s^{(\infty)}$ (black dotted) compared with $s^{(2)}$ as a function of $\dot{\phi}_0/m_2 r_{\rm min}$, respectively.
  The parameters are assumed to be $\tilde{\lambda}_4=1$ and $\tilde{\lambda}_3=10^2$.
}
\label{fig:lam31p5}
\end{figure}

    \item $\tilde{\lambda}_4=1$ and $\dot{\phi}_0/m_2 r_{\rm min}=10^{-1}$ --- 
    Figure~\ref{fig:lam310} depicts a plot similar to Fig.~\ref{fig:lam31p5}, assuming $\tilde{\lambda}_4=1$ and $\dot{\phi}_0/m_2 r_{\rm min}=10^{-1}$ with a horizontal axis of $\tilde{\lambda}_3-1$. 
    Compared to Fig.~\ref{fig:lam31p5}, the relative entropy shows almost the same qualitative behavior, except for the difference in the horizontal axis. 
    In particular, it can be seen that, as shown in Fig.~\ref{fig:lam31p5}, the regions where the perturbation theory is valid are characterized by the point where a sign flip in the relative entropy occurs. 

    \begin{figure}
\centering
\includegraphics[width=0.48\textwidth]{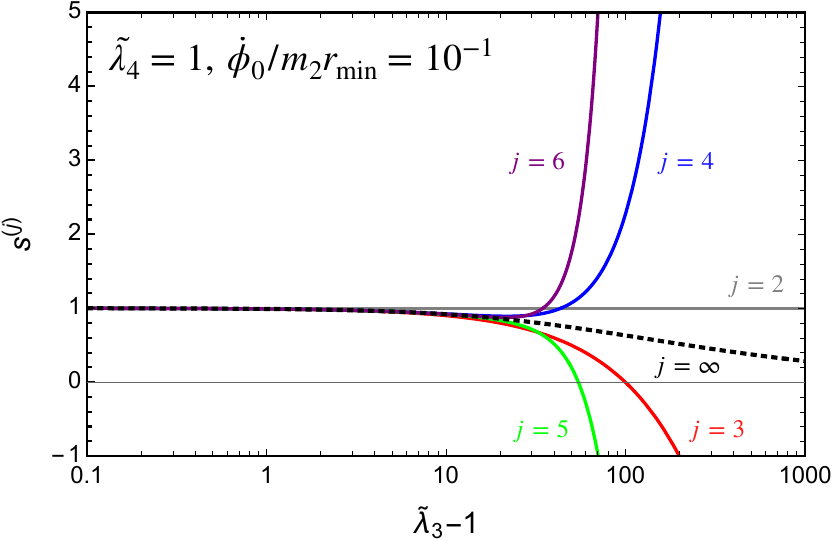}
\includegraphics[width=0.49\textwidth]{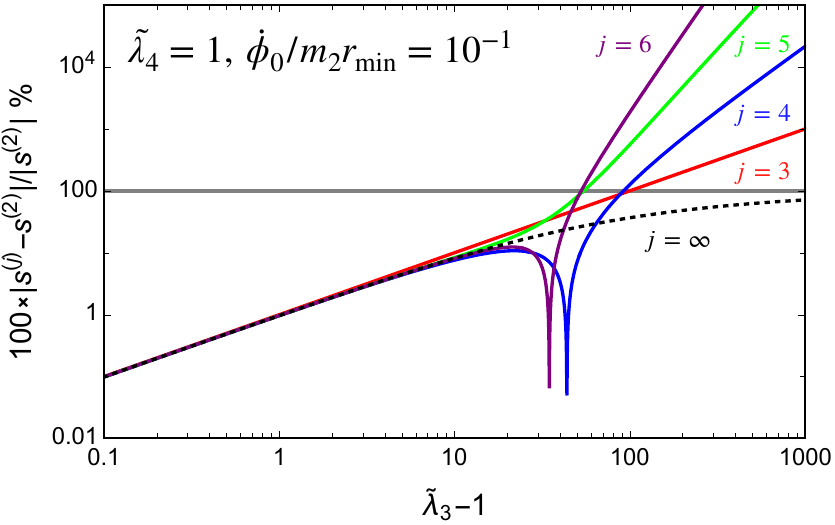}
\caption{
    Left panel: Normalized relative entropies, $s^{(2)}$ (black), $s^{(3)}$ (red), $s^{(4)}$ (blue), $s^{(5)}$ (green), $s^{(6)}$ (purple), and $s^{(\infty)}$ (black dotted) as a function of $\tilde{\lambda}_3-1$.
  Right panel: The size of normalized relative entropies, $s^{(3)}$ (red), $s^{(4)}$ (blue), $s^{(5)}$ (green), $s^{(6)}$ (purple), and $s^{(\infty)}$ (black dotted) compared with $s^{(2)}$ as a function of $\tilde{\lambda}_3-1$, respectively.
  The parameters are assumed to be $\tilde{\lambda}_4=1$ and $\dot{\phi}_0/m_2 r_{\rm min}=10^{-1}$.
  The blue and purple curves in the right panel touch zero value.
  However, because of the plot's resolution, those curves disappear in small-value of $100\times|s^{(j)}-s^{(2)}|/|s^{(2)}|$ regions.
}
\label{fig:lam310}
\end{figure}
    
\end{itemize}

    Through the numerical calculations, it has been observed that there are parameter regions in which the relative entropy does not maintain a consistent sign in the perturbative calculations. 
In this example, the non-negativity of relative entropy is preserved in the full-order tree-level calculations. Therefore, as outlined in Sec.~\ref{sec:weak}, if the non-negativity of relative entropy is perturbatively violated, it indicates a breaking down of the operator expansion in the EFT.
The critical point is that the sign flip of the perturbative relative entropy represents that the operator expansion (EFT description) is no longer a faithful representation of physics. This is because its non-negativity is connected with fundamental properties such as the unitarity of the theory (see also Ref.~\cite{Cao:2022iqh,Cao:2022ajt}).
These numerical results suggest that the parameter regions where the EFT description is valid can be identified using relative entropy. Theories that violate the non-negativity of relative entropy are invalid for the EFT description, indicating they are strongly coupled.
In Sec.~\ref{sec:non-G}, we apply this property to analyze the relation between the non-Gaussianity of the single-field inflation EFT~\eqref{eq:infEFT0} and the parameter regions where its EFT description is valid.

\subsection{Three scalar fields theory}
\label{sec:three}
We consider a scenario where a new heavy field generates the displacements of $\tilde{\lambda}_3$ and $\tilde{\lambda}_4$ from unity. 
Consider the following three scalar fields theory,
\begin{align}
    {I}^{(3)}:&=\int d^4 x\sqrt{-g}\left[\frac{1}{2}M^2_{\rm Pl} R
    -\sum_{i=1}^3 \frac{1}{2}(\partial_{\mu}\phi_i)^2-V^{(3)}(\phi_i)\right],\label{eq:actthr}
\end{align}
where $\phi_1$ and $\phi_2$ are the same degrees of freedom as the scalar fields appearing in eq.~\eqref{eq:3:1}, and $\phi_3$ is a real scalar field with a mass heavier than $\phi_1$ and $\phi_2$, and it behaves as a new particle in the viewpoint of the theory~\eqref{eq:3:1}, and $V^{(3)}(\phi_i)$ is a potential of this theory. 
To preserve the flatness of the inflaton potential as in eq.~\eqref{eq:3:3}, we assume U(1) symmetry under a phase shift of $\phi_1+i\phi_2=r e^{i\theta}$ in the potential $V^{(3)}(\phi_i)$. 
By using the fields $r, \theta$ and $\chi:=\phi_3$, the action \eqref{eq:actthr} can be specified at renormalizable level as follows:
\begin{align}
    I^{(3)}&=\int d^4x \sqrt{-g}\Bigg[
    -\frac{1}{2}\left(1+\frac{\sigma}{r_{\rm min}}\right)^2 (\partial_{\mu}\phi)^2
    -\frac{1}{2}(\partial_{\mu}\sigma)^2-\frac{m^2_2 r^2_{\rm min}}{2}\left[
    \left(\frac{\sigma}{r_{\rm min}}\right)^2+ \left(\frac{\sigma}{r_{\rm min}}\right)^3 +\frac{1}{4}\left(\frac{\sigma}{r_{\rm min}}\right)^4 
    \right]\notag
    \\
    &\quad\quad\quad\quad\quad\quad-\frac{1}{2}(\partial_{\mu}\chi)^2-\frac{m^2_3 r_{\rm min}^2}{2} \left[\left(\frac{\chi}{r_{\rm min}}\right)^2+\frac{\kappa_3}{3}\left(\frac{\chi}{r_{\rm min}}\right)^3+\frac{\kappa_4}{12}\left(\frac{\chi}{r_{\rm min}}\right)^4\right]\notag
    \\
&\quad\quad\quad\quad\quad\quad
    -\frac{\beta r_{\rm min}^3}{2}\left[2\left(\frac{\sigma}{r_{\rm min}}\right)+\left(\frac{\sigma}{r_{\rm min}}\right)^2\right]\chi
-\frac{\gamma r_{\rm min}^2}{2}\left[2\left(\frac{\sigma}{r_{\rm min}}\right)+\left(\frac{\sigma}{r_{\rm min}}\right)^2\right]\chi^2
-V_{\rm soft}(\phi)
\Bigg],\label{eq:I33}
\end{align}
where $V_{\rm soft}(\phi)$ is a term that softly breaks the shift symmetry of $V^{(3)}(\phi_i)$, $\kappa_3, \kappa_4, \beta$ and $\gamma$ are dimensionless parameters, and $m_3$ is the mass of $\chi$ and is assumed to be $m_2\leq  m_3$. 
Here, for brevity, we omit the term proportional to the Ricci scalar.
From eq.~\eqref{eq:I33}, for $N=3$, the coefficients appearing in eq.~\eqref{eq:LN} are given as follows: 
\begin{align}
    &a_0^{(3)}\left(\phi,\sigma\right)=\frac{1}{2}\left(1+\frac{\sigma}{r_{\rm min}}\right)^2(\partial_{\mu}\phi)^2+V_{\rm soft}(\phi)+\frac{1}{2}(\partial_{\mu}\sigma)^2+\frac{m^2_2 r^2_{\rm min}}{2}\left[
    \left(\frac{\sigma}{r_{\rm min}}\right)^2+ \left(\frac{\sigma}{r_{\rm min}}\right)^3 +\frac{1}{4}\left(\frac{\sigma}{r_{\rm min}}\right)^4 
    \right],\notag
    \\
    &a_1^{(3)}\left(\phi,\sigma\right)=\frac{\beta r_{\rm min}^3}{2}\left[2\left(\frac{\sigma}{r_{\rm min}}\right)+\left(\frac{\sigma}{r_{\rm min}}\right)^2\right],~~
    a_2^{(3)}\left(\phi,\sigma\right)=\frac{m^2_3}{2}+\frac{\gamma r_{\rm min}^2}{2}\left[2\left(\frac{\sigma}{r_{\rm min}}\right)+\left(\frac{\sigma}{r_{\rm min}}\right)^2\right],\notag
    \\
    &a_3^{(3)}\left(\phi,\sigma\right)=\frac{\kappa_3}{3!}\frac{m_3^2}{r_{\rm min}},~~a_4^{(3)}\left(\phi,\sigma\right)=\frac{\kappa_4}{4!}\frac{m_3^2}{r_{\rm min}^2},~~a_k^{(3)}\left(\phi,\sigma\right)=0~{\rm for}~k\geq 5.\label{eq:coethree}
\end{align}
Upon substituting eq.~\eqref{eq:coethree} into eq.~\eqref{eq:NWg}, the Euclidean effective action, up to the fourth order of the perturbative expansion, is calculated as follows: 
\begin{align}
{I}^{(2)}:&=\int (d^4x)_{\rm E}\sqrt{g}\Bigg[
\frac{1}{2}\left(1+\frac{\sigma}{r_{\rm min}}\right)^2 (\partial_{\mu}\phi)^2 +\frac{1}{2}(\partial_{\mu}\sigma)^2
+\frac{m^2_2 r_{\rm min}^2}{2}\left[
\left(\frac{\sigma}{r_{\rm min}}\right)^2
+\left(\frac{\sigma}{r_{\rm min}}\right)^3
+\frac{1}{4}\left(\frac{\sigma}{r_{\rm min}}\right)^4
\right]\notag
\\
&\quad\quad\quad\quad-\frac{\beta^2 r_{\rm min}^6}{8m_3^2}\left[
2\left(\frac{\sigma}{r_{\rm min}}\right)+\left(\frac{\sigma}{r_{\rm min}}\right)^2
\right]^2
-\frac{\beta^2 r_{\rm min}^8}{48m_3^4}\left(
-6\gamma +\beta \kappa_3
\right)\left[
2\left(\frac{\sigma}{r_{\rm min}}\right)
+\left(\frac{\sigma}{r_{\rm min}}\right)^2
\right]^3\notag
\\
&\quad\quad\quad\quad-\frac{\beta^2 r_{\rm min}^{10}}{384 m_3^6}
\left\{48 \gamma^2-24 \beta \gamma\kappa_3+\beta^2 \left(3\kappa_3^2-\kappa_4\right)\right\}
\left[
2\left(\frac{\sigma}{r_{\rm min}}\right)+\left(\frac{\sigma}{r_{\rm min}}\right)^2
\right]^4+V_{\rm soft}(\phi)
\Bigg].
\end{align}
Then, for $N=2$, the coefficients appearing in eq.~\eqref{eq:LN} are given as follows: 
\begin{align}
    &a_0^{(2)}(\phi)=\frac{1}{2} (\partial_{\mu}\phi)^2+V_{\rm soft}(\phi),~~a_1^{(2)}(\phi)=\frac{1}{r_{\rm min}} (\partial_{\mu}\phi)^2,~~a_2^{(2)}(\phi)=\frac{m^2_{\sigma}}{2}+\frac{1}{2r^2_{\rm min}} (\partial_{\mu}\phi)^2,\notag
    \\
    &a_3^{(2)}(\phi)= 
    \left\{
    1-\frac{\beta^2 r_{\rm min}^6}{3m^2_{\sigma} m_3^4}\left(-6\gamma+\beta\kappa_3\right)
    \right\}\frac{m^2_{\sigma}}{2r_{\rm min}},\notag
    \\
    &a_4^{(2)}(\phi)=
\left[
1+\frac{\beta^2 r_{\rm min}^6}{3m^2_{\sigma}m_3^4}\left\{
6\left(6\gamma-\beta\kappa_3\right)
    +\left\{-3 \left(-4\gamma+\beta\kappa_3\right)^2+\beta^2\kappa_4\right\}\frac{r_{\rm min}^2}{m_3^2}
\right\}
\right]\frac{m^2_{\sigma}}{8r_{\rm min}^2},\label{eq:threeN}
\end{align}
where $m^2_{\sigma}:=m^2_2-\beta^2 r_{\rm min}^4/m_3^2$ denotes the squared mass of $\sigma$.
By comparing eq.~\eqref{eq:threeN} with eq.~\eqref{eq:co2}, we find that the parameters $\tilde{\lambda}_3$ and $\tilde{\lambda}_4$ are generated from the new particle $\chi$ as follows: 
\begin{align}
    &\tilde{\lambda}_3= 1-\frac{\beta^2 r_{\rm min}^6}{3m^2_{\sigma} m_3^4}\left(-6\gamma+\beta\kappa_3\right),\label{eq:lam3three}
    \\
    &\tilde{\lambda}_4=1+\frac{\beta^2 r_{\rm min}^6}{3m^2_{\sigma}m_3^4}\left[
6\left(6\gamma-\beta\kappa_3\right)
    +\left\{-3 \left(-4\gamma+\beta\kappa_3\right)^2+\beta^2\kappa_4\right\}\frac{r_{\rm min}^2}{m_3^2}
\right],\label{eq:lam4three}
\end{align}
Then, according to Sec.~\ref{sec:3.1.2}, the low-energy EFT~\eqref{eq:infEFT0} is generated by integrating the scalar field $\sigma$. 
The Wilson coefficients of the EFT~\eqref{eq:infEFT0} (or effective action~\eqref{eq:WgU1br}) is then obtained as follows: 
\begin{align}
    c_2=\lambda^2 \frac{2}{m^2_{\sigma} r^2_{\rm min}},~~~c_3=\lambda^3 \frac{4}{m^4_{\sigma}r^4_{\rm min}}\left[\frac{\beta^2 r_{\rm min}^6}{3m^2_{\sigma} m_3^4}\left(-6\gamma+\beta\kappa_3\right)\right]
.
\end{align}
The relative entropy can also be calculated by substituting eqs.~\eqref{eq:lam3three} and \eqref{eq:lam4three} into eq.~\eqref{eq:relBR}.
From these examples, we see that, in weakly coupled UV theories, the relative entropy is generally a linear combination of higher-order operators arising in the EFT.

\subsection{An EFT of inflation generated from more general UV theories }
The above discussions have focused on cases where particular UV theories, {\it e.g.,} eqs.~\eqref{eq:acBR} or \eqref{eq:actthr}, are assumed. 
In such situations, even if the description of physical phenomena by the EFT~\eqref{eq:infEFT0} is no longer possible due to truncation at finite orders of the operator expansion, it is still possible to describe the physical phenomena with the particular UV theories. 
Therefore, the importance of assessing the validity of the perturbative calculations in terms of the relative entropy might be phenomenologically unclear. 
However, when conducting UV model-independent analyses using the EFT~\eqref{eq:infEFT0}, as primarily discussed in this subsection, it is phenomenologically meaningful to assess the validity of the perturbative calculation and the EFT description in terms of relative entropy. 
Because, as mentioned around the beginning of Sec.~\ref{sec:quasi}, EFT-based analysis of physical phenomena has the advantage of making quantitative predictions independent of UV theories and revealing parameter regions where the validity of the EFT is broken has a role in ensuring the reliability of the predictions made by the EFT.

Similar to the two subsections above, consider the case where the higher dimensional operators ($X^{j+1}$ for $j>1$) in eq.~\eqref{eq:infEFT0} are generated by integrating the heavy fields at the tree-level through the interaction between heavy and light fields. 
It is important to emphasize that we assume that the EFT~\eqref{eq:infEFT0} is generated through the interactions from some unknown UV theory, but we are not considering any specific UV theory. 
The Euclidean effective actions at the tree-level are then obtained as follows: 
\begin{align}
    W_{\lambda=1}[\phi]&=\int (d^4x)_{\rm E}\sqrt{g}\left[
    -X-\sum_{j=1}c_{j+1}X^{j+1}+V_{\rm soft}(\phi)
    \right],\label{eq:EFTWg}
    \\
    W_0[\phi]&=\int (d^4x)_{\rm E}\sqrt{g}\left[
    -X+V_{\rm soft}(\phi)
    \right],
\end{align}
where we consider the inflationary background~\eqref{eq:inBG}, and $X=\dot{\phi}_0^2/2$.
From eq.~\eqref{eq:SscalBG}, the relative entropy is given by, 
\begin{align}
    S(P_{\rm R}||P_{\rm T})&=W_0[\phi]-W_{\lambda=1}[\phi]=\int (d^4x)_{\rm E}\sqrt{g} \sum_{j=1}c_{j+1} X^{j+1}.
\end{align}
Here, we assume that EFT~\eqref{eq:EFTWg} is generated from the weakly coupled theory and that truncating at any order of the $X$ expansion is quantitatively valid. 
The non-negativity of relative entropy then implies, for instance, that the following inequalities hold: 
\begin{align}
    &c_{2}\int (d^4x)_{\rm E} \sqrt{g} X^{2}\gtrsim 0\Rightarrow c_2\gtrsim 0,\label{eq:bo1}
    \\
    &c_{2}\int (d^4x)_{\rm E} \sqrt{g} X^{2}+c_{3}\int (d^4x)_{\rm E}\sqrt{g}  X^{3}\gtrsim 0 \Rightarrow c_2+\frac{c_3}{2}\dot{\phi}_0^2\gtrsim 0,\label{eq:bo2}
\end{align}
where the first line represents that the main contribution to the relative entropy of the $X$ expansion is non-negative, and the second line says that the relative entropy is non-negative up to the third order of $X$.
Although we have focused on the operator expansions of the relative entropy up to the third order, similar inequalities hold for any finite order of $X$ if the truncation at any order of the $X$ expansion is quantitatively valid.

To consider the phenomenological implications of the above inequalities~\eqref{eq:bo1} and \eqref{eq:bo2}, let us imagine the following situation. 
Consider the case where the Wilson coefficients $c_{2}$, $c_{3}$, etc., are determined by evaluating the observables and comparing them with experimental data. 
If those values of the Wilson coefficients do not satisfy the above inequalities~\eqref{eq:bo1} and \eqref{eq:bo2}, considerations based on the relative entropy suggest the following possibilities: 
(i) the physical phenomena are not described by the EFT~\eqref{eq:infEFT0}, including the breakdown of the operator expansion; 
(ii) the physical phenomena are described by the EFT~\eqref{eq:infEFT0}, but the loop-level contribution to the Wilson coefficients dominates; or 
(iii) the EFT~\eqref{eq:infEFT0} describe the phenomena, but the EFT~\eqref{eq:infEFT0} is not generated through the interaction of heavy and light fields. 
For each of the three cases above, we mention some possibilities. 
In case (i), for instance, the phenomenon is described by a theory with entirely different degrees of freedom and symmetries than the EFT~\eqref{eq:infEFT0}, or the operator expansion breaks down, as seen in the sections above, meaning that UV theory, which also includes heavy degrees of freedom, is needed to describe the phenomenon.
In case (ii), the UV theory of EFT~\eqref{eq:infEFT0} is the strongly coupled theory, meaning that the perturbation theory does not represent physics. 
In particular, this case implies that the UV theory is the strongly coupled theory, in the same sense as the $S$-matrix-based perturbative unitarity argument. 
Case (iii) includes, for instance, the possibility that there is no UV theory corresponding to the EFT~\eqref{eq:infEFT0}, which could lead to pathological behavior, such as unitarity violation. 
The fact that physical models can belong to these three cases is expected to provide valuable insights into identifying models describing physical phenomena. 
In the next section, we will see that there are parameter regions allowed in non-Gaussianity experimental data that suggest the possibility of (i), (ii), and (iii).

\section{Non-Gaussianity and non-negativity of relative entropy}
\label{sec:non-G}
In this section, we analyze the non-linear parameter in the EFT of inflation~\eqref{eq:infEFT0} and show a relation between it and the regions where the relative entropy suggests the EFT~\eqref{eq:infEFT0} is valid.
First of all, we define the fluctuation around the inflationary background of \eqref{eq:inBG} as $\pi:=(\phi-\phi_0)/\dot{\phi}_0$ to list the power spectrum and nonlinear parameter. 
Then, in the spatially flat
gauge, the action~\eqref{eq:infEFT0} up to the third order of $X$ can be rewritten as follows: 
\begin{align}
    I_{X}&=\int dtd^3x a^3 \Bigg[
    -\frac{1}{2}\dot{\phi}_0^2\left(1+c_2 \dot{\phi}_0^2+\frac{3}{4}c_3 \dot{\phi}_0^4\right)\left(-\dot{\pi}^2+\frac{( \partial_i \pi)^2}{a^2}\right)\notag
    \\
    &+\frac{1}{4}\left(
    c_2 \dot{\phi}_0^4 +\frac{3}{2}c_3 \dot{\phi}_0^6
    \right)\left(
    -2\dot{\pi}-\dot{\pi}^2 +\frac{(\partial_i \pi)^2}{a^2}
    \right)^2
    -\frac{1}{8} c_3 \dot{\phi}_0^6\left(
    -2\dot{\pi}-\dot{\pi}^2 +\frac{(\partial_i \pi)^2}{a^2}
    \right)^3
    \Bigg]+\mathcal{O}(X^4),\label{eq:3:15}
\end{align}
where we omit fluctuation-independent terms and $V_{\rm soft}$ for the sake of brevity. 
To evaluate the non-linear parameter, we perform matching of eq.~\eqref{eq:3:15} to the following EFT of inflation~\cite{Cheung:2007st}, 
\begin{align}
    I_{\rm EFT}:=\int dt d^3 x a^3 \bigg[
    &-\frac{M^2_{\rm Pl}\dot{H}}{c^2_s}
    \left(\dot{\pi}^2-c_s^2\frac{\left(\partial_i \pi\right)^2}{a^2}\right)\notag
    \\
   & +
    M^2_{\rm Pl} \dot{H} \left(c^{-2}_s-1\right)
    \left(
    \dot{\pi}\frac{\left(\partial_i \pi\right)^2}{a^2}
    -\left(1+\frac{2}{3} \frac{\tilde{c}_3}{c^2_s}\right)\dot{\pi}^3
    \right)+\mathcal{O}(\pi^4)
    \bigg],
\end{align}
where the parameter $M^2_{\rm Pl} \dot{H}$ is given as,
\begin{align}
	&M^2_{\rm Pl} \dot{H}=-\frac{1}{2}\dot{\phi}_0^2  \left(
	1+c_2 \dot{\phi}_0^2 +\frac{3}{4}c_3\dot{\phi}_0^4
	\right),
\end{align}	
The speed of sound $c_s$, and a dimensionless parameter $\tilde{c}_3$ are given by \cite{Kim:2021pbr}, 
\begin{align}
c^{-2}_s =1+2\dot{\phi}_0^2 \frac{c_2 +\frac{3}{2} c_3 \dot{\phi}_0^2}{1+c_2 \dot{\phi}_0^2+\frac{3}{4}c_3 \dot{\phi}_0^4}
,~~~
    \tilde{c}_3 \left(c^{-2}_s -1\right)=\frac{3}{2}\frac{c_3\dot{\phi}_0^2}{c_2+\frac{3}{2}c_3 \dot{\phi}_0^2}c_s^2(c_s^{-2}-1)    
    .\label{eq:dmc3}
\end{align}
    \begin{figure}
\centering
\includegraphics[width=0.50\textwidth]{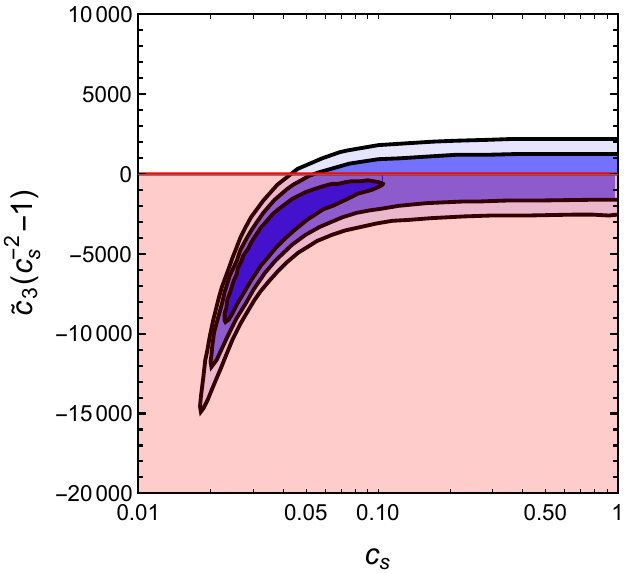}
\caption{
  The observationally 68\% (dark blue), 95\% (blue), and 99.7\% (light blue) confidence regions in the $(c_s, \tilde{c}_3(c_s^{-2}-1))$ parameter space~\cite{Planck:2019kim}. 
  In the red-shaded regions corresponding to eq.~\eqref{eq:ineqnon}, the non-negativity of relative entropy is violated in perturbation theory up to the third order in $X$, and the validity of the EFT~\eqref{eq:infEFT0} is lost, {\it i.e.}, the operator or loop expansion becomes invalid when the EFT is generated by the interaction between heavy and light fields.
}
\label{fig:con_EFT}
\end{figure}
For instance, for the two scalar fields theory defined in eq.~\eqref{eq:acBR}, these parameters are 
\begin{align}
c^{-2}_s =1+4 \left(\frac{\dot{\theta}_0}{m}\right)^2,~~~\tilde{c}_3 \left(c^{-2}_s -1\right)=-\frac{3}{4}\left(-1+\lambda_3\right)c_s^2 \left(c_s^{-2}-1\right)^2.\label{eq:two_mat}
\end{align}
where $\dot{\theta}_0:=\dot{\phi}_0/r_{\rm min}$, and these quantities are expressed in terms of parameters defined by the mass eigenstates of the heavy field in the theory with the interactions between the heavy and light fields.
$m$ denotes the mass of the heavy particle~\eqref{eq:ratmm2} in that theory involving the interactions, and $\lambda_{3}$ is defined in eq.~\eqref{eq:pot_mass} of Appendix~\ref{app:bases}. These parameters were introduced for convenient comparisons with previous studies~\cite{Kim:2021pbr}. 
For a detailed derivation of eq.~\eqref{eq:two_mat}, see Appendix~\ref{app:bases} or Ref.~\cite{Kim:2021pbr}. 
Also, the power spectrum is expressed as,
\begin{align}
    P_{\zeta}=\frac{1}{c_s}\frac{H^4}{(2\pi)^2 (-2 M^2_{\rm Pl}\dot{H})}.
    \label{eq:PzeEFT}
\end{align}
In addition, the non-linear parameter is given as follows: 
\begin{align}
    f_{\rm NL}=-\left(c^{-2}_s-1\right)\left[
    \frac{85}{324}+\frac{10}{243}\left(
    \tilde{c}_3 +\frac{3}{2}c^2_s
    \right)
    \right].\label{eq:EFTfNL}
\end{align}
We are now ready to discuss the relation between the validity of EFT and the non-linear parameter. 
In the following, after looking at their relation analytically, numerical results are presented.

From the inequalities~\eqref{eq:bo1} and \eqref{eq:bo2} arising from the non-negativity of relative entropy, and the expression~\eqref{eq:dmc3}, a condition under which the EFT description is valid can be written as follows: 
\begin{align}
	\tilde{c}_3 (c_s^{-2}-1)&\simeq \frac{3}{2}\frac{c_3\dot{\phi}_0^2}{c_2}c_s^2(c_s^{-2}-1)\gtrsim -3 c_s^2(c_s^{-2}-1),  \label{eq:ineqnon}  
\end{align}
where, from the second condition of the weakly coupled theory defined in Sec.~\ref{sec:weak} (validity of operator expansion), we used $(c_2+\frac{3}{2}c_3 \dot{\phi}_0^2)^{-1}\simeq c_2^{-1}(1-\frac{3}{2}c_3 \dot{\phi}_0^2/c_2+\cdots)$ and focused on the main contribution of the operator expansion. 
Figure.~\ref{fig:con_EFT} shows allowed regions on $(c_s, \tilde{c}_3(c_s^{-2}-1))$ plane from the current experiment~\cite{Planck:2019kim}.
In red shaded regions in Fig.~\ref{fig:con_EFT}, the inequality~\eqref{eq:ineqnon} does not hold, and the validity of the EFT is broken. 
Also, by combining eqs.~\eqref{eq:dmc3}, \eqref{eq:EFTfNL} and \eqref{eq:ineqnon}, an upper bound on $f_{\rm NL}$ is obtained as follows: 
\begin{align}
    -\left(c^{-2}_s-1\right)\frac{85}{324}\left(1-\frac{4}{17}c^2_s\right)\gtrsim f_{\rm NL}.
\end{align}
We see that when the EFT is generated through the weakly coupled theory (see Sec.~\ref{sec:weak} for its definition) with the interactions between heavy and light fields, the observationally allowed parameter region is limited. 
It should be emphasized that the red shaded regions in Fig.~\ref{fig:con_EFT} are parameter regions that can be realized by the strongly coupled theory and is not physically inadmissible.

    \begin{figure}
\centering
\includegraphics[width=0.49\textwidth]{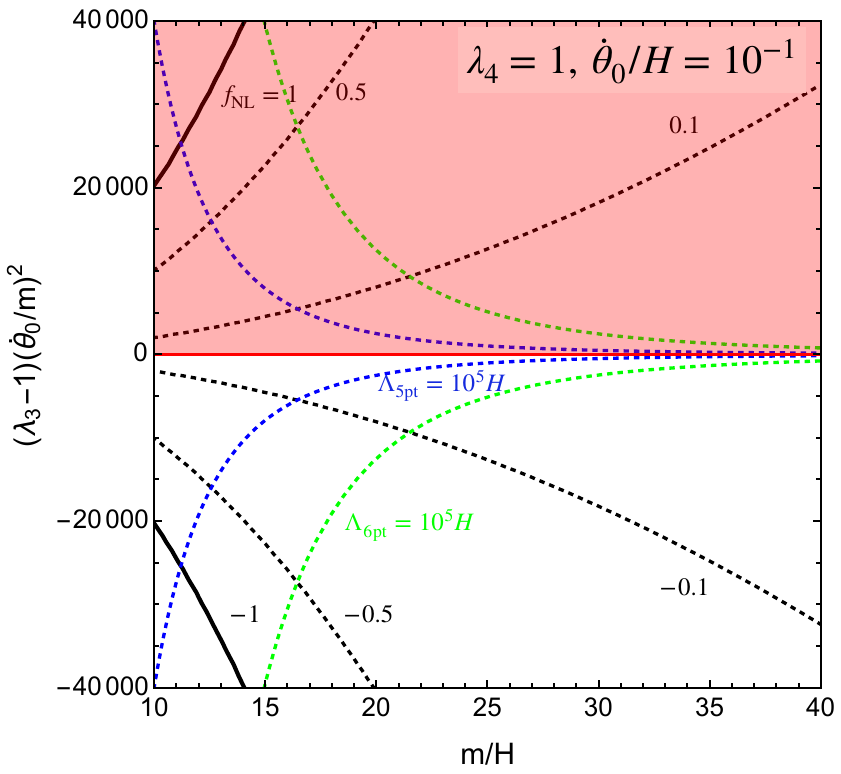}
\caption{
  Contour curves of $f_{\rm NL}=\pm1$ (black solid lines), $f_{\rm NL}=\pm0.5$, and $\pm0.1$ (black dotted lines) in the two scalar fields theory~\eqref{eq:acBR}, shown in the plane of $m/H$ and $(\lambda_3-1)(\dot{\theta}_0/m)^2$, for $\dot{\theta}_0/H=10^{-1}$.
  Please also see Appendix~\ref{app:bases} and Ref.~\cite{Kim:2021pbr} for the definition of $\lambda_3$ and $\lambda_4$. 
  The colored dotted curves represent the perturbative unitarity bounds~\eqref{eq:La5p} (blue), and \eqref{eq:La6p} (green).
  We choose $\Lambda_{\rm 5pt}=\Lambda_{\rm 6pt}=10^{5}H$ for the blue and green curves.
  In the red-shaded regions corresponding to eq.~\eqref{eq:plot_ine}, the non-negativity of relative entropy is perturbatively violated, and the theory~\eqref{eq:acBR} behaves as the strongly coupled theory.
}
\label{fig:con_EFT2}
\end{figure}

Before closing this section, we compare our method of characterizing the perturbation validity with conventional scattering amplitude-based perturbative unitarity results.
Here, we compare the two methods, in particular, taking as an example the theory of two scalar fields in Sec.~\ref{sec:3.1.2}.
First, from the inequality~\eqref{eq:nomrel3} up to the third order of the perturbative expansion in the relative entropy and relations presented in Appendix~\ref{app:bases}, we obtain 
\begin{align}
	1+\left(1-\lambda_3\right)\left(\frac{\dot{\theta}_0}{m}\right)^2\gtrsim  0.\label{eq:plot_ine}
\end{align}
Upon substituting eq.~\eqref{eq:two_mat} into this inequality, we obtain 
\begin{align}
	\tilde{c}_3\left(c_s^{-2}-1\right) \gtrsim -3 c_s^{2}\left(c_s^{-2}-1\right). 
\end{align}
This result represents that the inequality~\eqref{eq:ineqnon}, which holds for the EFT of inflation generated from general UV theories, can be confirmed using a concrete example. 
Next, we consider inequalities arising from the unitarity of the $S$-matrix, which has been studied in previous literature~\cite{Kim:2021pbr}. 
In theories for which the perturbative calculations are valid, inequalities arising from the unitarity of the $S$-matrix should also be satisfied at the tree-level for high-energy scatterings. 
As studied in Ref.~\cite{Kim:2021pbr}, deviations of $\lambda_3$ and $\lambda_4$\footnote{Note that $\lambda_3$ and $\lambda_4$ are the parameters defined in Appendix~\ref{app:bases} for the mass eigenstate of the heavy field in the theory with the interactions between heavy and light fields and are different from $\tilde{\lambda}_3$ and $\tilde{\lambda}_4$. See Appendix~\ref{app:bases} for the relation between $\tilde{\lambda}_{3,4}$ and ${\lambda}_{3,4}$.} from one break the perturbative unitarity of scattering amplitudes, yielding cutoff scales of the theory if the perturbation theory represents physics faithfully.
In the theory~\eqref{eq:acBR}, the cutoff scales calculated from 5 and 6-point scatterings are provided as~\cite{Kim:2021pbr}
\begin{align}
    &\Lambda_{\rm 5pt}=\frac{64 \sqrt{2}\pi^2}{9}\frac{r_0^3}{m^2 |\lambda_3-1|},\label{eq:La5p}
    \\
    &\Lambda_{\rm 6pt}=\frac{16 \sqrt{2}\pi^{3/2}}{\sqrt{15}}\frac{r_0^2}{m|\lambda_3-1|^{1/2}}.\label{eq:La6p}
\end{align}
From eqs.~\eqref{eq:La5p} and \eqref{eq:La6p}, we also see that the perturbative calculation of the scattering amplitude is valid for $\lambda_3=\lambda_4=1$, even in high energy regions. 
This fact reminds us that the non-negativity of relative entropy is unbreakable for $\lambda_3=\lambda_4=1$. 
It should be noted here that the scattering amplitude method characterizes the condition under which loop-level contributions do not exceed tree-level ones, whereas the relative entropy method can, in addition, characterize the validity of the operator expansion (see Sec.~\ref{sec:weak} for details).
The $S$-matrix and the relative entropy are quantities related to the unitarity, but due to these differences, different constraints would represent the validity of the perturbation theory. 
From eqs.~\eqref{eq:La5p} and \eqref{eq:La6p}, we obtain
\begin{align}
    &|\lambda_3-1|\left(\frac{\dot{\theta}_0}{m}\right)^2\simeq 4\times 10^{13} \left(\frac{H}{\Lambda_{\rm 5pt}}\right)\left(\frac{\dot{\theta}_0/H}{0.1}\right)^{-1}
    \left(\frac{m}{H}\right)^{-4}
    \left(\frac{P_{\zeta}}{2\times 10^{-9}}\right)^{-3/2},\label{eq:Lam5B}
    \\
    &|\lambda_3-1|\left(\frac{\dot{\theta}_0}{m}\right)^2\simeq 2\times 10^{19} \left(\frac{H}{\Lambda_{\rm 6pt}}\right)^2
    \left(\frac{\dot{\theta}_0/H}{0.1}\right)^{-2}\left(\frac{m}{H}\right)^{-4} \left(\frac{P_{\zeta}}{2\times 10^{-9}}\right)^{-2}.\label{eq:Lam6B}
\end{align}
As these relations depend on the cutoff scale of the theory, we show these bounds\footnote{By assuming the cutoff to be $\Lambda_{\rm 5pt}\leq 10^5 H$, and $\Lambda_{\rm 6pt}\leq 10^5 H$, regions inside the blue or green curves are allowed from the perturbative unitarity of scattering amplitudes.} in Fig.~\ref{fig:con_EFT2} for some benchmark values of the cutoff scale. 
Please look at the caption of Fig.~\ref{fig:con_EFT2} for more information.  
As shown in Fig.~\ref{fig:con_EFT2}, considerations based on the relative entropy can reveal regions of the non-linear parameter where the perturbative calculations are valid and are complementary to constraints from the conventional perturbative unitarity for a given cutoff scale. 
It should be emphasized that while constraints from the perturbative unitarity are highly dependent on the unknown cut-off scale, the relative entropy method is not.

\section{Summary and outlook}
\label{sec:summary}
The relative entropy is a non-negative quantity in the absence of approximations, and negative relative entropy in a given theory implicitly indicates that the theory is pathological, {\it e.g.,} a violation of unitarity or the second law of thermodynamics. 
In this paper, we studied the relative entropy in scalar field theories and showed that the non-negativity of perturbatively calculated relative entropy is potentially violated in strongly coupled theories. 
Since the relative entropy should be non-negative, the occurrence of negative relative entropy in perturbative calculations implies that the perturbative calculation is invalid. 
In short, we have found a way to quantify whether a given theory is weakly or strongly coupled by using the relative entropy.

We revisited an EFT of single-field inflation as an application of this relative entropy method.  
The experimentally allowed regions of the non-linear parameter in EFT~\eqref{eq:infEFT0} were shown to be limited potentially when the EFT description is valid in low-energy regions. 
The relative entropy of the two scalar field theories (one of the UV theories of the EFT of single-field inflation) was evaluated in detail by analytical and numerical calculations for two typical cases: the U(1) symmetry and the modified theory. 
In the perturbative calculations, it is found that the U(1) symmetry theory in subsubsection~\ref{sec:U(1)} respects the non-negativity of relative entropy, while the modified theories may not. 
Although the conventional method of perturbative unitarity based on scattering amplitude has the advantage of constraining the absolute values of the non-linear parameters (in two scalar fields theory), our study showed that a novel advantage, namely a severe constraint on its sign, can be exposed by shedding new light on the relative entropy.
In addition, we presented the regions of the non-linear parameter where the EFT description is valid in a UV theory-independent way, which would provide valuable insights into identifying models describing physical phenomena complementary to the experimental data.

We will mention a couple of things to the prospects of this work at the end.
First, constraints on the EFT of single-field inflation might be generalized to EFTs of multi-field inflation, which have also been studied in the context of the positivity bounds~\cite{Freytsis:2022aho}. 
Second, although our work has in mind the particular class of UV theories presented in Sec.~\ref{sec:entropy}, it would be interesting to discuss possible extensions to UV theory independent arguments (more specifically, an approach based on calculations of the relative entropy from the EFT). 
Third, a natural extension of this work would be to generalize it to open systems. 
In cosmology, non-unitary time evolution may play a significant role~\cite{Brahma:2022yxu,Brahma:2024yor}, and its effects on EFT could be explored by considering relative entropy.
Forth, our relative entropy perspective can be applied to other EFTs describing Lorentz symmetry-breaking phenomena, such as superfluids. 
Since conventional positivity bounds rely on the Lorentz symmetry, we expect our relative entropy perspective, which does not depend on Lorentz symmetry, to produce more diverse physics results. 

\section*{Acknowledgements}
We are grateful to Toshifumi Noumi for his helpful comments and the opportunity to initiate this work.
DU would like to thank the CERN Theory Department for their financial support and hospitality during this work.
DU is supposed by grants from the ISF (No.~1002/23 and 597/24) and the BSF (No.~2021800).

\appendix

\section{Field redefinition of heavy fields}
\label{app:red}
We here consider redefinitions of the heavy field such that Eq.~\eqref{eq:par0BG} holds.
From eq.~\eqref{eq:lag0}, we generally obtain the following: 
\begin{align}
    \left(\frac{\partial \mathcal{L}(0,\phi_2)}{\partial \phi_2}\right)_{\phi_2=0}=a _1(0),~~~\left(\frac{\partial^2 \mathcal{L}(0,\phi_2)}{\partial \phi_2^2}\right)_{\phi_2=0}=2 a_2(0).
\end{align}
Here, we assume that $v$ is a solution of $\phi_2$ of the Lagrangian $\mathcal{L}(0,\phi_2)$.
By rewriting $\phi_2$ as $\phi_2=v+\eta$, a condition that $v$ is the solution is given by
\begin{align}
        \left(\frac{\partial \mathcal{L}(0,\phi_2)}{\partial \eta}\right)_{\eta=0}=\left(\frac{\partial \mathcal{L}(0,\phi_2)}{\partial \phi_2}\right)_{\phi_2=v}=0.\label{eq:loc1}
\end{align}
Thus, by a redefinition of the heavy field as $\phi_2\to \eta$, we obtain
\begin{align}
\left(\frac{\partial^2 \mathcal{L}(0,\phi_2)}{\partial \eta^2}\right)_{\eta=0}=\left(\frac{\partial^2 \mathcal{L}(0,\phi_2)}{\partial \phi_2^2}\right)_{\phi_2=v}:=m^2_2.\label{eq:sec1}
\end{align}
From eqs.~\eqref{eq:loc1} and \eqref{eq:sec1}, we found that eq.~\eqref{eq:par0BG} holds regarding the field $\eta$.

\section{Perturbative calculation of $\widetilde{\phi}_{1,\lambda}$ in eq.~\eqref{eq:Zgdy}}
\label{sec:phi1}
Let us consider $\lambda$ dependencies in $W_{\lambda}[\widetilde{\phi}_{1,\lambda}]$ and $\widetilde{\phi}_{1,\lambda}$.
The classical solution $\widetilde{\phi}_{1,\lambda}$ is defined so that $\delta W_{\lambda}[\phi]/\delta \phi|_{\phi=\widetilde{\phi}_{1,\lambda}}=0$, where $W_{\lambda}[\phi]$ is given, up to the third order of $\lambda$, as
\begin{align}
    W_{\lambda}[\phi]&=\int (d^4x)_{\rm E} \Bigg[
    a_0\left(\phi\right)
    -\lambda^2 \frac{1}{2m_2^2\left(1-\partial^2_I /m_2^2\right)}
    \left(a_1\left(\phi\right)\right)^2
    +\lambda^3 \frac{1}{m_2^6 \left(1-\partial^2_I/m_2^2\right)^3}\notag
    \\
    &\times \bigg\{
    -a_3\left(0\right)
    \left(a_1\left(\phi\right)\right)^3
    +m_2^2 \left(1-\frac{\partial^2_I}{m_2^2}\right)
    \left(a_1\left(\phi_1\right)\right)^2
    \left(
    a_2\left(\phi\right)
    -\frac{m_2^2}{2}
    \right)
    \bigg\}\Bigg].\label{eq:appWlambda}
\end{align}
For ease of considerations, from eq.~\eqref{eq:appWlambda}, we introduce two independent parameters $\lambda^{(2)}$ and $\lambda^{(3)}$, and define as,
\begin{align}
    W[\phi;\lambda^{(2)},\lambda^{(3)}]:&=
    \int (d^4x)_{\rm E} \Bigg[
    a_0\left(\phi\right)
    -\lambda^{(2)} \frac{1}{2m_2^2\left(1-\partial^2_I /m_2^2\right)}
    \left(a_1\left(\phi\right)\right)^2
    +\lambda^{(3)} \frac{1}{m_2^6 \left(1-\partial^2_I/m_2^2\right)^3}\notag
    \\
    &\times \bigg\{
    -a_3\left(0\right)
    \left(a_1\left(\phi\right)\right)^3
    +m_2^2 \left(1-\frac{\partial^2_I}{m_2^2}\right)
    \left(a_1\left(\phi_1\right)\right)^2
    \left(
    a_2\left(\phi\right)
    -\frac{m_2^2}{2}
    \right)
    \bigg\}\Bigg].
\end{align}
Regarding the two parameters $\lambda^{(2)}$ and $\lambda^{(3)}$ as expansion parameters, the solution of $W[\phi;\lambda^{(2)},\lambda^{(3)}]$ can be expressed as,
\begin{align}
    \widetilde{\phi}_{1,\lambda}=\widetilde{\phi}_{1,0}+\delta \phi_{\lambda}, \label{eq:appsol_phi1}
\end{align}
where $\widetilde{\phi}_{1,0}$ is the classical solution of $W_0[\phi]=W[\phi;0,0]=\int (d^4x)_{\rm E} a_0\left(\phi\right)$, and $\delta \phi_{\lambda}$ denotes a perturbative effect coming from $\lambda^{(2)}$ and $\lambda^{(3)}$ and is expressed as a power expansion of $\lambda^{(2)}$ and $\lambda^{(3)}$.
Since we can obtain the classical solution of eq.~\eqref{eq:appWlambda} from eq.~\eqref{eq:appsol_phi1} by replacing as $\lambda^{(2)}\to \lambda^2$ and $\lambda^{(3)}\to \lambda^3$, we find $\delta \phi_{\lambda}=\mathcal{O}\left(\lambda^2\right)$.

Next, let us substitute eq.~\eqref{eq:appsol_phi1} into eq.~\eqref{eq:appWlambda}, and then we obtain as
\begin{align}
    W_{\lambda}[\widetilde{\phi}_{1,\lambda}]&=\int (d^4x)_{\rm E} \Bigg[
    a_0\left(\widetilde{\phi}_{1,\lambda}\right)
    -\lambda^2 \frac{1}{2m_2^2\left(1-\partial^2_I /m_2^2\right)}
    \left(a_1\left(\widetilde{\phi}_{1,\lambda}\right)\right)^2
    +\lambda^3 \frac{1}{m_2^6 \left(1-\partial^2_I/m_2^2\right)^3}\notag
    \\
    &\times \bigg\{
    -a_3\left(0\right)
    \left(a_1\left(\widetilde{\phi}_{1,\lambda}\right)\right)^3
    +m_2^2 \left(1-\frac{\partial^2_I}{m_2^2}\right)
    \left(a_1\left(\widetilde{\phi}_{1,\lambda}\right)\right)^2
    \left(
    a_2\left(\widetilde{\phi}_{1,\lambda}\right)
    -\frac{m_2^2}{2}
    \right)
    \bigg\}\Bigg],\label{eq:appWlambda2}
\end{align}
where three functions $a_0\left(\widetilde{\phi}_{1,\lambda}\right)$, $a_1\left(\widetilde{\phi}_{1,\lambda}\right)$, and $a_2\left(\widetilde{\phi}_{1,\lambda}\right)$ can be expressed as
\begin{align}
    a_0\left(\widetilde{\phi}_{1,\lambda}\right)&=a_0\left(\widetilde{\phi}_{1,0}\right)+\frac{\delta a_0}{\delta \phi}\bigg|_{\phi=\widetilde{\phi}_{1,0}}\delta \phi_{\lambda}+\mathcal{O}\left(\delta \phi_{\lambda}^2\right)=a_0\left(\widetilde{\phi}_{1,0}\right)+\mathcal{O}\left(\delta \phi_{\lambda}^2\right),\label{eq:appa0}
    \\
    a_1\left(\widetilde{\phi}_{1,\lambda}\right)&=a_1\left(\widetilde{\phi}_{1,0}\right)+\frac{\delta a_1}{\delta \phi}\bigg|_{\phi=\widetilde{\phi}_{1,0}}\delta \phi_{\lambda}+\mathcal{O}\left(\delta \phi_{\lambda}^2\right),\label{eq:appa1}
    \\
    a_2\left(\widetilde{\phi}_{1,\lambda}\right)&=a_2\left(\widetilde{\phi}_{1,0}\right)+\frac{\delta a_2}{\delta \phi}\bigg|_{\phi=\widetilde{\phi}_{1,0}}\delta \phi_{\lambda}+\mathcal{O}\left(\delta \phi_{\lambda}^2\right),\label{eq:appa2}
\end{align}
where $\delta a_0/\delta \phi |_{\phi=\widetilde{\phi}_{1,0}}=0$ holds because $\widetilde{\phi}_{1,0}$ is the classical solution of $W_0[\phi]=\int (d^4x)_{\rm E} a_0\left(\phi\right)$.
Note here that $\mathcal{O}\left(\delta \phi_{\lambda}^2\right)=\mathcal{O}\left(\lambda^4\right)$ because of $\delta \phi_{\lambda}=\mathcal{O}\left(\lambda^2\right)$.
Combining eqs.~\eqref{eq:appa0}\text{-}\eqref{eq:appa2} with eq.~\eqref{eq:appWlambda2}, we end up with
\begin{align}
    W_{\lambda}[\widetilde{\phi}_{1,\lambda}]&=\int (d^4x)_{\rm E} \Bigg[
    a_0\left(\widetilde{\phi}_{1,0}\right)
    -\lambda^2 \frac{1}{2m_2^2\left(1-\partial^2_I /m_2^2\right)}
    \left(a_1\left(\widetilde{\phi}_{1,0}\right)\right)^2
    +\lambda^3 \frac{1}{m_2^6 \left(1-\partial^2_I/m_2^2\right)^3}\notag
    \\
    &\times \bigg\{
    -a_3\left(0\right)
    \left(a_1\left(\widetilde{\phi}_{1,0}\right)\right)^3
    +m_2^2 \left(1-\frac{\partial^2_I}{m_2^2}\right)
    \left(a_1\left(\widetilde{\phi}_{1,0}\right)\right)^2
    \left(
    a_2\left(\widetilde{\phi}_{1,0}\right)
    -\frac{m_2^2}{2}
    \right)
    \bigg\}\Bigg]+\mathcal{O}\left(\lambda^4\right).
\end{align}

\section{Higher order of perturbative expansion in the effective action}
\label{app:B}
We evaluate the Euclidean effective action up to the sixth order of $\lambda$.
Since higher-derivative terms are irrelevant in this work, we neglect derivative terms coming from the kinetic term of the heavy field in the following calculations.
From eq.~\eqref{eq:redN}, the classical solution is expressed as follows:
\begin{align}
\widetilde{\phi}_{2,\lambda}
&=\sum_{i=1}^{\infty}\lambda^i \phi_2^{(i)},\label{eq:Phig4}
\end{align}
where the coefficients are defined as follows:
\begin{align}
    \phi_2^{(1)}:&=-\frac{a_1(\phi_1)}{m_2^2},~~~\phi_2^{(2)}:=\frac{a_1(\phi_1)\left(2 a_2(\phi_1)-m^2_2\right)}{m^4_2}-\frac{3 \left(a_1(\phi_1)\right)^2a_3(0)}{m_2^6},
    \\
    \phi_2^{(3)}:&=
-\frac{a_1(\phi_1)}{m_2^2}
+\frac{4 a_1(\phi_1)a_2(\phi_1)}{m_2^4}-\frac{a_1(\phi_1)\left\{4\left(a_2(\phi_1)\right)^2+3a_1(\phi_1)\left(2 a_3(0)+a_3(\phi_1)\right)\right\}}{m_2^6}\notag
\\
&+\frac{2 \left(a_1(\phi_1)\right)^2 \left(9 a_2(\phi_1) a_3(0)+2 a_1(\phi_1)a_4(0)\right)}{m_2^8}-\frac{18 \left(a_1(\phi_1)\right)^3 \left(a_3(0)\right)^2}{m_2^{10}},
\\
\phi_2^{(4)}:&=
-\frac{a_1(\phi_1)}{m_2^2}
+\frac{6 a_1(\phi_1)a_2(\phi_1)}{m_2^4}
-\frac{3 a_1(\phi_1)\left\{4\left(a_2(\phi_1)\right)^2+3a_1(\phi_1)\left(a_3(0)+a_3(\phi_1)\right)\right\}}{m_2^6}\notag
\\
&+\frac{2 a_1(\phi_1)\left\{4 \left(a_2(\phi_1)\right)^3+9 a_1(\phi_1)a_2(\phi_1)\left(3 a_3(0)+a_3(\phi_1)\right)+2 \left(a_1(\phi_1)\right)^2\left(3 a_4(0)+a_4(\phi_1)\right)\right\}}{m_2^8}\notag
\\
&-\frac{\left(a_1(\phi_1)\right)^2}{m_2^{10}}\bigg\{72 \left(a_2(\phi_1)\right)^2a_3(0)+32 a_1(\phi_1)a_2(\phi_1)a_4(0)\notag
\\
&+a_1(\phi_1)\left(54 (a_3(0))^2+36a_3(0)a_3(\phi_1)+5 a_1(\phi_1)a_5(0)\right)\bigg\}\notag
\\
&+\frac{60 \left(a_1(\phi_1)\right)^3 a_3(0)\left(3 a_2(\phi_1)a_3(0)+a_1(\phi_1)a_4(0)\right)}{m_2^{12}}
-\frac{135 \left(a_1(\phi_1)\right)^4 \left(a_3(0)\right)^3}{m_2^{14}},
\\
\phi_2^{(5)}:&=-\frac{a_1(\phi_1)}{m_2^2}+\frac{8 a_1(\phi_1)a_2(\phi_1)}{m_2^4}
-\frac{6 a_1(\phi_1)\left(4 \left(a_2(\phi_1)\right)^2+2 a_1(\phi_1) a_3(0)+3 a_1(\phi_1) a_3(\phi_1)\right)}{m_2^6}\notag
\\
&+\frac{4 a_1(\phi_1)\left\{ 8 \left(a_2(\phi_1)\right)^3+9 a_1(\phi_1)a_2(\phi_1)\left(3 a_3(0) +2 a_3(\phi_1)\right)+2 \left(a_1(\phi_1)\right)^2\left(3 a_4(0)+2 a_4(\phi_1)\right)\right\}}{m_2^8}\notag
\\
&-\frac{a_1(\phi_1)}{m_2^{10}}\bigg[
16 \left(a_2(\phi_1)\right)^4+72 a_1(\phi_1) \left(a_2(\phi_1)\right)^2 \left(4 a_3(0)+a_3(\phi_1)\right)+32 \left(a_1(\phi_1)\right)^2 a_2(\phi_1) \left(4 a_4(0)+a_4(\phi_1)\right)\notag
\\
&+\left(a_1(\phi_1)\right)^2
\left\{
108 \left(a_3(0)\right)^2
+144 a_3(0) a_3(\phi_1)
+18 \left(a_3(\phi_1)\right)^2
+5 a_1(\phi_1) \left(
4 a_5(0)+ a_5(\phi_1)
\right)
\right\}
\bigg]\notag
\\
&+\frac{10 \left(a_1(\phi_1)\right)^2}{m_2^{12}}\bigg[
24 \left(a_2(\phi_1)\right)^3 a_3(0)+16 a_1(\phi_1) \left(a_2(\phi_1)\right)^2 a_4(0)\notag
\\
&+
6 \left(a_1(\phi_1)\right)^2 \left\{
a_3(\phi_1) a_4(0)+
a_3(0)\left(
4 a_4(0)+a_4(\phi_1)
\right)
\right\}\notag
\\
&+a_1(\phi_1) a_2(\phi_1)\left(
72 \left(a_3(0)\right)^2 +36 a_3(0) a_3(\phi_1) +5 a_1(\phi_1) a_5(0)
\right)
\bigg]\notag
\\
&-\frac{3 \left(a_1(\phi_1)\right)^3}{m_2^{14}}\bigg[
360 \left(a_2(\phi_1)\right)^2 \left(a_3(0)\right)^2+
240 a_1(\phi_1) a_2(\phi_1) a_3(0) a_4(0)\notag
\\
&+a_1(\phi_1)\left(
180 \left(a_3(0)\right)^3+135 (a_3(0))^2 a_3(\phi_1)
+16 a_1(\phi_1) (a_4(0))^2 +30 a_1(\phi_1) a_3(0) a_5(0)
\right)\notag
\\
&+\frac{378\left(a_1(\phi_1)\right)^4 \left(a_3(0)\right)^2 \left(5 a_2(\phi_1) a_3(0)+ 2a_1(\phi_1)a_4(0) \right)}{m_2^{16}}
\bigg]\notag
\\
&-\frac{1134 \left(a_1(\phi_1)\right)^5 \left(a_3(0)\right)^4}{m_2^{18}}
,
\end{align}
Upon substituting eq.~\eqref{eq:Phig4} into eq.~\eqref{eq:Eucaction}, we end up with
\begin{align}
    W_\lambda[\phi_1]&=\int (d^4 x)_{\rm E}\Bigg[
    W^{(0)}+\sum_{i=2}^{\infty}\lambda^i W^{(i)} 
    \Bigg],\label{eq:Wg5}
\end{align}
where the coefficients are defined as follows:
\begin{align}
    W^{(0)}:&=a_0(\phi_1),~~W^{(2)}:=-\frac{\left(a_1(\phi_1)\right)^2}{2m_2^2} ,~~W^{(3)}:=\frac{1}{m_2^6}
    \left[
    -a_3(0)\left(a_1(\phi_1)\right)^3
    +m_2^2 \left(a_1(\phi_1)\right)^2
    \left\{a_2(\phi_1)-\frac{m_2^2}{2}\right\}
    \right],
    \\
    W^{(4)}:&=-\frac{\left(a_1(\phi_1)\right)^2}{2m_2^{10}}
    \Bigg[
    -12 a_1(\phi_1)a_2(\phi_1)a_3(0)m_2^2
    +2 a_1(\phi_1)\left\{2 a_3(0)+a_3(\phi_1)\right\}m_2^4
    +\left\{-2 a_2(\phi_1)+m_2^2\right\}^2 m_2^4\notag
    \\
    &+(a_1(\phi_1))^2 \left\{9 (a_3(0))^2 - 2 a_4(0) m_2^2 \right\}
    \Bigg],
    \\
    W^{(5)}:&=- \frac{\left(a_1(\phi_1)\right)^2}{2 m_2^{14}} 
    \Bigg[
    m_2^6 \left(-2 a_2(\phi_1)+m_2^2\right)^3
    +2 \left(a_1(\phi_1)\right)^3 \left\{27 \left(a_3(0)\right)^3-12 a_3(0) a_4(0) m_2^2 +a_5(0) m_2^4 \right\}\notag
    \\
    &+6 m_2^4 a_1(\phi_1)  \left(-2 a_2(\phi_1)+m_2^2\right)
    \left\{-4 a_2(\phi_1) a_3(0)+ \left(a_3(0)+a_3(\phi_1)\right)m_2^2\right\}\notag
    \\
    &+\left(a_1(\phi_1)\right)^2 \bigg\{
    a_2(\phi_1)\bigg(-90 \left(a_3(0)\right)^2 m_2^2+16 a_4(0) m_2^4\bigg)\notag
    \\
    &+m_2^4 \left(27 \left(a_3(0)\right)^2+18 a_3(0) a_3(\phi_1)
    -2 m_2^2 \left(3 a_4(0)+a_4(\phi_1)\right)
    \right)
    \bigg\}
    \Bigg],
    \\
    W^{(6)}:&=-\frac{\left(a_1(\phi_1)\right)^2}{2 m_2^{18}}\Bigg[
    m_2^8 \left(-2 a_2(\phi_1)+m_2^2\right)^4
    +4 a_1(\phi_1) m_2^6 \left(-2 a_2(\phi_1)+m_2^2\right)^2
    \bigg\{
    -10 a_2(\phi_1) a_3(0)\notag
    \\
    &+\left(2 a_3(0)+ 3a_3(\phi_1)\right)m_2^2
    \bigg\}
    +2 \left(a_1(\phi_1)\right)^4
    \bigg(
    189 \left(a_3(0)\right)^4-126 \left(a_3(0)\right)^2 a_4(0) m_2^2+ 8 \left(a_4(0)\right)^2 m_2^4\notag
    \\
    &+15a_3(0) a_5(0)m_2^4 -a_6(0) m_2^6
    \bigg)+
    \left(a_1(\phi_1)\right)^2\bigg\{
    20 \left(a_2(\phi_1)\right)^2 m_2^4 \left(27 \left(a_3(0)\right)^2-4 a_4(0) m_2^2\right)\notag
    \\
    &+4 a_2(\phi_1) m_2^6 \left(-90 \left(a_3(0)\right)^2-45 a_3(0) a_3(\phi_1)+4 \left(4 a_4(0)+a_4(\phi_1)\right)m_2^2\right)\notag
    \\
    &+
    m_2^8 \left(
    54 \left(a_3(0)\right)^2+72 a_3(0)a_3(\phi_1) +9 \left(a_3(\phi_1)\right)^2
    -4 \left(3 a_4(0)+2 a_4(\phi_1)\right)m_2^2
    \right)
    \bigg\}
    \\
    &+\left(a_1(\phi_1)\right)^3
    \bigg\{
    2 m^4_2 \bigg(
    108 \left(a_3(0)\right)^3+81 \left(a_3(0)\right)^2 a_3(\phi_1)-12 a_3(\phi_1) a_4(0) m_2^2 \notag
    \\
    &-12 a_3(0) \left(4 a_4(0)+a_4(\phi_1)\right)m_2^2+\left(4 a_5(0)+ a_5(\phi_1)\right) m_2^4
    \bigg)\notag
    \\
    &-4 a_2(\phi_1) \left(
    189 \left(a_3(0)\right)^3 m_2^2-72 a_3(0)a_4(0)m_2^4+5 a_5(0)m_2^6
    \right)
    \bigg\}
    \Bigg].
\end{align}
Now, let us apply the above formulae to the two typical cases as follows:
\begin{enumerate}
    \item U(1) symmetric potential in subsubsection~\ref{sec:U(1)} --- Upon substituting eq.~\eqref{eq:coU(1)} into eq.~\eqref{eq:Wg5}, we obtain the Euclidean effective actions, up to the sixth order of $\lambda$, as
\begin{align}
    W_\lambda^{\rm U(1)}[\phi]&=\int (d^4x)_{\rm E}\sqrt{g}\bigg[
    \frac{1}{2} \left(\partial_{\mu}\phi\right)^2+V_{\rm soft}(\phi)-\lambda^2\frac{1}{2m_2^2 r^2_{\rm min} }(\partial_{\mu}\phi)^4+\mathcal{O}(\lambda^7)
    \bigg],\label{eq:ApWgU1}
    \\
    W_0^{\rm U(1)}[\phi]&=\int (d^4x)_{\rm E}\sqrt{g}
    \left[\frac{1}{2} \left(\partial_{\mu}\phi\right)^2+V_{\rm soft}(\phi)\right].\label{eq:ApW0U1}
\end{align}
From eqs.~\eqref{eq:relFI}, \eqref{eq:ApWgU1}, and \eqref{eq:ApW0U1}, the relative entropy is calculated,  up to the sixth order of $\lambda$, as
\begin{align}
    S(P_{\rm R}||P_{\rm T})_{U(1)}=\lambda^2 \frac{1}{2m_2^2 r^2_{\rm min}}\int (d^4 x)_{\rm E}\sqrt{g}(\partial_{\mu}\phi)^4+\mathcal{O}(\lambda^7).
\end{align}
This is consistent with the result of eq.~\eqref{eq:relfull} in the full order of $\lambda$.

    \item A modified version of potential in eq.~\eqref{eq:U1sym} --- Upon substituting eq.~\eqref{eq:co2} into eq.~\eqref{eq:Wg5}, we obtain the Euclidean effective actions, up to the sixth order of $\lambda$, as
\begin{align}
    W_\lambda^{\cancel{\rm U(1)} }[\phi]&=\int (d^4 x)_{\rm E}\sqrt{g}\Bigg[
    \frac{1}{2}\left(\partial_{\mu}\phi\right)^2+V_{\rm soft}(\phi)
    -\lambda^2 \frac{\left(\partial_{\mu}\phi\right)^4}{2m^2_2 r^2_{\rm min}} 
    +\lambda^3 \left(1-\tilde{\lambda}_3\right)\frac{\left(\partial_{\mu}\phi\right)^6}{2m_2^4 r^4_{\rm min}}\notag
    \\
    &\quad\quad-\lambda^4 \left\{
    4-12\tilde{\lambda}_3 + (9\tilde{\lambda}_3^2-\tilde{\lambda}_4)
    \right\} \frac{\left(\partial_{\mu}\phi\right)^8}{8 m_2^6 r^6_{\rm min}}\notag
    \\
    &\quad\quad+\lambda^5 \left(
4-4\tilde{\lambda}_4-6 \tilde{\lambda}_3 (4-\tilde{\lambda}_4)+45\tilde{\lambda}_3^2 -27 \tilde{\lambda}_3^3
    \right) \frac{\left(\partial_{\mu}\phi\right)^{10}}{8 m_2^8 r^8_{\rm min}}
    \notag
    \\
    &\quad\quad -\lambda^6\bigg\{
    8+ \left(-20 \tilde{\lambda}_4 +2 \tilde{\lambda}_4^2\right)+\tilde{\lambda}_3 \left(-80+72 \tilde{\lambda}_4\right)
    +\tilde{\lambda}_3^2 \left(270-63 \tilde{\lambda}_4\right)\notag
    \\
    &\quad\quad-378 \tilde{\lambda}_3^3 +189 \tilde{\lambda}_3^4
    \bigg\} \frac{\left(\partial_{\mu}\phi\right)^{12}}{16 m_2^{10} r^{10}_{\rm min}}
    \Bigg]+\mathcal{O}(\lambda^7),\label{eq:2Wg5}
    \\
    W_0^{\cancel{\rm U(1)} }[\phi]&=\int (d^4 x)_{\rm E}\sqrt{g}
    \left[\frac{1}{2} \left(\partial_{\mu}\phi\right)^2+V_{\rm soft}(\phi)\right].\label{eq:2W05}
\end{align}
From eqs.~\eqref{eq:relFI}, \eqref{eq:2Wg5} and \eqref{eq:2W05}, the relative entropy is calculated, up to the sixth order of $\lambda$, as
\begin{align}
    S(P_{\rm R}||P_{\rm T})_{\cancel{\rm U(1)}}&=\lambda^2\frac{1}{2m_2^2 r^2_{\rm min}}\int (d^4x)_{\rm E} \sqrt{g} \left(\partial_{\mu}\phi\right)^4
    -\lambda^3 \frac{1}{2m_2^4 r^4_{\rm min}}\left(1-\tilde{\lambda}_3\right)\int (d^4x)_{\rm E} \sqrt{g} \left(\partial_{\mu}\phi\right)^6\notag
    \\
    &+\lambda^4 \frac{1}{8 m_2^6 r^6_{\rm min}} \left\{
    4-12 \tilde{\lambda}_3 +\left(9\tilde{\lambda}_3^2-\tilde{\lambda}_4\right)
    \right\}\int (d^4x)_{\rm E} \sqrt{g}  \left(\partial_{\mu}\phi\right)^8\notag
    \\
    &-\lambda^5 \frac{1}{8 m_2^8 r^8_{\rm min}}
    \left\{
    4-4\tilde{\lambda}_4 -6 \tilde{\lambda}_3 \left(4-\tilde{\lambda}_4\right) +45 \tilde{\lambda}_3^2 -27 \tilde{\lambda}_3^3
    \right\}\int (d^4 x)_{\rm E}
    \left(\partial_{\mu}\phi\right)^{10}\notag
    \\
    &+\lambda^6 \frac{1}{16 m_2^{10} r_{\rm min}^{10}}\bigg\{
    8 + \left(-20 \tilde{\lambda}_4 +2 \tilde{\lambda}_4^2\right) + \tilde{\lambda}_3 \left(-80 + 72 \tilde{\lambda}_4\right)
    +\tilde{\lambda}_3^2 \left(270-63 \tilde{\lambda}_4\right)\notag
    \\
    &-378 \tilde{\lambda}_3^3 +189 \tilde{\lambda}_3^4
    \bigg\} \int (d^4x)_{\rm E} \left(\partial_{\mu}\phi\right)^{12}
    +\mathcal{O}(\lambda^7).
\end{align}
The relative entropy up to the third order of $\lambda$ is controlled only by one parameter $\tilde{\lambda}_3$.
On the other hand, for the higher order of $\lambda$ than the fourth, the parameter $\tilde{\lambda}_4$ also contributes.
Thus, the validity of perturbative calculation of the relative entropy is generally determined by combinations of the two parameters, $\tilde{\lambda}_3$, and $\tilde{\lambda}_4$.

\end{enumerate}

\section{Derivation of eq.~\eqref{eq:delWgN}}
\label{sec:delv}
Returning to the definition of $\left({\partial W_{\lambda}^{(N)}}/{\partial \lambda}\right)_{\lambda=0}$ helps us understand how to calculate eq.~\eqref{eq:delWgN}.
The left-hand side of eq.~\eqref{eq:delWgN} is given by
\begin{align}
    \left(\frac{\partial W_{\lambda}^{(N)}}{\partial \lambda}\right)_{\lambda=0}&= 
    -\left(\frac{\partial \ln Z_{\lambda}^{(N)}[\widetilde{\phi}_{1,\lambda}^{(N)},\cdots,\widetilde{\phi}^{(N)}_{N,\lambda}]}{\partial \lambda}\right)_{\lambda=0}\label{eq:first_line_W_lambda}
    \\
    &=-\left(\frac{1}{Z_{\lambda}^{(N)}[\widetilde{\phi}_{1,\lambda}^{(N)},\cdots,\widetilde{\phi}^{(N)}_{N,\lambda}]}\frac{\partial }{\partial \lambda}Z_{\lambda}^{(N)}[\widetilde{\phi}_{1,\lambda}^{(N)},\cdots,\widetilde{\phi}^{(N)}_{N,\lambda}]\right)_{\lambda=0}
    \\
    &=-\left(\frac{1}{Z_{\lambda}^{(N)}[\widetilde{\phi}_{1,\lambda}^{(N)},\cdots,\widetilde{\phi}^{(N)}_{N,\lambda}]}\frac{\partial }{\partial \lambda}\int \prod _{i=1}^N d[\phi_i] e^{-I^{(N)}_{\lambda}[\phi_1,\cdots,\phi_N]}\right)_{\lambda=0}\label{eq:third_line}
    \\
    &=\left(\frac{1}{Z_{\lambda}^{(N)}[\widetilde{\phi}_{1,\lambda}^{(N)},\cdots,\widetilde{\phi}^{(N)}_{N,\lambda}]}\int \prod _{i=1}^N d[\phi_i] e^{-I^{(N)}_{\lambda}[\phi_1,\cdots,\phi_N]}
    \frac{\partial}{\partial\lambda}I_{\lambda}^{(N)}[\phi_1,\cdots,\phi_N]
    \right)_{\lambda=0}\label{eq:forth_line}
    \\
    &=\int \prod _{i=1}^N d[\phi_i] P_{\rm R}[\phi_1,\cdots,\phi_N] I_{\rm I}[\phi_1,\cdots,\phi_N],\label{eq:appEXpath}
\end{align}
where $W_{\lambda}^{(N)}:= -\ln Z_{\lambda}^{(N)}[\widetilde{\phi}_{1,\lambda}^{(N)},\cdots,\widetilde{\phi}^{(N)}_{N,\lambda}]$, and the following relations,
\begin{align}
&\frac{\partial}{\partial \lambda}e^{-I^{(N)}_{\lambda}[\phi_1,\cdots,\phi_N]}=-e^{-I^{(N)}_{\lambda}[\phi_1,\cdots,\phi_N]}\frac{\partial}{\partial\lambda}I_{\lambda}[\phi_1,\cdots,\phi_N] =- I_{\rm I}[\phi_1,\cdots,\phi_N]e^{-I^{(N)}_{\lambda}[\phi_1,\cdots,\phi_N]},
\\
&P_{\rm R}[\phi_1,\cdots,\phi_N]=e^{-I^{(N)}_{0}[\phi_1,\cdots,\phi_N]}/Z_0^{(N)}[\widetilde{\phi}_{1,0}^{(N)},\cdots,\widetilde{\phi}^{(N)}_{N,0}],
\\
&Z_{\lambda}^{(N)}[\widetilde{\phi}_{1,\lambda}^{(N)},\cdots,\widetilde{\phi}^{(N)}_{N,\lambda}]:=\int \prod _{i=1}^N d[\phi_i] e^{-I^{(N)}_{\lambda}[\phi_1,\cdots,\phi_N]},
\end{align}
are used.
Equation~\eqref{eq:appEXpath} was used to derive eq.~\eqref{eq:relFI}, {\it i.e.}, the right-hand side of eq.~\eqref{eq:appEXpath} is the definition of $\left({\partial W_{\lambda}^{(N)}}/{\partial \lambda}\right)_{\lambda=0}$ in this work.
%
%
%
At the tree-level (corresponding to saddle point approximation), eq.~\eqref{eq:appEXpath} is calculated as follows: 
\begin{align}
     \left(\frac{\partial W_{\lambda}^{(N)}}{\partial \lambda}\right)_{\lambda=0}&= \int \prod _{i=1}^N d[\phi_i] \frac{1}{Z_0^{(N)}[\widetilde{\phi}_{1,0}^{(N)},\cdots,\widetilde{\phi}^{(N)}_{N,0}]} 
     e^{-I^{(N)}_{0}[\phi_1,\cdots,\phi_N]}
     I_{\rm I}[\phi_1,\cdots,\phi_N]\notag
     \\
     &=\frac{1}{Z_0^{(N)}[\widetilde{\phi}_{1,0}^{(N)},\cdots,\widetilde{\phi}^{(N)}_{N,0}]} e^{-I^{(N)}_{0}[\widetilde{\phi}_{1,0}^{(N)},\cdots,\widetilde{\phi}^{(N)}_{N,0}]}
     I_{\rm I}[\widetilde{\phi}_{1,0}^{(N)},\cdots,\widetilde{\phi}^{(N)}_{N,0}]\notag
     \\
     &=I_{\rm I}[\widetilde{\phi}_{1,0}^{(N)},\cdots,\widetilde{\phi}^{(N)}_{N,0}],\label{eq:Wlambda_II}
\end{align}
where we used the following relation:
\begin{align}
    Z_0^{(N)}[\widetilde{\phi}_{1,0}^{(N)},\cdots,\widetilde{\phi}^{(N)}_{N,0}]=\int \prod _{i=1}^N d[\phi_i] 
     e^{-I^{(N)}_{0}[\phi_1,\cdots,\phi_N]}= e^{-I^{(N)}_{0}[\widetilde{\phi}_{1,0}^{(N)},\cdots,\widetilde{\phi}_{N,0}^{(N)}]},
\end{align}
with the saddle point approximation.
From eq.~\eqref{eq:LIN} and the solution $\widetilde{\phi}^{(N)}_{N,0}=0$, we find $I_{\rm I}[\widetilde{\phi}_{1,0}^{(N)},\cdots,\widetilde{\phi}^{(N)}_{N,0}]=0$.
And then, from eq.~\eqref{eq:Wlambda_II}, we end up with
\begin{align}
    \left(\frac{\partial W^{(N)}_{\lambda}}{\partial \lambda}\right)_{\lambda=0}=0.
\end{align}

\section{Relation between mass bases and interaction bases of the fields}
\label{app:bases}
In Sec.~\ref{sec:quasi}, we performed calculations utilizing the interaction bases of the heavy field.
On the other hand, for instance, in Ref.~\cite{Kim:2021pbr}, the perturbative unitarity in the same theory as eq.~\eqref{eq:acBR} was assessed, focusing on the mass bases of the heavy field. 
Hence, we need to clarify the relation between mass bases and interaction bases of the fields to compare the results of this work with those of Ref.~\cite{Kim:2021pbr}.
First, consider the classical solution of eq.~\eqref{eq:acBR} presented as 
\begin{align}
\left(\frac{\dot{\theta}_0}{m_2}\right)^2&=\frac{1}{2}\Bigg[
2\left(1-\frac{r_{\rm min}}{r_0}\right)+3\tilde{\lambda}_3 \frac{r_{\rm min}}{r_0}\left(\frac{r_0}{r_{\rm min}}-1\right)^2+\tilde{\lambda}_4\frac{r_{\rm min}}{r_0}\left(\frac{r_0}{r_{\rm min}}-1\right)^3
\Bigg],\label{eq:thm2}
\end{align}
where $r_0:= r_{\rm min} +\tilde{\sigma}_0$ is defined with the solution $\tilde{\sigma}_0$, and $\phi=r_{\rm min}\theta$ is used. 
The mass of $\tilde{\sigma}$ in eq.~\eqref{eq:acBR} is expressed as $m^2:=\partial^2V_r/\tilde{\sigma}^2-\dot{\theta}_0^2$, and is rewritten as follows:
\begin{align}
\left(\frac{m}{m_2}\right)^2&=1+3\tilde{\lambda}_3 \left(-1+\frac{r_0}{r_{\rm min}}\right)+
\frac{3}{2}\tilde{\lambda}_4 \left(-1+\frac{r_0}{r_{\rm min}}\right)^2\notag
\\
&\quad\quad\quad\quad- \left[
1-\frac{r_{\rm min}}{r_0}+\frac{3}{2}\tilde{\lambda}_3 \frac{r_{\rm min}}{r_0}\left(-1+\frac{r_0}{r_{\rm min}}\right)^2
+\frac{1}{2}\tilde{\lambda}_4 \frac{r_{\rm min}}{r_0}\left(-1+\frac{r_0}{r_{\rm min}}\right)^3
\right].\label{eq:ratmm2}
\end{align}
Then, the mass basis of the field $\tilde{\sigma}$ is given by $\sigma:=r-r_0=r_{\rm min}-r_0+\tilde{\sigma}$, and eq.~\eqref{eq:Ubpo} can be expressed using the mass basis $\sigma$ as 
\begin{align}
    V_r(r)\supset \frac{m^2 r_0^2}{2}\left[
    \left(\frac{\sigma}{r_0}\right)^2
    +\lambda_3 \left(\frac{\sigma}{r_0}\right)^3
    +\frac{\lambda_4}{4} \left(\frac{\sigma}{r_0}\right)^4
    \right],\label{eq:pot_mass}
\end{align}
where the zero-th order of $\sigma$ is omitted, and $\lambda_3$ and $\lambda_4$ are defined as follows:
\begin{align}
\lambda_3&=  \left(\frac{m_2}{m}\right)^2\left(\frac{r_0}{r_{\rm min}}\right)\left[\tilde{\lambda}_3+\tilde{\lambda}_4 \left(-1+\frac{r_0}{r_{\rm min}}\right)\right],\label{eq:lam3def}
    \\
    \lambda_4&=\tilde{\lambda}_4 \left(\frac{m_2}{m}\right)^2\left(\frac{r_0}{r_{\rm min}}\right)^2.\label{eq:lam4def}
\end{align}
From eqs.~\eqref{eq:lam3def} and \eqref{eq:lam4def}, we obtain
\begin{align}
	\tilde{\lambda}_3&=\left(\frac{m}{m_2}\right)^2 \left[
	\lambda_3 \left(\frac{r_0}{r_{\rm min}}\right)^{-1}
	- \lambda_4 \left(\frac{r_0}{r_{\rm min}}\right)^{-2} \left(-1+\frac{r_0}{r_{\rm min}}\right)
	\right],\label{eq:lam3tilre}
	\\
	\tilde{\lambda}_4&=\lambda_4 \left(\frac{m}{m_2}\right)^2  \left(\frac{r_0}{r_{\rm min}}\right)^{-2}.\label{eq:lam4tilre}
\end{align}
Utilizing eqs.~\eqref{eq:lam3tilre} and \eqref{eq:lam4tilre}, eq.~\eqref{eq:ratmm2} is rewritten as  
\begin{align}
	\left(\frac{m_2}{m}\right)^2 &=\frac{1}{2} \left(\frac{r_0}{r_{\rm min}}\right)^{-2}
	\bigg[
	2  \left(\frac{r_0}{r_{\rm min}}\right)^{3}
	-3 \lambda_3  \left(\frac{r_0}{r_{\rm min}}\right) \left(
	-1+\left(\frac{r_0}{r_{\rm min}}\right)^2
	\right)\notag
	\\
	&\quad\quad\quad\quad\quad\quad\quad\quad\quad\quad\quad\quad\quad\quad\quad\quad+\lambda_4 \left(
	2-3 \left(\frac{r_0}{r_{\rm min}}\right)+\left(\frac{r_0}{r_{\rm min}}\right)^3
	\right)
	\bigg]\label{eq:memre}
	.
\end{align}
Upon substituting eqs.~\eqref{eq:lam3tilre}, \eqref{eq:lam4tilre}, and \eqref{eq:memre} into eq.~\eqref{eq:thm2}, we obtain
\begin{align}
 	\left(\frac{\dot{\theta}_0}{m}\right)^2=\frac{1}{2} \left(\frac{r_0}{r_{\rm min}}\right)^{-2} \left(-1+ \frac{r_0}{r_{\rm min}} \right)
	\left[
	\lambda_4 \left(-1+ \frac{r_0}{r_{\rm min}} \right)^2
	+\left(\frac{r_0}{r_{\rm min}}\right) \left(
	2 \left(\frac{r_0}{r_{\rm min}}\right)-3 \lambda_3 \left(-1+ \frac{r_0}{r_{\rm min}} \right)
	\right)
	\right].\label{eq:there}
\end{align}
From eq.~\eqref{eq:there}, $r_0/r_{\rm min}$ is expressed, up to the sixth order of $\dot{\theta}_0$, as
\begin{align}
\frac{r_0}{r_{\rm min}}&\simeq 1+\left(\frac{\dot{\theta}_0}{m}\right)^2 X^{(1)}+\left(\frac{\dot{\theta}_0}{m}\right)^4 X^{(2)}+\left(\frac{\dot{\theta}_0}{m}\right)^6 X^{(3)},\label{eq:r0rmin}
\end{align}
where we defined the coefficients as follows:
\begin{align}
    X^{(1)}&=1,
    \\
    X^{(2)}&=\frac{3}{2}\lambda_3,
    \\
    X^{(3)}&=\frac{1}{2}\left(
    -3 \lambda_3 +9 \lambda_3^2 -\lambda_4
    \right).
\end{align}
From eq.~\eqref{eq:memre}, $(m/m_2)^2$ is also given, up to the sixth order of $\dot{\theta}_0$, as
\begin{align}
    \left(\frac{m}{m_2}\right)^2&\simeq 1+\left(\frac{\dot{\theta}_0}{m}\right)^2 Y^{(1)}+\left(\frac{\dot{\theta}_0}{m}\right)^4 Y^{(2)}+\left(\frac{\dot{\theta}_0}{m}\right)^6 Y^{(3)},\label{eq:mm2}
\end{align}
where
\begin{align}
    Y^{(1)}&=-1+3 \lambda_3,
    \\
    Y^{(2)}&=1-9 \lambda_3 +\frac{27}{2}\lambda_3^2 -\frac{3}{2} \lambda_4,
    \\
    Y^{(3)}&=-1-\frac{135}{2}\lambda_3^2 +\frac{135}{2}\lambda_3^3 + 6\lambda_4 -3 \lambda_3 \left(-6+5 \lambda_4\right).
\end{align}
Upon substituting eqs.~\eqref{eq:r0rmin} and \eqref{eq:mm2} into eqs.~\eqref{eq:lam3tilre} and \eqref{eq:lam4tilre}, we obtain
\begin{align}
	\tilde{\lambda}_3&\simeq \lambda_3 + \left(\frac{\ dot{\theta}_0}{m}\right)^2 Z^{(1)} +\left(\frac{\dot{\theta}_0}{m}\right)^4 Z^{(2)}+\left(\frac{\dot{\theta}_0}{m}\right)^6 Z^{(3)},
	\\
	\tilde{\lambda}_4&\simeq \lambda_4 \left[
	1+  \left(\frac{\dot{\theta}_0}{m}\right)^2 W^{(1)}+\left(\frac{\dot{\theta}_0}{m}\right)^4 W^{(2)}+\left(\frac{\dot{\theta}_0}{m}\right)^6 W^{(3)}
	\right],
\end{align}
with
\begin{align}
	Z^{(1)}&=\lambda_3 \left(-2+3 \lambda_3\right)-\lambda_4,
	\\
	Z^{(2)}&=\frac{3}{2}\left(
	-9 \lambda_3^2 +9 \lambda_3^3 +\lambda_3 \left(2-4 \lambda_4\right)+2 \lambda_4
	\right),
	\\
	Z^{(3)}&=-90 \lambda_3^3 +\frac{135}{2}\lambda_3^4 +\lambda_3^2 \left(36-\frac{75}{2}\lambda_4\right)+2 \left(-3+\lambda_4\right)\lambda_4 +4 \lambda_3 \left(-1+8 \lambda_4 \right),
	\\
	W^{(1)}&=-3+3 \lambda_3,
	\\
	W^{(2)}&=\frac{3}{2}\left( 4-12 \lambda_3 +9 \lambda_3^2 -\lambda_4\right),
	\\
	W^{(3)}&=\frac{5}{2}\left(-2+3 \lambda_3\right)\left(2-9 \lambda_3 +9 \lambda_3^2 -2 \lambda_4 \right).
\end{align}
According to these, we will rewrite eq.~\eqref{eq:dmc3} using the parameter $\lambda_3$. 
Upon substituting eq.~\eqref{eq:twoscl_c2_c3} into eq.~\eqref{eq:dmc3}, we obtain
\begin{align}
	c_s^{-2}&=\frac{m_2^4 +6 m_2^2 \dot{\theta}_0^2 -15 (-1+\tilde{\lambda}_3)\dot{\theta}_0^4}{m_2^4 +2 m_2^2\dot{\theta}_0^2 -3 (-1+\tilde{\lambda}_3)\dot{\theta}_0^4 },
	\\
	\tilde{c}_3(c_s^{-2}-1)&=\frac{3(-1+\tilde{\lambda}_3)\dot{\theta}_0^2}{3(-1+\tilde{\lambda}_3) \dot{\theta}_0^2 -m_2^2}c_s^2 (c_s^{-2}-1).
\end{align}
Utilizing the formulae presented above, we also obtain
\begin{align}
	c_s^{-2}&=1+4 \left(\frac{\dot{\theta}_0}{m}\right)^2+\mathcal{O}(\dot{\theta}_0^6),
	\\
	\tilde{c}_3(c_s^{-2}-1)&=-\frac{3}{4}\left(-1+\lambda_3\right)c_s^2 (c_s^{-2}-1)^2+\mathcal{O}(\dot{\theta}_0^6).
\end{align}
This result is completely consistent with the results of Ref.~\cite{Kim:2021pbr}.

\clearpage
\bibliographystyle{JHEP}
\bibliography{entropy.bib}

\providecommand{\href}[2]{#2}\begingroup\raggedright\begin{thebibliography}{10}

\bibitem{Lee:1977eg}
B.~W. Lee, C.~Quigg, and H.~B. Thacker, {\it {Weak Interactions at Very
  High-Energies: The Role of the Higgs Boson Mass}},  {\em Phys. Rev. D} {\bf
  16} (1977) 1519.

\bibitem{PhysRevLett.38.883}
B.~W. Lee, C.~Quigg, and H.~B. Thacker, {\it Strength of weak interactions at
  very high energies and the higgs boson mass},  {\em Phys. Rev. Lett.} {\bf
  38} (Apr, 1977) 883--885.

\bibitem{PhysRevD.7.3111}
D.~A. Dicus and V.~S. Mathur, {\it Upper bounds on the values of masses in
  unified gauge theories},  {\em Phys. Rev. D} {\bf 7} (May, 1973) 3111--3114.

\bibitem{Chanowitz:1985hj}
M.~S. Chanowitz and M.~K. Gaillard, {\it {The TeV Physics of Strongly
  Interacting W's and Z's}},  {\em Nucl. Phys. B} {\bf 261} (1985) 379--431.

\bibitem{Adams:2006sv}
A.~Adams, N.~Arkani-Hamed, S.~Dubovsky, A.~Nicolis, and R.~Rattazzi, {\it
  {Causality, analyticity and an IR obstruction to UV completion}},  {\em JHEP}
  {\bf 10} (2006) 014, [\href{http://arxiv.org/abs/hep-th/0602178}{{\tt
  hep-th/0602178}}].

\bibitem{Corbett:2014ora}
T.~Corbett, O.~J.~P. \'Eboli, and M.~C. Gonzalez-Garcia, {\it {Unitarity
  Constraints on Dimension-Six Operators}},  {\em Phys. Rev. D} {\bf 91}
  (2015), no.~3 035014, [\href{http://arxiv.org/abs/1411.5026}{{\tt
  arXiv:1411.5026}}].

\bibitem{Chang:2019vez}
S.~Chang and M.~A. Luty, {\it {The Higgs Trilinear Coupling and the Scale of
  New Physics}},  {\em JHEP} {\bf 03} (2020) 140,
  [\href{http://arxiv.org/abs/1902.05556}{{\tt arXiv:1902.05556}}].

\bibitem{Remmen:2019cyz}
G.~N. Remmen and N.~L. Rodd, {\it {Consistency of the Standard Model Effective
  Field Theory}},  {\em JHEP} {\bf 12} (2019) 032,
  [\href{http://arxiv.org/abs/1908.09845}{{\tt arXiv:1908.09845}}].

\bibitem{Remmen:2020vts}
G.~N. Remmen and N.~L. Rodd, {\it {Flavor Constraints from Unitarity and
  Analyticity}},  {\em Phys. Rev. Lett.} {\bf 125} (2020), no.~8 081601,
  [\href{http://arxiv.org/abs/2004.02885}{{\tt arXiv:2004.02885}}]. [Erratum:
  Phys.Rev.Lett. 127, 149901 (2021)].

\bibitem{Remmen:2020uze}
G.~N. Remmen and N.~L. Rodd, {\it {Signs, spin, SMEFT: Sum rules at dimension
  six}},  {\em Phys. Rev. D} {\bf 105} (2022), no.~3 036006,
  [\href{http://arxiv.org/abs/2010.04723}{{\tt arXiv:2010.04723}}].

\bibitem{Pham:1985cr}
T.~N. Pham and T.~N. Truong, {\it {Evaluation of the Derivative Quartic Terms
  of the Meson Chiral Lagrangian From Forward Dispersion Relation}},  {\em
  Phys. Rev. D} {\bf 31} (1985) 3027.

\bibitem{Ananthanarayan:1994hf}
B.~Ananthanarayan, D.~Toublan, and G.~Wanders, {\it {Consistency of the chiral
  pion pion scattering amplitudes with axiomatic constraints}},  {\em Phys.
  Rev. D} {\bf 51} (1995) 1093--1100,
  [\href{http://arxiv.org/abs/hep-ph/9410302}{{\tt hep-ph/9410302}}].

\bibitem{Pennington:1994kc}
M.~R. Pennington and J.~Portoles, {\it {The Chiral Lagrangian parameters, l1,
  l2, are determined by the rho resonance}},  {\em Phys. Lett. B} {\bf 344}
  (1995) 399--406, [\href{http://arxiv.org/abs/hep-ph/9409426}{{\tt
  hep-ph/9409426}}].

\bibitem{Maldacena:2002vr}
J.~M. Maldacena, {\it {Non-Gaussian features of primordial fluctuations in
  single field inflationary models}},  {\em JHEP} {\bf 05} (2003) 013,
  [\href{http://arxiv.org/abs/astro-ph/0210603}{{\tt astro-ph/0210603}}].

\bibitem{Grall:2020tqc}
T.~Grall and S.~Melville, {\it {Inflation in motion: unitarity constraints in
  effective field theories with (spontaneously) broken Lorentz symmetry}},
  {\em JCAP} {\bf 09} (2020) 017, [\href{http://arxiv.org/abs/2005.02366}{{\tt
  arXiv:2005.02366}}].

\bibitem{Kim:2021pbr}
S.~Kim, T.~Noumi, K.~Takeuchi, and S.~Zhou, {\it {Perturbative unitarity in
  quasi-single field inflation}},  {\em JHEP} {\bf 07} (2021) 018,
  [\href{http://arxiv.org/abs/2102.04101}{{\tt arXiv:2102.04101}}].

\bibitem{Cheung:2007st}
C.~Cheung, P.~Creminelli, A.~L. Fitzpatrick, J.~Kaplan, and L.~Senatore, {\it
  {The Effective Field Theory of Inflation}},  {\em JHEP} {\bf 03} (2008) 014,
  [\href{http://arxiv.org/abs/0709.0293}{{\tt arXiv:0709.0293}}].

\bibitem{Baumann:2015nta}
D.~Baumann, D.~Green, H.~Lee, and R.~A. Porto, {\it {Signs of Analyticity in
  Single-Field Inflation}},  {\em Phys. Rev. D} {\bf 93} (2016), no.~2 023523,
  [\href{http://arxiv.org/abs/1502.07304}{{\tt arXiv:1502.07304}}].

\bibitem{Grall:2021xxm}
T.~Grall and S.~Melville, {\it {Positivity bounds without boosts: New
  constraints on low energy effective field theories from the UV}},  {\em Phys.
  Rev. D} {\bf 105} (2022), no.~12 L121301,
  [\href{http://arxiv.org/abs/2102.05683}{{\tt arXiv:2102.05683}}].

\bibitem{Freytsis:2022aho}
M.~Freytsis, S.~Kumar, G.~N. Remmen, and N.~L. Rodd, {\it {Multifield
  positivity bounds for inflation}},  {\em JHEP} {\bf 09} (2023) 041,
  [\href{http://arxiv.org/abs/2210.10791}{{\tt arXiv:2210.10791}}].

\bibitem{Cao:2022iqh}
Q.-H. Cao and D.~Ueda, {\it {Entropy constraints on effective field theory}},
  {\em Phys. Rev. D} {\bf 108} (2023), no.~2 025011,
  [\href{http://arxiv.org/abs/2201.00931}{{\tt arXiv:2201.00931}}].

\bibitem{Cao:2022ajt}
Q.-H. Cao, N.~Kan, and D.~Ueda, {\it {Effective field theory in light of
  relative entropy}},  {\em JHEP} {\bf 07} (2023) 111,
  [\href{http://arxiv.org/abs/2211.08065}{{\tt arXiv:2211.08065}}].

\bibitem{10.1214/aoms/1177729694}
S.~Kullback and R.~A. Leibler, {\it {On Information and Sufficiency}},  {\em
  The Annals of Mathematical Statistics} {\bf 22} (1951), no.~1 79 -- 86.

\bibitem{10.2996/kmj/1138844604}
H.~Umegaki, {\it {Conditional expectation in an operator algebra. IV. Entropy
  and information}},  {\em Kodai Mathematical Seminar Reports} {\bf 14} (1962),
  no.~2 59 -- 85.

\bibitem{RevModPhys.50.221}
A.~Wehrl, {\it General properties of entropy},  {\em Rev. Mod. Phys.} {\bf 50}
  (Apr, 1978) 221--260.

\bibitem{Vafa:2005ui}
C.~Vafa, {\it {The String landscape and the swampland}},
  \href{http://arxiv.org/abs/hep-th/0509212}{{\tt hep-th/0509212}}.

\bibitem{Arkani-Hamed:2006emk}
N.~Arkani-Hamed, L.~Motl, A.~Nicolis, and C.~Vafa, {\it {The String landscape,
  black holes and gravity as the weakest force}},  {\em JHEP} {\bf 06} (2007)
  060, [\href{http://arxiv.org/abs/hep-th/0601001}{{\tt hep-th/0601001}}].

\bibitem{Kats:2006xp}
Y.~Kats, L.~Motl, and M.~Padi, {\it {Higher-order corrections to mass-charge
  relation of extremal black holes}},  {\em JHEP} {\bf 12} (2007) 068,
  [\href{http://arxiv.org/abs/hep-th/0606100}{{\tt hep-th/0606100}}].

\bibitem{Cheung:2018cwt}
C.~Cheung, J.~Liu, and G.~N. Remmen, {\it {Proof of the Weak Gravity Conjecture
  from Black Hole Entropy}},  {\em JHEP} {\bf 10} (2018) 004,
  [\href{http://arxiv.org/abs/1801.08546}{{\tt arXiv:1801.08546}}].

\bibitem{Cheung:2019cwi}
C.~Cheung, J.~Liu, and G.~N. Remmen, {\it {Entropy Bounds on Effective Field
  Theory from Rotating Dyonic Black Holes}},  {\em Phys. Rev. D} {\bf 100}
  (2019), no.~4 046003, [\href{http://arxiv.org/abs/1903.09156}{{\tt
  arXiv:1903.09156}}].

\bibitem{Loges:2019jzs}
G.~J. Loges, T.~Noumi, and G.~Shiu, {\it {Thermodynamics of 4D Dilatonic Black
  Holes and the Weak Gravity Conjecture}},  {\em Phys. Rev. D} {\bf 102}
  (2020), no.~4 046010, [\href{http://arxiv.org/abs/1909.01352}{{\tt
  arXiv:1909.01352}}].

\bibitem{Reall:2019sah}
H.~S. Reall and J.~E. Santos, {\it {Higher derivative corrections to Kerr black
  hole thermodynamics}},  {\em JHEP} {\bf 04} (2019) 021,
  [\href{http://arxiv.org/abs/1901.11535}{{\tt arXiv:1901.11535}}].

\bibitem{Goon:2019faz}
G.~Goon and R.~Penco, {\it {Universal Relation between Corrections to Entropy
  and Extremality}},  {\em Phys. Rev. Lett.} {\bf 124} (2020), no.~10 101103,
  [\href{http://arxiv.org/abs/1909.05254}{{\tt arXiv:1909.05254}}].

\bibitem{Bellazzini:2019xts}
B.~Bellazzini, M.~Lewandowski, and J.~Serra, {\it {Positivity of Amplitudes,
  Weak Gravity Conjecture, and Modified Gravity}},  {\em Phys. Rev. Lett.} {\bf
  123} (2019), no.~25 251103, [\href{http://arxiv.org/abs/1902.03250}{{\tt
  arXiv:1902.03250}}].

\bibitem{Hamada:2018dde}
Y.~Hamada, T.~Noumi, and G.~Shiu, {\it {Weak Gravity Conjecture from Unitarity
  and Causality}},  {\em Phys. Rev. Lett.} {\bf 123} (2019), no.~5 051601,
  [\href{http://arxiv.org/abs/1810.03637}{{\tt arXiv:1810.03637}}].

\bibitem{Arkani-Hamed:2021ajd}
N.~Arkani-Hamed, Y.-t. Huang, J.-Y. Liu, and G.~N. Remmen, {\it {Causality,
  Unitarity, and the Weak Gravity Conjecture}},
  \href{http://arxiv.org/abs/2109.13937}{{\tt arXiv:2109.13937}}.

\bibitem{Bittar:2024xuc}
P.~Bittar, S.~Fichet, and L.~de~Souza, {\it {Gravity-Induced Photon
  Interactions and Infrared Consistency in any Dimensions}},
  \href{http://arxiv.org/abs/2404.07254}{{\tt arXiv:2404.07254}}.

\bibitem{Colas:2024ysu}
T.~Colas, J.~Grain, G.~Kaplanek, and V.~Vennin, {\it {In-in formalism for the
  entropy of quantum fields in curved spacetimes}},  {\em JCAP} {\bf 08} (2024)
  047, [\href{http://arxiv.org/abs/2406.17856}{{\tt arXiv:2406.17856}}].

\bibitem{Garriga:1999vw}
J.~Garriga and V.~F. Mukhanov, {\it {Perturbations in k-inflation}},  {\em
  Phys. Lett. B} {\bf 458} (1999) 219--225,
  [\href{http://arxiv.org/abs/hep-th/9904176}{{\tt hep-th/9904176}}].

\bibitem{Chen:2006nt}
X.~Chen, M.-x. Huang, S.~Kachru, and G.~Shiu, {\it {Observational signatures
  and non-Gaussianities of general single field inflation}},  {\em JCAP} {\bf
  01} (2007) 002, [\href{http://arxiv.org/abs/hep-th/0605045}{{\tt
  hep-th/0605045}}].

\bibitem{Planck:2019kim}
{\bf Planck} Collaboration, Y.~Akrami et~al., {\it {Planck 2018 results. IX.
  Constraints on primordial non-Gaussianity}},  {\em Astron. Astrophys.} {\bf
  641} (2020) A9, [\href{http://arxiv.org/abs/1905.05697}{{\tt
  arXiv:1905.05697}}].

\bibitem{Brahma:2022yxu}
S.~Brahma, A.~Berera, and J.~Calder\'on-Figueroa, {\it {Quantum corrections to
  the primordial tensor spectrum: open EFTs \& Markovian decoupling of UV
  modes}},  {\em JHEP} {\bf 08} (2022) 225,
  [\href{http://arxiv.org/abs/2206.05797}{{\tt arXiv:2206.05797}}].

\bibitem{Brahma:2024yor}
S.~Brahma, J.~Calder\'on-Figueroa, and X.~Luo, {\it {Time-convolutionless
  cosmological master equations: Late-time resummations and decoherence for
  non-local kernels}},  \href{http://arxiv.org/abs/2407.12091}{{\tt
  arXiv:2407.12091}}.

\end{thebibliography}\endgroup

\end{document}